\documentclass[a4paper,11pt]{article}
\pdfoutput=1

\usepackage{jcappub}
\usepackage{bm}
\usepackage{mathtools}
\usepackage{tikz}
\usepackage{placeins}
\usepackage{booktabs}
\usepackage[T1]{fontenc}
\usepackage{lmodern}

\usepackage{xcolor}
\usepackage{tcolorbox}
\usepackage{amsmath,amssymb}
\usepackage{enumitem}
\newtcolorbox{derivationbox}{
    colback=gray!10,
    colframe=gray!65,
    boxrule=0.8pt,
    arc=0pt,
    outer arc=0pt,
    left=4mm,
    right=4mm,
    top=3mm,
    bottom=3mm,
    before skip=10pt,
    after skip=10pt
}
\newtcolorbox{equationbox}{
    colback=gray!10,
    colframe=gray!65,
    boxrule=0.8pt,
    arc=0pt,
    outer arc=0pt,
    boxsep=0pt,
    left=4mm,
    right=4mm,
    top=3mm,
    bottom=3mm,
    before skip=10pt,
    after skip=10pt,
    before upper={
        \setlength{\abovedisplayskip}{0pt}
        \setlength{\abovedisplayshortskip}{0pt}
        \setlength{\belowdisplayskip}{0pt}
        \setlength{\belowdisplayshortskip}{0pt}
    }
}

\newcommand{\de}{{\rm d}}
\newcommand{\ii}{{\rm i}}
\newcommand{\ee}{{\rm e}}
\newcommand{\HeunG}{\operatorname{HeunG}}

\newcommand{\Disc}{\operatorname{Disc}}

\newcommand{\cF}{\mathcal F}
\newcommand{\cS}{\mathcal S}
\newcommand{\cI}{\mathcal I}

\newcommand{\bPhi}{\bm\Phi}
\newcommand{\bPsi}{\bm\Psi}

\title{\boldmath Strongly mixed cosmological collider at unequal sound speeds}

\author[a]{Javier Huenupi,}
\author[b]{Gonzalo A. Palma,}
\author[c]{and Spyros Sypsas}

\affiliation[a]{Center for Particle Cosmology, Department of Physics and Astronomy,\\ University of Pennsylvania, Philadelphia, PA 19104, USA}
\affiliation[b]{Departamento de F\'isica, FCFM, Universidad de Chile,\\ Blanco Encalada 2008, Santiago, Chile}
\affiliation[c]{Centro de Ciencias Exactas, Facultad de Ciencias,\\ Universidad del B\'io-B\'io, Chill\'an, Chile}

\emailAdd{huenupi1@sas.upenn.edu}
\emailAdd{gapalma@uchile.cl}
\emailAdd{ssypsas@ubiobio.cl}

\hypersetup{
  pdftitle={Strongly mixed cosmological colliders at unequal sound speeds},
  pdfauthor={Javier Huenupi, Gonzalo A. Palma and Spyros Sypsas}
}

\abstract{ We derive the canonically normalized modes of a two-field inflationary system with arbitrary constant quadratic mixing and unequal sound speeds $c_\varphi$ and $c_\sigma$ for the curvature and isocurvature perturbations, respectively. A Laplace representation reduces the coupled dynamics to a second-order Heun equation and gives the exact late-time curvature power spectrum for arbitrary values of the mixing parameter and entropy mass. The sound-speed ratio $r_c=c_\sigma/c_\varphi$ controls the separation between the two sound horizons. At fixed nonzero mixing and $r_c\ll1$, the power exhibits power-law enhancement, growth as $\ln^2(1/r_c)$, or bounded oscillations in $\ln r_c$, depending on the mass and mixing. For $r_c\gg1$, the correction to the unmixed spectrum vanishes as $\ln r_c/r_c^2$ at fixed mass and mixing. Using the same normalized modes, we compute the tree-level bispectrum for three representative cubic interactions, with the mixing parameter treated exactly. For heavy entropy fields, unequal speeds modify the collider amplitude and phase while preserving the strict squeezed frequency set by the final entropy mass. The power-normalized amplitude can vary nonmonotonically with $r_c$, and a sound-speed hierarchy can delay the approach to this asymptotic regime.  }

\keywords{Inflation, cosmological collider physics, primordial non-Gaussianity}

\begin{document}

\maketitle
\flushbottom


\section{Introduction}
\label{sec:introduction}

Cosmic inflation may have involved a particle spectrum extending far beyond the minimal picture of a single self-interacting quantum scalar field. Fluctuations of additional fields can leave imprints in primordial curvature correlators, with linear mixing in the equations of motion providing the most direct mechanism for transferring their effects to the curvature perturbation~\cite{GarciaBellido:1995metric,Gordon:2000entropy,Bartolo:2001oscillations,GrootNibbelink:2001qt,Starobinsky:2001multifield,Wands:2002transfer,Tsujikawa:2002correlations,DiMarco:2002entropy,Byrnes:2006curvature,Lalak:2007curvature,Langlois:2008multifield,Gao:2009crosscorrelations,Peterson:2010twofield,Cremonini:2010strong,Achucarro:2010da,Achucarro:2016cumulative,Achucarro:2018ngj,Palma:2020ejf,Koh:2026GaussBonnet,Kolb:2026xfp}. Together with the nonlinear interactions that typically accompany it, this mixing lies at the heart of the cosmological collider programme~\cite{Chen:2009zp,Chen:2009we,Chen:2012ge,Noumi:2012vr,Arkani-Hamed:2015bza,Lee:2016vti,Chen:2016uwp,Chakraborty:2023qbp,Aoki:2026Continuum,Sundrum:2026Chemical,Aoki:2020zbj,Pinol:2021aun,Werth:2026dcj}, with no general principle requiring it to be weak. The requirements of stability, unitarity, and effective-theory validity constrain the allowed parameter space while permitting regimes in which strong linear mixing coexists with perturbatively controlled nonlinear interactions~\cite{Gwyn:2012eft,Bjorkmo:2019perturbative,Pinol:2023oux}.

Most cosmological-collider calculations and observational searches have nevertheless treated the mixing as a small perturbation. In this approximation, the additional scalars propagate almost independently, while perturbative mixing insertions transfer their imprint to the curvature perturbation. For fixed masses and interaction operators, varying the mixing at leading order changes the overall amplitude of the corresponding correlation functions while leaving their momentum-space shapes fixed. At strong mixing, the fields must be treated as a coupled system, which can be solved for the mixed mode functions or used to evolve the coupled correlators directly~\cite{An:2017hlx,Iyer:2017qzw,Werth:2023pfl}. Both approaches have been implemented numerically to study collider signals beyond weak mixing~\cite{Pinol:2023oux} and to construct full-shape templates for Planck searches~\cite{Kumar:2026ogn,Kumar:2026dih,Philcox:2026Dissecting}.

The observational importance of going beyond weak mixing is illustrated in Ref.~\cite{Philcox:2026rpn}: once theoretical bounds on the interactions are imposed, weakly mixed tree-level scalar signals fall below the sensitivity of the Planck bispectrum analysis across most of the parameter space studied, apart from certain low-mass double- and triple-exchange configurations. Strong mixing can bring several interaction channels within observational reach, making a consistent treatment of coupled propagation and nonlinear interactions essential for predicting these signals. In particular, the mixing changes both the normalization of the curvature power spectrum and the momentum dependence of the bispectrum, including the amplitude and phase of its oscillatory features. The phase itself can help distinguish the underlying particle dynamics and interactions, providing information beyond the oscillation frequency~\cite{Qin:2022Phase}.

These questions also extend beyond the cosmic microwave background. The sensitivity of large-scale structure to primordial collider signals has motivated forecasts for future 21-cm surveys~\cite{Meerburg:2016zdz} and searches in galaxy surveys~\cite{Cabass:2024wob,Yang:2026mip}, while recent simulations have followed their oscillatory imprint into halo bias in the nonlinear universe~\cite{Anbajagane:2026HaloBias}. Relating such observables to the inflationary particle content requires predictions for the primordial amplitude and phase that retain the effects of coupled propagation.

Recent work has substantially advanced our analytic understanding of how coupled fluctuations propagate in the strong-mixing regime. In Ref.~\cite{Huenupi:2026abj}, we derived the exact coupled mode functions and primordial power spectrum without expanding in the mixing parameter. In Ref.~\cite{Huenupi:2026aqc}, we used these solutions to derive cosmological-collider bispectra in the same regime. Complementary analytic treatments of strong-mixing collider signals have also appeared~\cite{Pinol:2026xnl,Wang:2026lff,Belrhali:2026uxn,Pinol:2026hey}, including extensions to spinning fields~\cite{Gorji:2026rvx}. Our earlier solutions assumed that the curvature and entropy fluctuations share a common sound speed. However, when these speeds differ, the two fluctuations cross their respective sound horizons at different times, and the relative importance of gradient, mass, and mixing terms changes over the interval between the crossings. This separation affects both the transfer of power and the evolution of the fields during their nonlinear interactions. Consequently, fields with the same final entropy mass can produce different collider amplitudes and phases, making the propagation speeds an essential part of the relation between the primordial signal and the underlying particle dynamics.

Analytic studies of collider correlations with unequal propagation speeds have already shown how a sound-speed hierarchy can enhance primordial non-Gaussianity and change its momentum dependence~\cite{Pimentel:2022Boostless,Jazayeri:2022kjy,Jazayeri:2023xcj}. In particular, a curvature perturbation that propagates more slowly than the additional massive field can produce characteristic resonances in mildly squeezed configurations, giving rise to the low-speed collider~\cite{Jazayeri:2022kjy,Jazayeri:2023xcj}. These signals can remain large beyond weak mixing~\cite{Jazayeri:2023xcj}, motivating an analytic treatment that follows the coupled evolution for either ordering of the sound horizons.

In this article we relax the equal-speed assumption and obtain the canonically normalized modes and primordial power spectrum at arbitrary constant mixing. We then use these solutions to determine the nonanalytic squeezed bispectrum for three representative cubic interactions, allowing us to follow how the sound-speed hierarchy affects both the primordial power and the collider signals.

\subsection{Basic setup and main results}

We consider a two-field inflationary system in which a canonically normalized curvature fluctuation $\varphi$ mixes with an isocurvature fluctuation $\sigma$. The two fields propagate with positive sound speeds $c_\varphi$ and $c_\sigma$, which need not coincide. Working at leading order in the slow-roll parameters, we take the Hubble expansion rate $H$ to be constant, so that in conformal time the scale factor is $a(\tau)=-1/(H\tau)$. With these considerations our starting point is the following quadratic action:
\begin{equation}
S^{(2)}=\frac12\int\de\tau\,\de^3x\,
\frac{1}{H^2\tau^2}\bigg[
\left(\varphi'+\frac{\lambda}{\tau}\sigma\right)^2
-c_\varphi^2(\boldsymbol\nabla\varphi)^2
+\sigma'^2
-c_\sigma^2(\boldsymbol\nabla\sigma)^2
-\frac{\mu^2}{H^2\tau^2}\sigma^2
\bigg].
\label{eq:action}
\end{equation}
Here $\lambda$ is the dimensionless quadratic mixing strength, and $\mu$ is the entropy mass associated to $\sigma$. At this order, we also treat $\lambda$, $\mu$, and the sound speeds as constant. We focus on $\mu^2>0$, for which the entropy fluctuation decays and the curvature perturbation approaches a constant at late times. Such an action is well motivated within an effective field theory perspective~\cite{Cheung:2007st, Cespedes:2012hu, Achucarro:2012sm, Achucarro:2012decoupling, Cespedes:2013rda, Castillo:2013sfa, GarciaSaenz:2019cubic, Pinol:2024EFT,Sun:2026iqd}. Varying the action gives
\begin{align}
\left(\varphi'+\frac{\lambda}{\tau}\sigma\right)'
-\frac{2}{\tau}
\left(\varphi'+\frac{\lambda}{\tau}\sigma\right)
-c_\varphi^2\nabla^2\varphi&=0,
\label{eq:eom-phi-intro}
\\
\sigma''-\frac{2}{\tau}\sigma'
-c_\sigma^2\nabla^2\sigma
+\frac{\mu^2}{H^2\tau^2}\sigma
&=\frac{\lambda}{\tau}
\left(\varphi'+\frac{\lambda}{\tau}\sigma\right).
\label{eq:eom-sigma-intro}
\end{align}
The system remains coupled whenever $\lambda\neq0$, while departures from the equal-speed case are characterized by the sound-speed ratio
\begin{equation}
r_c\equiv\frac{c_\sigma}{c_\varphi}.
\label{eq:kappa-definition-intro}
\end{equation}
When $r_c\neq1$, the two gradient terms become small at different times. Their relative importance during the coupled evolution is therefore sensitive to the ordering and separation of the two sound horizons.

In Ref.~\cite{Huenupi:2026abj}, we first obtained the full solution of the equal-speed system by combining Eqs.~\eqref{eq:eom-phi-intro} and \eqref{eq:eom-sigma-intro} into a single fourth-order equation and solving it using algebraic methods. More recently, Belrhali, Poisson and Renaux-Petel provided an alternative derivation of these solutions by rewriting the fourth-order equation in Laplace space~\cite{Belrhali:2026uxn}, where the system reduces to a second-order hypergeometric equation. Here we formulate the unequal-speed problem in a Laplace representation~\cite{Belrhali:2026ygh}, where the coupled dynamics reduces to a single second-order equation that can be brought to the general Heun form~\cite{DLMF:Heun}, recovering the hypergeometric equation when $r_c=1$. Reconstructing the field modes as single integrals over plane waves and imposing the Bunch--Davies initial conditions and canonical normalization gives the exact modes in Eqs.~\eqref{eq:F-Laplace-tau} and \eqref{eq:S-Laplace-tau}.

To present our results, we define the dimensionless power spectrum $\Delta_\zeta(k)$ of the geometrical curvature perturbation $\zeta$ through
\begin{equation}
\langle\zeta_{\boldsymbol k}\zeta_{\boldsymbol k'}\rangle
=(2\pi)^3\delta^{(3)}(\boldsymbol k+\boldsymbol k')P_\zeta(k),
\qquad
P_\zeta(k)=\frac{2\pi^2}{k^3}\Delta_\zeta(k).
\label{eq:intro-power-definition}
\end{equation}
In general, the curvature perturbation $\zeta$ is related to $\varphi$ appearing in \eqref{eq:action} through a linear relation of the form
\begin{equation}
\varphi=F_\zeta\zeta,
\end{equation}
where $F_\zeta$ is a background normalization factor taken to be a constant at the order considered here. For the usual two-derivative adiabatic action coupled to Einstein gravity, the normalization is $F_\zeta=\sqrt{2\epsilon}\,M_{\rm Pl}/c_\varphi$, where $\epsilon\equiv-\dot H/H^2$ and $M_{\rm Pl}$ is the reduced Planck mass~\cite{Cheung:2007st,Pinol:2024EFT}. Neglecting slow-roll corrections it follows that, in the absence of mixing ($\lambda = 0$), the dimensionless power spectrum becomes~\cite{Garriga:1999vw}
\begin{equation}
\Delta_\zeta(k) = \Delta_0^{(c_\varphi)} , \qquad
\Delta_0^{(c_\varphi)}\equiv
\frac{H^2}{4\pi^2F_\zeta^2c_\varphi^3}.
\label{eq:unmixed-power}
\end{equation}
In the case where the mixing is turned on $\lambda \neq 0$, the exact late-time power spectrum can be written in terms of the local Heun function $\HeunG$, normalized to one at zero argument~\cite{DLMF:Heun}:
\begin{equation}
\begin{aligned}
\frac{\Delta_\zeta}{\Delta_0^{(c_\varphi)}}
={}&\frac{\left|\Gamma(\frac34-\frac\nu2)
\Gamma(\frac34+\frac\nu2)\right|^2}{\pi} \Bigg[
\left|\HeunG\!\left(1-r_c^{-2},Q_1;
\frac14-\frac\nu2,\frac14+\frac\nu2,1,\frac12;1\right)\right|^2\\
&\qquad+\frac{\lambda^2r_c}{4(r_c^2-1)^2}
\left|\HeunG\!\left(1-r_c^2,Q_2;
\frac54-\frac\nu2,\frac54+\frac\nu2,2,\frac12;1\right)\right|^2
\Bigg],
\end{aligned}
\label{eq:intro-power-Heun}
\end{equation}
where the parameters $\nu$, $Q_1$ and $Q_2$ are fixed by the mass, mixing, and sound-speed ratio:
\begin{equation}
\nu\equiv\sqrt{\frac94-\frac{\mu^2}{H^2}} , \qquad
Q_1=\frac{\lambda^2+(\nu^2-\frac14)(1-r_c^2)}{4r_c^2},
\qquad
Q_2=\frac{\lambda^2+6-2r_c^2}{4}.
\label{eq:intro-Heun-parameters}
\end{equation}
Equation~\eqref{eq:intro-power-Heun} applies to both light and heavy entropy fields with $\mu^2>0$, without expanding in the mixing $\lambda$ or the sound-speed ratio $r_c$. It recovers the unmixed spectrum \eqref{eq:unmixed-power} at $\lambda=0$ and approaches the equal-speed result continuously as $r_c\to1$, where the Heun representation degenerates and the spectrum can be expressed through gamma functions. 

The effects of the sound-speed hierarchy become particularly clear when the two speeds are widely separated. For $r_c\ll1$, the entropy fluctuation crosses its sound horizon first, and its subsequent conversion into curvature retains the evolution between the two gradient scales. At fixed $\lambda\ne0$ and $\mu$, with $\nu^2+\lambda^2>0$, the leading enhanced spectrum is
\begin{equation}
\frac{\Delta_\zeta}{\Delta_0^{(c_\varphi)}}
\sim\lambda^2\left|\mathcal C(\lambda,\nu)\right|^2
r_c^{-2\sqrt{\nu^2+\lambda^2}},
\qquad r_c\ll1,
\label{eq:intro-small-kappa}
\end{equation}
where $\mathcal C$ is the gamma-function coefficient given in Eq.~\eqref{eq:C-small-kappa}. This expression describes the leading total power once the transferred contribution dominates. The enhancement can occur even when the entropy fluctuation decays between the two horizon crossings, since its amplitude at the earlier crossing also depends on its propagation speed. At $\nu^2+\lambda^2=0$, the enhancement instead grows as $\ln^2(1/r_c)$, while negative values of this combination give bounded oscillations in $\ln r_c$.

In the opposite hierarchy, the entropy fluctuation propagates more rapidly than the curvature fluctuation, and its effect on the final curvature power becomes small. At fixed mass and mixing, we obtain
\begin{equation}
\frac{\Delta_\zeta}{\Delta_0^{(c_\varphi)}}
=1+\frac{\lambda^2}{r_c^2}
\left[2\ln r_c+\mathcal B(\nu)\right]+\cdots,
\qquad r_c\gg1,
\label{eq:intro-large-kappa}
\end{equation}
where the finite mass-dependent term $\mathcal B$ is given in Eq.~\eqref{eq:B-large-kappa}. The correction therefore vanishes as $\lambda^2\ln r_c/r_c^2$, despite the increasing separation between the two sound horizons. These limits, as well as the exact spectrum between them, are tested against independent numerical solutions of the coupled field equations.

The same normalized modes can be used to compute the bispectrum at tree level in the cubic interactions, without expanding in the quadratic mixing. We consider the three operators studied in Ref.~\cite{Huenupi:2026aqc}, with couplings $\alpha$, $\beta$, and $\gamma$. In terms of cosmic time, their field dependence is, respectively, $\sigma(\dot\varphi-\lambda H\sigma)^2$, $\sigma^2(\dot\varphi-\lambda H\sigma)$, and $\sigma^3$, with their precise normalization specified in Eq.~\eqref{eq:cubic-action-squeezed}. Normalizing the contribution $B_I$ of each interaction to the curvature power, we find that for squeezed isosceles configurations, with $q\equiv k_L/k_S\ll1$:
\begin{equation}
\frac{B_I^{\rm NA}(k_S,k_S,k_L)}{P_\zeta(k_S)P_\zeta(k_L)}
={} b_{I,+}\,q^{\frac32-\nu}
+b_{I,-}\,q^{\frac32+\nu} +\mathcal O\!\left(q^{7/2-\operatorname{Re}\nu}\right),
\label{eq:intro-squeezed-bispectrum}
\end{equation}
where $I=\alpha,\beta,\gamma$, and where the coefficients $b_{I,\pm}$ are specified explicitly in Eq.~\eqref{eq:dimensional-squeezed-coefficients}. At generic masses, the next term in each entropy branch carries two additional powers of $q$. The regular contribution, which starts at order $q^2$ in this normalization, is excluded by the superscript ${\rm NA}$. For a heavy entropy field, $\mu>3H/2$, write $\nu=\ii\rho$, with $\rho\equiv\sqrt{\mu^2/H^2-9/4}$. The two coefficients obey $b_{I,-}=b_{I,+}^*$, so the branches combine into the oscillatory signal
\begin{equation}
\frac{B_I^{\rm NA}(k_S,k_S,k_L)}{P_\zeta(k_S)P_\zeta(k_L)} 
=2|b_{I,+}|\,q^{3/2}
\cos\!\left[\rho\ln\!\left(\frac{k_S}{k_L}\right)
+\arg b_{I,+}\right]+\mathcal O(q^{7/2}).
\label{eq:intro-heavy-bispectrum}
\end{equation}
The same coefficient thus fixes both the amplitude and the phase, while the logarithmic frequency remains fixed by $\mu/H$. Together with the exact power spectrum in the denominator, these ingredients retain the dependence on the mass, quadratic mixing, sound-speed ratio, and the corresponding cubic coupling. 

\begin{figure}[t!]
\centering
\includegraphics[width=\textwidth]{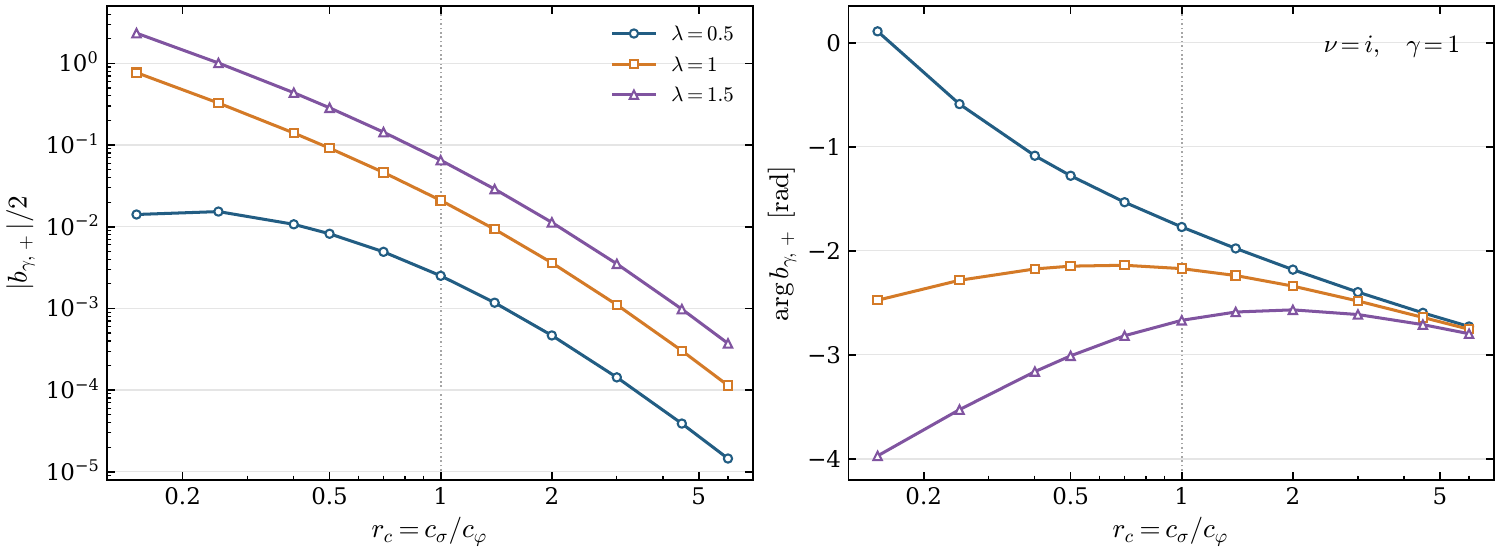}
\caption{Amplitude and phase of the squeezed signal from the $\sigma^3$ interaction as a function of $r_c=c_\sigma/c_\varphi$, at fixed $\gamma=1$ and $\nu=\ii$, for three quadratic mixing strengths. The left panel shows $|b_{\gamma,+}|/2$, while the coefficient of $q^{3/2}$ multiplying the cosine in Eq.~\eqref{eq:intro-heavy-bispectrum} is $2|b_{\gamma,+}|$. The right panel shows $\arg b_{\gamma,+}$, continued smoothly from its principal value at $r_c=1$. Each marker is an independent numerical evaluation, and the lines guide the eye; the vertical dotted line marks equal sound speeds. All curves have the same final entropy mass $\mu/H=\sqrt{13}/2$ and logarithmic frequency $\rho=1$, while their amplitudes and phases respond differently to the sound-speed hierarchy.}
\label{fig:gamma-sound-speed-dependence}
\end{figure}

Figure~\ref{fig:gamma-sound-speed-dependence} illustrates this dependence for the $\sigma^3$ interaction at fixed $\gamma=1$ and $\nu=\ii$, comparing three mixing strengths.
At $\lambda=1$, reducing $r_c$ from $2$ to $1/2$ increases the power-normalized oscillation amplitude by a factor of approximately $25$. The dependence need not be monotonic, as the weaker-mixing example with $\lambda=1/2$ exhibits a turnover. The phase also varies with the sound-speed ratio, although all three examples have the same logarithmic frequency. These results reflect the evolution during the interaction as well as the conversion into final curvature, so the hierarchy dependence of the collider amplitude cannot be inferred from the power spectrum alone. The distinction between the two stages of evolution also explains the different mass combinations appearing in these results. The combination $\sqrt{\nu^2+\lambda^2}$ controls the small-$r_c$ dependence accumulated between the two sound horizons, whereas $\nu$ controls the final entropy behavior and hence the strict squeezed exponents. The squeezed expressions above apply as $q\to0$ at fixed finite $r_c$, once the long mode has reached its late-time form during the interaction of the short modes. A large sound-speed hierarchy can delay the approach to this regime. 

\FloatBarrier

\subsection{Outline}

Section~\ref{sec:equal-speed} introduces the Laplace representation and obtains the normalized equal-speed solutions. Section~\ref{sec:early-time-propagation} explains how the early-time propagation changes when the sound speeds differ, and Section~\ref{sec:exact} constructs the normalized unequal-speed modes. Section~\ref{sec:power-exact} derives the exact power spectrum displayed in Eq.~\eqref{eq:intro-power-Heun}, while Section~\ref{sec:sound-speed-hierarchies} extracts its two sound-speed limits and explains their physical interpretation. Section~\ref{sec:squeezed} applies the normalized modes to the isosceles bispectrum and its squeezed limit, including the amplitude and phase illustrated in Fig.~\ref{fig:gamma-sound-speed-dependence}. Section~\ref{sec:discussion} concludes. Appendix~\ref{app:supporting-derivations} supplies the integral identity and limiting connection formulas entering the power spectrum, together with the long-mode coefficients, convergence arguments and numerical comparisons for the bispectrum.


\section{Equal sound speeds in the Laplace representation}
\label{sec:equal-speed}

In this section we revisit the equal-speed solutions obtained in Ref.~\cite{Huenupi:2026abj}, expressing the mode functions as integrals over plane waves. In this representation, the integration contours select the frequency sector, while the behavior near the spectral singularities determines the normalization and early-time evolution of the fields. The two positive-frequency solutions have distinct mixed polarizations, which provide a natural basis for the quantum-field expansion. Keeping the normalization and phases of our earlier works, we use these modes to describe the coupled evolution and obtain the curvature power spectrum.

\subsection{Linear dynamics in conformal time}
\label{sec:setup}

We begin this discussion by expanding the quantum fields in Fourier space in terms of two independent pairs of mode functions, labeled by $b=1,2$:
\begin{align}
\varphi(\boldsymbol x,\tau)
&=\sum_{b=1}^2\int_{\boldsymbol k}
\left[\varphi_b(k,\tau)\hat a_b(\boldsymbol k)
+\varphi_b^*(k,\tau)\hat a_b^\dagger(-\boldsymbol k)\right]
\ee^{\ii\boldsymbol k\cdot\boldsymbol x}, \label{eq:field-expansion-1}
\\
\sigma(\boldsymbol x,\tau)
&=\sum_{b=1}^2\int_{\boldsymbol k}
\left[\sigma_b(k,\tau)\hat a_b(\boldsymbol k)
+\sigma_b^*(k,\tau)\hat a_b^\dagger(-\boldsymbol k)\right]
\ee^{\ii\boldsymbol k\cdot\boldsymbol x},
\label{eq:field-expansion-2}
\end{align}
where $\int_{\boldsymbol k}\equiv\int\de^3k/(2\pi)^3$. Here $\varphi_b(k,\tau)$ and $\sigma_b(k,\tau)$ are the mode functions associated with the $b$th independent solution, while $\hat a_b$ and $\hat a_b^\dagger$ are annihilation and creation operators, respectively. These operators satisfy the standard commutation relation
\begin{equation}
[\hat a_a(\boldsymbol k),\hat a_b^\dagger(\boldsymbol k')]
=(2\pi)^3\delta_{ab}\delta^{(3)}(\boldsymbol k-\boldsymbol k'),
\end{equation}
with all other commutators vanishing. To obtain the equations obeyed by these modes, we insert Eqs.~\eqref{eq:field-expansion-1} and \eqref{eq:field-expansion-2} into Eqs.~\eqref{eq:eom-phi-intro} and \eqref{eq:eom-sigma-intro}. Suppressing the label $b$ and writing $\varphi_k$ and $\sigma_k$ for a mode with comoving wavenumber $k$, the mode equations take the form
\begin{align}
P_k'-\frac{2}{\tau}P_k+k^2c_\varphi^2\varphi_k&=0,
\label{eq:eom-P}
\\
\sigma_k''-\frac{2}{\tau}\sigma_k'
+\left(k^2c_\sigma^2+
\frac{\frac94-\nu^2}{\tau^2}\right)\sigma_k
&=\frac{\lambda}{\tau}P_k,
\label{eq:eom-sigma-tau}
\end{align}
where we have introduced the mixed velocity
\begin{equation}
P_k\equiv\varphi_k'+\frac{\lambda}{\tau}\sigma_k .
\label{eq:P-definition}
\end{equation}
Having written the evolution equations, we can now specify the normalization of the mode functions appearing in the field expansion. The canonical momenta following from Eq.~\eqref{eq:action} are $a^2P$ and $a^2\sigma'$, which give a conserved inner product between mode solutions. Arranging the fields as $\bPhi_b=(\varphi_b,\sigma_b)^T$ and leaving the common wavenumber implicit, we write this product as
\begin{equation}
(\bPhi_a,\bPhi_b)
\equiv \ii a^2\big[
\varphi_a^* P_b
- P_a^*\varphi_b +\sigma_a^*\sigma_b'-\sigma_a'^*\sigma_b\big].
\label{eq:symplectic-product}
\end{equation}
The Bunch--Davies condition selects the positive-frequency solutions as $\tau\to-\infty$, while canonical quantization requires $(\bPhi_a,\bPhi_b)=\delta_{ab}$ and $(\bPhi_a,\bPhi_b^*)=0$. Since the product is conserved, imposing these conditions in the asymptotic past fixes the normalization throughout the evolution. To make our phase convention explicit, let us consider the unmixed limit $\lambda=0$. The two independent positive-frequency modes are then $(u_\varphi,0)^T$ and $(0,u_\sigma)^T$, with canonically normalized components given by
\begin{align}
u_\varphi(k,\tau)
&=\frac{\ii H}{\sqrt{2c_\varphi^3k^3}}
\left(1+\ii kc_\varphi\tau\right)
\ee^{-\ii kc_\varphi\tau},
\label{eq:u-phi}
\\
u_\sigma(k,\tau)
&=-\frac{H}{\sqrt{2k^3}}\sqrt{\frac\pi2}\,
\ee^{\ii\theta_\nu}(-k\tau)^{3/2}
H_\nu^{(1)}(-kc_\sigma\tau),
\label{eq:u-sigma}
\end{align}
where $\theta_\nu=\frac\pi2(\nu-\frac32)$. In the asymptotic past, these expressions approach
\begin{equation}
u_\varphi\sim
\frac{H(-\tau)}{\sqrt{2kc_\varphi}}
\ee^{-\ii kc_\varphi\tau},
\qquad
u_\sigma\sim
\frac{H(-\tau)}{\sqrt{2kc_\sigma}}
\ee^{-\ii kc_\sigma\tau}.
\label{eq:free-UV}
\end{equation}
The plane-wave factors fix our positive-frequency convention. Once the mixing is included, the remaining time dependence must be obtained from the coupled equations, while the amplitudes are constrained by the conserved inner product in Eq.~\eqref{eq:symplectic-product}.

\subsection{Differential equation in Laplace space}
\label{sec:equal-transform}
\label{sec:equal-kernel}
\label{sec:equal-Laplace-comparison}

We now proceed to solve Eqs.~\eqref{eq:eom-P} and \eqref{eq:eom-sigma-tau} for $r_c=1$, that is, taking the two sound speeds to be equal:
\begin{equation}
c_\varphi=c_\sigma\equiv c>0.
\label{eq:equal-speed}
\end{equation}
As a first step, notice that for $\lambda\neq0$, both fields $\varphi_k$ and $\sigma_k$ can be reconstructed once $P_k$ is known. Indeed, suppressing the mode subscript $k$, we first use Eq.~\eqref{eq:eom-P} to write the curvature fluctuation as
\begin{equation}
\varphi=\frac{1}{k^2c^2}
\left(\frac{2P}{\tau}-P'\right),
\label{eq:F-from-P}
\end{equation}
while the definition of $P$ gives the entropy fluctuation as
\begin{equation}
\sigma=\frac{\tau}{\lambda}(P-\varphi').
\label{eq:S-from-P}
\end{equation}
These two relations also reveal a useful way of organizing the remaining time dependence. Differentiating Eq.~\eqref{eq:F-from-P} and substituting the result into Eq.~\eqref{eq:S-from-P}, we obtain
\begin{equation}
\left(\partial_\tau^2+k^2c^2\right)
\left(\frac{P}{\tau}\right)
=\lambda k^2c^2\frac{\sigma}{\tau^2}.
\label{eq:equal-P-sigma-wave}
\end{equation}
The operator acting on $P/\tau$ now has constant coefficients, with $\sigma/\tau^2$ playing the role of a source. This suggests expressing $\sigma/\tau^2$ as a superposition of plane waves, on which the operator acts by multiplication. We can then obtain the contribution of each wave to $P/\tau$ by dividing by the corresponding factor. To implement this construction, we introduce the Laplace representation through the contour integral
\begin{equation}
\cI_{\Gamma,k}[K](\tau)
\equiv\frac{1}{2\pi\ii}
\int_\Gamma\de u\,
\ee^{-\ii kc\tau u}K(u).
\label{eq:I-definition}
\end{equation}
Here $u$ is a dimensionless spectral variable, $K(u)$ is the kernel of the transform, and $\Gamma$ is an oriented, piecewise smooth contour in the complex $u$-plane, chosen independently of $\tau$. An admissible contour avoids the singularities of the kernel and ensures convergence of the integral and its required time derivatives. Integration by parts also requires the endpoint contribution $[\ee^{-\ii kc\tau u}K(u)]_{\partial\Gamma}$ to vanish. Under these conditions, differentiating with respect to $\tau$ and integrating by parts with respect to $u$, respectively, gives the identities
\begin{equation}
\partial_\tau\cI_{\Gamma,k}[K]
=-\ii kc\cI_{\Gamma,k}[uK],
\qquad
\tau \cI_{\Gamma,k}[K]
=-\frac{\ii}{kc}\cI_{\Gamma,k}[K'].
\label{eq:I-identities}
\end{equation}
These relations allow us to exchange time derivatives and explicit powers of $\tau$ for operations on the spectral kernel. Returning to the mode label $b$, we denote the kernel associated with each entropy mode by $Y_b(u)$ and write
\begin{equation}
\sigma_b(k,\tau)=\cS_{\Gamma,k}[Y_b](\tau),
\label{eq:sigma_b-Y_b}
\end{equation}
where the reconstruction operator $\cS_{\Gamma,k}$ includes the explicit time dependence suggested by Eq.~\eqref{eq:equal-P-sigma-wave}:
\begin{equation}
\cS_{\Gamma,k}[Y](\tau)
\equiv -\frac{2\ii N_k\,k^2c^2\tau^2}{\lambda}
\cI_{\Gamma,k}[Y](\tau).
\label{eq:S-Laplace}
\end{equation}
Here $N_k\equiv H/\sqrt{2c^3k^3}$ carries the free curvature normalization. To reconstruct $P$, notice that the operator in Eq.~\eqref{eq:equal-P-sigma-wave} acts on each plane wave as
\begin{equation}
\left(\partial_\tau^2+k^2c^2\right)\ee^{-\ii kc\tau u}
=k^2c^2(1-u^2)\ee^{-\ii kc\tau u}.
\label{eq:equal-wave-eigenvalue}
\end{equation}
With the entropy mode written as in Eq.~\eqref{eq:S-Laplace}, we can therefore satisfy Eq.~\eqref{eq:equal-P-sigma-wave} by taking $P_b(k,\tau)={\cal P}_{\Gamma,k}[Y_b](\tau)$, where
\begin{equation}
{\cal P}_{\Gamma,k}[Y](\tau)
\equiv -2\ii N_k\,k^2c^2\tau\,
\cI_{\Gamma,k}\!\left[\frac{Y(u)}{1-u^2}\right].
\label{eq:P-general}
\end{equation}
Substituting this into Eq.~\eqref{eq:F-from-P} then determines the curvature mode through
\begin{equation}
\varphi_b(k,\tau)=\cF_{\Gamma,k}[Y_b](\tau),
\label{eq:phi_b-Y_b}
\end{equation}
where the curvature reconstruction operator is
\begin{equation}
\cF_{\Gamma,k}[Y](\tau)
\equiv -2\ii N_k\,
\cI_{\Gamma,k}\!\left[
\frac{1+\ii kc\tau u}{1-u^2}Y(u)
\right].
\label{eq:F-Laplace}
\end{equation}

To complete the solution, we substitute Eqs.~\eqref{eq:S-Laplace} and \eqref{eq:P-general} into the entropy equation, Eq.~\eqref{eq:eom-sigma-tau}. The time derivatives act on the explicit factor $\tau^2$ and on the plane waves, while the identities in Eq.~\eqref{eq:I-identities} allow us to remove the remaining powers of $\tau$ by integrating by parts with respect to $u$. After two such integrations, the entropy equation is satisfied provided that $Y$ obeys the single second-order equation
\begin{equation}
\frac{\de}{\de u}\left[(1-u^2)Y'(u)\right]
+\left[\nu^2-\frac14+
\frac{\lambda^2}{1-u^2}\right]Y(u)=0.
\label{eq:Y-equation-equal}
\end{equation}
This is the associated Legendre equation, with degree $\nu-\frac12$ and orders $\pm\ii\lambda$. Its regular singular points lie at $u=1$, $u=-1$, and infinity, identifying the locations where the kernel may develop branches.

\subsection{A basis of solutions in Laplace space}

To understand the solutions of Eq.~\eqref{eq:Y-equation-equal}, let us first examine their leading behavior near the finite singular points. Near $u=1$, the two independent local solutions are proportional to $(1-u)^{\pm\ii\lambda/2}$, while near $u=-1$ they are proportional to $(1+u)^{\pm\ii\lambda/2}$, in each case multiplied by a function analytic at the corresponding endpoint. For $\lambda\neq0$, we may choose a basis that selects one of these powers at $u=1$:
\begin{equation}
Y^{(w)}(u)
=\frac{\ee^{\pi w\lambda/4}}{\sqrt2}\,
\Gamma\!\left(1-\frac{\ii w\lambda}{2}\right)
\Gamma(1+\ii w\lambda)
P_{\nu-\frac12}^{-\ii w\lambda}(u),
\qquad w=\pm1.
\label{eq:equal-Y-solution}
\end{equation}
Here $P_\ell^m(u)$ denotes the analytic continuation of the Ferrers function of the first kind from the interval $-1<u<1$~\cite{DLMF:Legendre}. The constant prefactor has been included to match the normalization and phase convention of the field modes in Refs.~\cite{Huenupi:2026abj,Huenupi:2026aqc}.

Let us now analyze the structure of these solutions more carefully. For this basis, the behavior near the positive endpoint follows from the expansion
\begin{equation}
P_{\nu-\frac12}^{-\ii w\lambda}(u)
=\frac{1}{\Gamma(1+\ii w\lambda)}
\left(\frac{1-u}{2}\right)^{\frac{\ii w\lambda}{2}}
\bigl[1+\mathcal O(1-u)\bigr].
\label{eq:equal-legendre-positive-endpoint}
\end{equation}
The two values of $w$ therefore select the two independent local powers allowed by the differential equation. For real nonzero $\lambda$, continuing a solution once counterclockwise around $u=1$ multiplies it by $\ee^{-\pi w\lambda}$. This point is therefore a branch point, and evaluating the Laplace integral requires a definite choice of continuation around it. To continue, expanding $P_{\nu-\frac12}^{-\ii w\lambda}(u)$ around $u=-1$ one finds
\begin{equation}
\begin{aligned}
P_{\nu-\frac12}^{-\ii w\lambda}(u)
={}&\frac{\Gamma(\ii w\lambda)}
{\Gamma(\frac12-\nu+\ii w\lambda)
\Gamma(\frac12+\nu+\ii w\lambda)}
\left(\frac{1+u}{2}\right)^{-\frac{\ii w\lambda}{2}}
\bigl[1+\mathcal O(1+u)\bigr]
\\
&+\frac{\Gamma(-\ii w\lambda)}
{\Gamma(\frac12-\nu)\Gamma(\frac12+\nu)}
\left(\frac{1+u}{2}\right)^{\frac{\ii w\lambda}{2}}
\bigl[1+\mathcal O(1+u)\bigr].
\end{aligned}
\label{eq:equal-legendre-negative-endpoint}
\end{equation}
Equation~\eqref{eq:equal-legendre-negative-endpoint} shows how the basis chosen at the positive endpoint is connected to the local basis at the negative endpoint, with the relative weights determined by the mass and mixing. In particular, the label $w$ specifies a single local power at $u=1$, while its continuation to $u=-1$ is generally a superposition of the two allowed powers. The remaining regular singular point is infinity. Keeping the leading terms of Eq.~\eqref{eq:Y-equation-equal} at large $|u|$ gives the two characteristic powers $u^{-\frac12+\nu}$ and $u^{-\frac12-\nu}$. At parameter values where the asymptotic solutions resonate, logarithms can accompany these powers; for example, at $\nu=0$ the independent behaviors are $u^{-1/2}$ and $u^{-1/2}\ln u$. Consequently, the kernels have at most algebraic growth, possibly multiplied by logarithms, along rays on a fixed sheet. Their contribution at infinity is therefore suppressed wherever the plane-wave factor in the Laplace integral is exponentially damped.

We can now specify the branches by taking the two cuts to extend vertically upwards from $u=1$ and $u=-1$, continuing the Legendre functions from the real interval $-1<u<1$, where both $1-u$ and $1+u$ are positive. The logarithms entering the local powers are accordingly chosen with
\begin{equation}
-\frac{\pi}{2}<\arg(1-u)<\frac{3\pi}{2},
\qquad
-\frac{3\pi}{2}<\arg(1+u)<\frac{\pi}{2}.
\label{eq:equal-legendre-branch-convention}
\end{equation}
These choices fix the values on both sides of each upward cut. To see how they are related, consider the positive cut, parametrized by $u=1+\ii s$ with $s>0$. Approaching it from the right gives $\arg(1-u)=3\pi/2$, whereas approaching from the left gives $\arg(1-u)=-\pi/2$. Since $Y^{(w)}$ contains a single local branch at $u=1$, the difference of $2\pi$ between these arguments implies
\begin{equation}
\lim_{\delta\to0^+}Y^{(w)}(1+\delta+\ii s)
=\ee^{-\pi w\lambda}
\lim_{\delta\to0^+}Y^{(w)}(1-\delta+\ii s),
\qquad s>0.
\label{eq:equal-legendre-positive-cut}
\end{equation}
At the negative cut, $u=-1+\ii s$, the corresponding arguments are $\arg(1+u)=\pi/2$ on the right and $\arg(1+u)=-3\pi/2$ on the left. The two local powers $(1+u)^{\pm\ii w\lambda/2}$ therefore acquire factors $\ee^{\mp\pi w\lambda}$ when their right-hand boundary values are compared with their left-hand values. Because a fixed-$w$ solution generally contains both powers in Eq.~\eqref{eq:equal-legendre-negative-endpoint}, its discontinuity across this cut involves both local branches with different continuation factors.

\subsection{Choosing the contour integrals}

\begin{figure}[tbp]
\centering
\includegraphics[width=\textwidth]{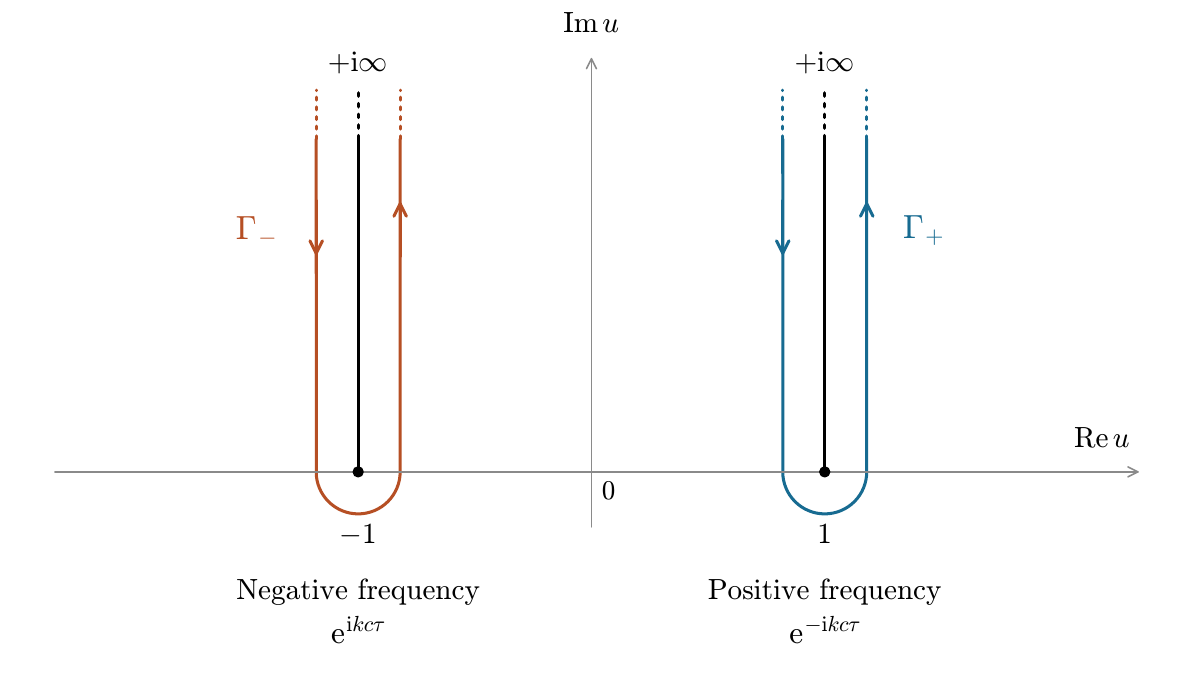}
\caption{Equal-speed spectral contours in the complex $u$-plane. The cuts extend vertically upwards from $u=1$ and $u=-1$, and the Hankel contours $\Gamma_+$ and $\Gamma_-$ turn counterclockwise around the corresponding branch points. The kernel is analytically continued along each contour, whose two infinite ends generally lie on different sheets. For real $\tau<0$, the exponential $\exp(-\ii kc\tau u)$ decays along both upward tails. The endpoints supply the early-time factors $\exp(-\ii kc\tau)$ and $\exp(\ii kc\tau)$, selecting the positive- and negative-frequency sectors, respectively.}
\label{fig:equal-laplace-contours}
\end{figure}

With the branches specified, we return to the plane waves in Eq.~\eqref{eq:I-definition} and choose a Hankel contour around each upward cut. Each contour approaches the branch point from positive imaginary infinity along one side of the cut, turns counterclockwise around it, and returns to positive imaginary infinity along the other side. The kernel is continued along the entire path, so the two ends generally lie on different sheets of its Riemann surface. We denote these contours by $\Gamma_+$ and $\Gamma_-$ according to whether they surround the cut originating at $u=1$ or $u=-1$, as illustrated in Fig.~\ref{fig:equal-laplace-contours}. They surround different branch points and cannot be deformed into one another while avoiding the singularities and preserving the analytic continuation of the kernel.
The advantage of this choice follows directly from the exponential in the reconstruction integrals. Along the cuts, parametrized by $u=\pm1+\ii s$ with $s\geq0$, we have
\begin{equation}
\ee^{-\ii kc\tau u}
=\ee^{\mp\ii kc\tau}\ee^{kc\tau s}.
\label{eq:equal-vertical-plane-waves}
\end{equation}
For real $\tau<0$, the second factor decays exponentially as $s\to+\infty$. Since the spectral kernels grow at most algebraically, up to possible logarithms, this ensures convergence at both infinite ends of either contour and suppresses the boundary terms required by integration by parts. The same damping also allows the time derivatives entering the field equations to be taken under the integral.

We can also express these integrals directly in terms of the discontinuities across the cuts, using the same parametrization to combine the contributions from their two sides. A counterclockwise Hankel contour approaches the branch point along the left side and returns to infinity along the right side, so the corresponding oriented discontinuity is
\begin{equation}
\Disc_{\Gamma_\pm}K(\pm1+\ii s)
\equiv\lim_{\delta\to0^+}
\left[
K(\pm1+\delta+\ii s)
-K(\pm1-\delta+\ii s)
\right].
\label{eq:oriented-discontinuity}
\end{equation}
Using $\de u=\ii\,\de s$, the contour representation becomes
\begin{equation}
\cI_{\Gamma_\pm,k}[K](\tau)
=\frac{\ee^{\mp\ii kc\tau}}{2\pi}
\int_0^\infty\de s\,
\ee^{kc\tau s}
\Disc_{\Gamma_\pm}K(\pm1+\ii s).
\label{eq:discontinuity-transform}
\end{equation}
The factor of $\ii$ from the change of variable cancels the one in the normalization of the contour integral. For real $\tau<0$, the remaining exponential is a decaying Laplace weight with positive parameter $-kc\tau$, while the factor outside the integral carries the frequency associated with the endpoint.\footnote{Written in this form, the integral also makes direct contact with the representation used in Ref.~\cite{Belrhali:2026uxn}. Their spectral variable coincides with $u$, and their time variable is $z_{\rm L}=kc\tau$; the kernel appearing in their real-ray integrals corresponds to the discontinuity density $(2\pi\ii)^{-1} \Disc_{\Gamma_+} K(u)$, with the overall factors accounted for in the field reconstruction and normalization. Their mass and mixing parameters are related to ours by
$$\widehat\rho=-\lambda,
\qquad
\widehat m^{\,2}=\frac{\mu^2}{H^2}-\lambda^2,
\qquad
\nu_{\rm eff}=\nu.
$$
With this dictionary, the discontinuity of $Y^{(w)}$ obeys their second-order kernel equation. The contour and real-ray expressions therefore describe the same equal-speed solutions, with the contour form keeping the analytic kernel and its branch structure explicit.}

For $K=Y^{(w)}$ on the positive cut, Eq.~\eqref{eq:equal-legendre-positive-cut} gives the discontinuity as the left-hand boundary value multiplied by $\ee^{-\pi w\lambda}-1$. The local powers have zero real part, so the integral of $Y^{(w)}$ converges at either endpoint and the small connecting arc can be shrunk to zero. Reconstruction kernels containing $Y/(1-u^2)$ have an additional endpoint pole, and their separate integrals along the two sides of a cut require the prescription inherited from the complete Hankel contour. For those kernels, Eq.~\eqref{eq:discontinuity-transform} is understood by analytic continuation from a domain where the endpoint integrals converge, including the contribution of the connecting arc.

\subsection{Early-time behavior}

We can use Eq.~\eqref{eq:equal-vertical-plane-waves} to understand how the endpoints determine the early-time behavior. As $\tau\to-\infty$, the damping confines the contribution along each cut to distances $s\sim(-kc\tau)^{-1}$ from its branch point, so the leading asymptotics depend only on the local form of the kernel. For $\Gamma_+$, the behavior $Y^{(w)}(u)\propto(1-u)^{\ii w\lambda/2}$ becomes a constant times $s^{\ii w\lambda/2}$ on either side of the cut, with the constants related by Eq.~\eqref{eq:equal-legendre-positive-cut}. The remaining time dependence follows from
\begin{equation}
\int_0^\infty\de s\,
\ee^{kc\tau s}s^{\ii w\lambda/2}
=
\Gamma\!\left(1+\frac{\ii w\lambda}{2}\right)
(-kc\tau)^{-1-\ii w\lambda/2}.
\label{eq:equal-endpoint-Laplace-integral}
\end{equation}
The endpoint location therefore supplies the plane-wave factor, while its local exponent supplies the logarithmic phase. In particular, apart from a time-independent prefactor, the leading positive-frequency transform behaves as
\begin{equation}
\cI_{\Gamma_+,k}[Y^{(w)}](\tau)
\propto
\frac{\ee^{-\ii kc\tau}}{-kc\tau}
\exp\!\left[-\frac{\ii w\lambda}{2}\ln(-kc\tau)\right],
\qquad \tau\to-\infty.
\label{eq:equal-transform-early-time}
\end{equation}
The reconstruction operators change the algebraic envelope but preserve these phases, giving both fields a leading dependence proportional to $(-kc\tau)\ee^{-\ii kc\tau}\exp[-\ii w\lambda\ln(-kc\tau)/2]$. The contour $\Gamma_-$ instead supplies the factor $\ee^{\ii kc\tau}$ and gives the negative-frequency sector. A basis of this sector can equivalently be obtained by complex conjugating the reconstructed positive-frequency modes. As Eq.~\eqref{eq:equal-legendre-negative-endpoint} shows, a fixed-$w$ kernel generally contains both local powers at the negative endpoint, but the two kernels together span the two independent negative-frequency solutions.

\subsection{Normalized mode functions and the vacuum basis}
\label{sec:equal-normalized-modes}

We now reconstruct the field modes and determine their normalization from the conserved product. To reconstruct the Bunch-Davis mode functions $\varphi_b$ and $\sigma_b$ out of the basis functions $Y^{(w)}(u)$ we first write
\begin{equation}
Y_b(u)=\sum_{w=\pm1}A_b^{(w)}Y^{(w)}(u),\qquad b=1,2.
\label{eq:equal-kernel-mode-expansion}
\end{equation}
Here the coefficients $A_b^{(w)}$ are coefficients independent of $u$ and $\tau$ (which we fix below) that specify how the two basis solutions are combined into the modes labeled by $b$. Then, $Y_b(u)$ gives back the entropy mode directly through Eq.~\eqref{eq:S-Laplace}. For the curvature mode, it is useful to make the behavior near $u=1$ explicit by rearranging the numerator in Eq.~\eqref{eq:F-Laplace}, which gives the equivalent expression
\begin{equation}
\cF_{\Gamma,k}[Y]
=N_k\bigg[
2kc\tau\cI_{\Gamma,k}\!\left[\frac{Y}{1-u}\right]
-2\ii(1-\ii kc\tau)
\cI_{\Gamma,k}\!\left[\frac{Y}{1-u^2}\right]
\bigg].
\label{eq:F-Laplace-threshold}
\end{equation}
Keeping the explicit prefactor in Eq.~\eqref{eq:equal-Y-solution}, the same endpoint calculation used above yields
\begin{equation}
\cI_{\Gamma_+,k}\!\left[\frac{Y^{(w)}}{1-u}\right]
\underset{-kc\tau\to\infty}{\sim}
-\frac{\ee^{\pi w\lambda/4-\ii kc\tau}}{\sqrt2}
\left(2\ii kc\tau\right)^{-\frac{\ii w\lambda}{2}}.
\label{eq:equal-endpoint-asymptotic}
\end{equation}
The endpoint prescription is the one specified after Eq.~\eqref{eq:discontinuity-transform}. At real negative conformal time, $\arg(2\ii kc\tau)=-\pi/2$, so the power appearing in this expression is
\begin{equation}
\left(2\ii kc\tau\right)^{-\frac{\ii w\lambda}{2}}
=\ee^{-\pi w\lambda/4}
\ee^{-\frac{\ii w\lambda}{2}\ln(-2kc\tau)}.
\label{eq:equal-endpoint-phase}
\end{equation}
The first factor cancels the real exponential in Eq.~\eqref{eq:equal-endpoint-asymptotic}, leaving a phase of unit modulus and the factor $1/\sqrt2$. Near the same endpoint, the second integral in Eq.~\eqref{eq:F-Laplace-threshold} is one half of the first at leading order. Together with the entropy reconstruction, these relations determine both the amplitude and the relative polarization of the two fields.

Following the notation of our earlier solutions, we write the resulting field doublets as
\begin{equation}
\bPsi_U^{(w)}(k,\tau)
\equiv
\begin{pmatrix}
\cF_U^{(w)}(k,\tau)\\[1mm]
\cS_U^{(w)}(k,\tau)
\end{pmatrix}
=
\begin{pmatrix}
\cF_{\Gamma_+,k}[Y^{(w)}](\tau)\\[1mm]
\cS_{\Gamma_+,k}[Y^{(w)}](\tau)
\end{pmatrix},
\qquad w=\pm1.
\label{eq:equal-mode-doublet}
\end{equation}
Their early-time behavior is
\begin{equation}
\bPsi_U^{(w)}(k,\tau)
\underset{-kc\tau\to\infty}{\sim}
\frac{H(-\tau)\ee^{-\ii kc\tau}}{\sqrt{2kc}}
\frac{\ee^{-\frac{\ii w\lambda}{2}\ln(-2kc\tau)}}{\sqrt2}
\begin{pmatrix}1\\ \ii w\end{pmatrix}.
\label{eq:equal-mode-UV}
\end{equation}
The leading oscillation has positive frequency $kc$, while the logarithm contributes a mixing-dependent phase. Although the mixing terms decay as $1/\tau$ in the asymptotic equations, their accumulated effect is logarithmic and must be retained in the modes. This phase leaves the leading amplitude unchanged, and its constant part, including the factor of $2$ inside the logarithm, agrees with the convention of Refs.~\cite{Huenupi:2026abj,Huenupi:2026aqc}.

The two polarization vectors in Eq.~\eqref{eq:equal-mode-UV} are orthonormal. Evaluating the conserved product in the asymptotic past therefore gives
\begin{equation}
\left(\bPsi_U^{(w)},\bPsi_U^{(w')}\right)=\delta_{ww'},
\qquad
\left(\bPsi_U^{(w)},\bPsi_U^{(w')*}\right)=0.
\label{eq:equal-mode-normalization}
\end{equation}
For real physical parameters, the field equations have real coefficients, so the conjugate doublets
\begin{equation}
\bigl[\bPsi_U^{(w)}(k,\tau)\bigr]^*
=
\begin{pmatrix}
\cF_U^{(w)}(k,\tau)^*\\[1mm]
\cS_U^{(w)}(k,\tau)^*
\end{pmatrix},
\qquad w=\pm1,
\label{eq:equal-negative-basis}
\end{equation}
provide the negative-frequency modes, with inner products
\begin{equation}
\left(\bPsi_U^{(w)*},\bPsi_U^{(w')*}\right)=-\delta_{ww'}.
\label{eq:equal-negative-normalization}
\end{equation}
Together, the two positive-frequency doublets and their conjugates form a complete basis of solutions. By linearity of the reconstruction operators, the kernel expansion in Eq.~\eqref{eq:equal-kernel-mode-expansion} gives the field modes
\begin{equation}
\bPhi_b(k,\tau)
=\sum_{w=\pm1}A_b^{(w)}\bPsi_U^{(w)}(k,\tau),
\qquad b=1,2.
\label{eq:equal-rotated-basis}
\end{equation}
Using Eq.~\eqref{eq:equal-mode-normalization}, the canonical normalization conditions become
\begin{equation}
\left(\bPhi_a,\bPhi_b\right)
=\sum_{w=\pm1}A_a^{(w)*}A_b^{(w)}=\delta_{ab},
\qquad
\left(\bPhi_a,\bPhi_b^*\right)=0.
\label{eq:equal-vacuum-normalization}
\end{equation}
The coefficients therefore form a unitary matrix, which also satisfies
\begin{equation}
\sum_{b=1}^2 A_b^{(w)}A_b^{(w')*}=\delta_{ww'}.
\label{eq:equal-coefficient-unitarity}
\end{equation}
The label $w$ distinguishes the two mixed polarizations $(1,\ii)^T$ and $(1,-\ii)^T$, while $b$ labels their orthonormal combinations in the quantum-field expansion. Any time-independent unitary choice of $A_b^{(w)}$ preserves the conserved products and describes the same Bunch--Davies vacuum.

To make contact with the field-aligned modes used in Refs.~\cite{Huenupi:2026abj,Huenupi:2026aqc}, we choose a reference time $\tau_0$ in the asymptotic subhorizon regime, with $z_0=-kc\tau_0\gg1$ at fixed mass and mixing, and require the leading field amplitudes to lie along $(1,0)^T$ for $b=1$ and $(0,1)^T$ for $b=2$. Inserting Eq.~\eqref{eq:equal-mode-UV} into Eq.~\eqref{eq:equal-rotated-basis} then fixes
\begin{equation}
A_1^{(w)}=\frac{\ee^{\frac{\ii w\lambda}{2}\ln(2z_0)}}{\sqrt2},\qquad
A_2^{(w)}=-\frac{\ii w}{\sqrt2}
\ee^{\frac{\ii w\lambda}{2}\ln(2z_0)}.
\label{eq:equal-original-field-basis}
\end{equation}
These are the leading Bunch--Davies coefficients of the earlier works, with corrections to the alignment suppressed by $z_0^{-1}$ at fixed parameters. The logarithmic phases in Eq.~\eqref{eq:equal-mode-UV} make the leading ultraviolet field alignment depend on time, so $\tau_0$ specifies the time at which the basis is aligned with the individual fields. For each choice of $\tau_0$, the coefficients remain constant throughout the evolution and satisfy Eq.~\eqref{eq:equal-coefficient-unitarity} exactly. Changing the reference time therefore changes the field alignment of the basis within the same Bunch--Davies vacuum.

\subsection{Tricomi representation of the mode functions}
\label{sec:equal-original-solutions}

We can now evaluate the reconstruction integrals in terms of the Tricomi functions appearing in our earlier solutions. To make this connection, it is useful to express the associated Legendre function through its hypergeometric representation,
\begin{equation}
\Gamma(1+\ii w\lambda)P_{\nu-\frac12}^{-\ii w\lambda}(u)
=\left(\frac{1-u}{1+u}\right)^{\frac{\ii w\lambda}{2}}
F_w\!\left(\frac{1-u}{2}\right),
\label{eq:equal-legendre-hypergeometric}
\end{equation}
where the hypergeometric factor is
\begin{equation}
F_w(p)
\equiv{}_2F_1\!\left(
\frac12-\nu,\frac12+\nu;
1+\ii w\lambda;p\right).
\label{eq:Fw}
\end{equation}
The powers in Eq.~\eqref{eq:equal-legendre-hypergeometric} inherit the branch convention of Eq.~\eqref{eq:equal-legendre-branch-convention}; in particular, the logarithm of the ratio is the difference of the logarithms of its numerator and denominator. This preserves the continuation already fixed for the Legendre functions.

Keeping the upward cut and rescaling its parameter as $u=1+2\ii s$, with $s>0$, we insert this representation into the first integral of Eq.~\eqref{eq:F-Laplace-threshold}. Combining the two boundary values according to Eq.~\eqref{eq:equal-legendre-positive-cut}, and using the gamma-function reflection identity, gives
\begin{equation}
\begin{aligned}
\cI_{\Gamma_+,k}\!\left[\frac{Y^{(w)}}{1-u}\right]
={}&-\frac{\ee^{\pi w\lambda/4-\ii kc\tau}}
{\sqrt2\,\Gamma(\frac{\ii w\lambda}{2})}
\int_0^\infty\frac{\de s}{s}\,
\ee^{2kc\tau s}
\left(\frac{\ii s}{1+\ii s}\right)^{\frac{\ii w\lambda}{2}}
F_w(-\ii s).
\end{aligned}
\label{eq:equal-curvature-ray}
\end{equation}
Here the power of $\ii s/(1+\ii s)$ is continued from $s>0$, with $\arg(\ii s)=\pi/2$ and the principal argument of $1+\ii s$. The integral is exponentially damped at infinity for real $\tau<0$, while its endpoint value is obtained by analytic continuation from $\operatorname{Re}(\ii w\lambda/2)>0$, as prescribed by the original Hankel contour.

For $\nu=1/2$, the hypergeometric factor reduces to $F_w=1$. The remaining integral is a representation of Tricomi's confluent hypergeometric function~\cite{DLMF:Confluent}:
\begin{equation}
U\!\left(\frac{\ii w\lambda}{2},1,2\ii kc\tau\right)
=\frac{1}{\Gamma(\frac{\ii w\lambda}{2})}
\int_0^\infty\frac{\de s}{s}\,
\ee^{2kc\tau s}
\left(\frac{\ii s}{1+\ii s}\right)^{\frac{\ii w\lambda}{2}}.
\label{eq:equal-Tricomi-seed-integral}
\end{equation}
For general $\nu$, the factor $F_w(-\ii s)$ can be generated by a differential operator acting on the same function. Indeed, $-\ii\partial_\tau/(2kc)$ multiplies the exponential $\ee^{2kc\tau s}$ by $-\ii s$, which motivates the definition
\begin{equation}
\widehat{\mathcal D}_\nu^{(w)}
\equiv F_w\!\left(-\frac{\ii}{2kc}\partial_\tau\right).
\label{eq:equal-original-D-operator}
\end{equation}
The operator acts on the Tricomi integral by inserting $F_w(-\ii s)$ into its integrand. Applying the operator to Eq.~\eqref{eq:equal-Tricomi-seed-integral} evaluates the first reconstruction integral. The second follows by including an additional factor $1/[2(1+\ii s)]$, which changes the second parameter of $U$ from $1$ to $0$. We obtain
\begin{align}
\cI_{\Gamma_+,k}\!\left[\frac{Y^{(w)}}{1-u}\right]
&=-\frac{\ee^{\pi w\lambda/4-\ii kc\tau}}{\sqrt2}
\widehat{\mathcal D}_\nu^{(w)}
\!\left[U\!\left(\frac{\ii w\lambda}{2},1,2\ii kc\tau\right)\right],
\label{eq:equal-original-U-identity}
\\
\cI_{\Gamma_+,k}\!\left[\frac{Y^{(w)}}{1-u^2}\right]
&=-\frac{\ee^{\pi w\lambda/4-\ii kc\tau}}{2\sqrt2}
\widehat{\mathcal D}_\nu^{(w)}
\!\left[U\!\left(\frac{\ii w\lambda}{2},0,2\ii kc\tau\right)\right].
\label{eq:equal-original-U-second-identity}
\end{align}
Substituting these expressions into Eq.~\eqref{eq:F-Laplace-threshold} gives the normalized curvature mode:
\begin{equation}
\begin{aligned}
\cF_U^{(w)}(k,\tau)
={}&\frac{N_k\ee^{\pi w\lambda/4-\ii kc\tau}}{\sqrt2}
\bigg\{
-2kc\tau\,\widehat{\mathcal D}_\nu^{(w)}
\!\left[U\!\left(\frac{\ii w\lambda}{2},1,2\ii kc\tau\right)\right]
\\
&\quad+\ii(1-\ii kc\tau)\widehat{\mathcal D}_\nu^{(w)}
\!\left[U\!\left(\frac{\ii w\lambda}{2},0,2\ii kc\tau\right)\right]
\bigg\}.
\end{aligned}
\label{eq:equal-Tricomi-curvature}
\end{equation}
The entropy mode follows from Eq.~\eqref{eq:S-Laplace}. Relative to the first integral above, removing the denominator $1-u$ inserts a factor $-2\ii s$, giving
\begin{equation}
\cS_U^{(w)}(k,\tau)
=\sqrt2\,wN_k(kc\tau)^2\ee^{\pi w\lambda/4-\ii kc\tau}
\widehat{\mathcal D}_\nu^{(w)}
\!\left[U\!\left(1+\frac{\ii w\lambda}{2},2,2\ii kc\tau\right)\right].
\label{eq:equal-Tricomi-entropy}
\end{equation}
In these expressions, the operator acts only on the function inside the square brackets; the plane wave and the other explicit factors of $\tau$ remain outside its action. The two values of $w$ give the basis solutions entering Eq.~\eqref{eq:equal-rotated-basis}, so the curvature and entropy modes in the quantum-field expansion are
\begin{align}
\varphi_b(k,\tau)
&=\sum_{w=\pm1}A_b^{(w)}\cF_U^{(w)}(k,\tau),
\label{eq:equal-curvature-mode-expansion}
\\
\sigma_b(k,\tau)
&=\sum_{w=\pm1}A_b^{(w)}\cS_U^{(w)}(k,\tau).
\label{eq:equal-entropy-mode-expansion}
\end{align}

Equations~\eqref{eq:equal-Tricomi-curvature} and \eqref{eq:equal-Tricomi-entropy} reproduce, including their overall normalization and phase, the curvature and entropy mode functions given in Sec.~V of Ref.~\cite{Huenupi:2026aqc}, following the original solution of Ref.~\cite{Huenupi:2026abj}. In the notation of those references, $z=-kc\tau$, so that $\ii\partial_z/2$ becomes $-\ii\partial_\tau/(2kc)$, while the free normalization becomes $N_k=H/\sqrt{2c^3k^3}$ when the common sound speed is kept explicit. The contour construction therefore gives the same full field solutions in a representation where their frequency content is fixed by the integration path. The complex conjugates defined in Eq.~\eqref{eq:equal-negative-basis} complete the solution space. An equivalent basis involving Kummer functions was constructed in Refs.~\cite{Huenupi:2026abj,Huenupi:2026aqc}, where the corresponding expressions and their derivation can be found.

\subsection{Equal-speed power spectrum}
\label{sec:equal-power}

For $\operatorname{Re}\nu<3/2$, each curvature mode approaches a constant at late times. Taking this limit in the Tricomi solutions gives
\begin{equation}
\cF_U^{(w)}(k,0)
=
\frac{\ii N_k\ee^{\pi w\lambda/4}}{\sqrt{2\pi}}
\frac{
\Gamma(\frac12+\frac{\ii w\lambda}{2})
\Gamma(\frac34-\frac\nu2)
\Gamma(\frac34+\frac\nu2)
}{
\Gamma(\frac34-\frac\nu2+\frac{\ii w\lambda}{2})
\Gamma(\frac34+\frac\nu2+\frac{\ii w\lambda}{2})
}.
\label{eq:equal-late-amplitude}
\end{equation}
The modes in Eq.~\eqref{eq:equal-curvature-mode-expansion} therefore approach
\begin{equation}
\varphi_b(k,0)
=
\sum_{w=\pm1}
A_b^{(w)}\cF_U^{(w)}(k,0),
\qquad b=1,2.
\label{eq:equal-late-mode}
\end{equation}
The dimensionless curvature spectrum follows by summing over the two modes $b=1,2$ appearing in the quantum-field expansion. Using the unitarity relation in Eq.~\eqref{eq:equal-coefficient-unitarity}, the coefficients cancel from this sum, giving
\begin{equation}
\Delta_\zeta(k)
=
\frac{k^3}{2\pi^2F_\zeta^2}
\sum_{b=1}^2|\varphi_b(k,0)|^2
=
\frac{k^3}{2\pi^2F_\zeta^2}
\sum_{w=\pm1}\left|\cF_U^{(w)}(k,0)\right|^2.
\label{eq:equal-power-definition}
\end{equation}
The spectrum is therefore independent of the choice of $A_b^{(w)}$, including the reference time $\tau_0$ used to align the modes with the individual fields. The same cancellation holds in the contracted mode sums entering higher correlation functions.

To evaluate the spectrum, we substitute Eq.~\eqref{eq:equal-late-amplitude} and use $|\Gamma(\frac12+\ii y)|^2=\pi/\cosh(\pi y)$. Measuring the result relative to the unmixed amplitude $\Delta_0^{(c)}=H^2/(4\pi^2F_\zeta^2c^3)$, we obtain
\begin{equation}
\frac{\Delta_\zeta}{\Delta_0^{(c)}}
=
\sum_{w=\pm1}
\frac{\ee^{\pi w\lambda/2}}
{2\cosh(\frac{\pi\lambda}{2})}
\left|
\frac{
\Gamma(\frac34-\frac\nu2)
\Gamma(\frac34+\frac\nu2)
}{
\Gamma(\frac34-\frac\nu2+\frac{\ii w\lambda}{2})
\Gamma(\frac34+\frac\nu2+\frac{\ii w\lambda}{2})
}
\right|^2.
\label{eq:equal-power-sum}
\end{equation}
For real entropy mass squared, $\nu$ is either real or purely imaginary, so the gamma-function ratio has the same modulus for both values of $w$. The remaining weights sum to one, and the spectrum reduces to
\begin{equation}
\frac{\Delta_\zeta}{\Delta_0^{(c)}}
=
\left|
\frac{
\Gamma(\frac34-\frac\nu2)
\Gamma(\frac34+\frac\nu2)
}{
\Gamma(\frac34-\frac\nu2+\frac{\ii\lambda}{2})
\Gamma(\frac34+\frac\nu2+\frac{\ii\lambda}{2})
}
\right|^2.
\label{eq:equal-speed-power}
\end{equation}
This reproduces the equal-speed result of Ref.~\cite{Huenupi:2026abj}, with unit enhancement at zero mixing. At nonzero mixing, a massless entropy perturbation, corresponding to $\nu=3/2$, continues to source the curvature mode outside the horizon, producing the cumulative growth discussed in Ref.~\cite{Achucarro:2016cumulative}. In this case the final time must be kept explicit, as is also necessary in the bispectra generated by massless mediators~\cite{Wang:2022LightScalars}.


\section{Early-time propagation and sound-speed degeneracy}
\label{sec:early-time-propagation}

Before extending the exact construction to unequal sound speeds, it is useful to understand how the distinction between the two cases appears directly in the equations of motion at early times. In particular, the logarithmic phases found in the equal-speed solutions have a simple origin in the coupled propagation of the fields, which becomes clear without introducing a Laplace representation.

To begin with, we may remove the common time dependence associated with the expansion by introducing the rescaled Fourier modes
\begin{equation}
\boldsymbol Q(k,\tau)\equiv
a(\tau)
\begin{pmatrix}
\varphi(k,\tau)\\
\sigma(k,\tau)
\end{pmatrix}.
\label{eq:early-time-rescaled-fields}
\end{equation}
Here the first component, $a\varphi=aF_\zeta\zeta$, is the adiabatic Mukhanov--Sasaki variable, while the second is the canonically rescaled entropy perturbation. At fixed $\lambda$ and $\mu/H$, their equations in the limit $\tau\to-\infty$ take the form
\begin{equation}
\boldsymbol Q''
+k^2
\begin{pmatrix}
c_\varphi^2&0\\
0&c_\sigma^2
\end{pmatrix}
\boldsymbol Q
+\frac{\lambda}{\tau}
\begin{pmatrix}
0&1\\
-1&0
\end{pmatrix}
\boldsymbol Q'
\simeq0,
\label{eq:early-time-coupled-equations}
\end{equation}
with neglected corrections of order $\mathcal O(\tau^{-2})\boldsymbol Q$ that contain the mass dependence and the remaining effects of the expansion and mixing. The gradient terms determine the leading oscillation frequencies, while the derivative coupling falls as $1/\tau$. Although this coupling becomes small sufficiently far in the past, whether its accumulated effect can be neglected depends on the relation between the two frequencies.

The derivative coupling suggests introducing a time-dependent rotation of the field coordinates. We therefore write
\begin{equation}
\boldsymbol Q=R\widetilde{\boldsymbol Q},
\qquad
R=
\begin{pmatrix}
\cos\theta&-\sin\theta\\
\sin\theta&\cos\theta
\end{pmatrix},
\qquad
\theta=\frac{\lambda}{2}\ln(-2kc_\varphi\tau).
\label{eq:early-time-field-rotation}
\end{equation}
With this choice, the rotation matrix satisfies
\begin{equation}
2R'
+\frac{\lambda}{\tau}
\begin{pmatrix}
0&1\\
-1&0
\end{pmatrix}R=0.
\label{eq:early-time-amplitude-rotation}
\end{equation}
Consequently, the contribution $2R'\widetilde{\boldsymbol Q}'$ generated by differentiating $\boldsymbol Q$ cancels the derivative mixing in Eq.~\eqref{eq:early-time-coupled-equations}. The remaining terms generated by the rotation are proportional to $1/\tau^2$ and belong to the same order already neglected in the early-time equations.

Consider now the equal-speed case $c_\varphi=c_\sigma=c$. The gradient matrix is then proportional to the identity and is unchanged by the rotation, so the equations for the new fields reduce to
\begin{equation}
\widetilde{\boldsymbol Q}''
+k^2c^2\widetilde{\boldsymbol Q}
\simeq0.
\label{eq:early-time-rotated-equations}
\end{equation}
Thus, at this order, the two rotated fields degenerate, propagating as independent oscillators with the same frequency $kc$. Their normalized positive-frequency solutions are plane waves $\ee^{-\ii kc\tau}/\sqrt{2kc}$ multiplying any pair of orthonormal constant polarization vectors. To connect with the solutions obtained in Section~\ref{sec:equal-speed}, we choose the circular polarizations $(1,\ii w)^{\mathsf T}/\sqrt2$, with $w=\pm1$, on which the rotation acts as
\begin{equation}
R\frac1{\sqrt2}
\begin{pmatrix}
1\\
\ii w
\end{pmatrix}
=
\ee^{-\ii w\theta}
\frac1{\sqrt2}
\begin{pmatrix}
1\\
\ii w
\end{pmatrix}.
\label{eq:early-time-polarization-rotation}
\end{equation}
Transforming back to the original field coordinates then gives
\begin{equation}
\boldsymbol Q^{(w)}(k,\tau)
\sim
\frac{\ee^{-\ii kc\tau}}{\sqrt{2kc}}\,
\exp\!\left[-\frac{\ii w\lambda}{2}\ln(-2kc\tau)\right]
\frac{1}{\sqrt2}
\begin{pmatrix}
1\\
\ii w
\end{pmatrix},
\qquad w=\pm1.
\label{eq:early-time-degenerate-modes}
\end{equation}
The constant inside the logarithm fixes the phase convention used in Section~\ref{sec:equal-speed}, and dividing by $a(\tau)$ recovers the early-time behavior of the original field modes. In this way, the logarithmic phases describe the rotation of the original field coordinates relative to the asymptotically free variables $\widetilde{\boldsymbol Q}$. Although the angular velocity $\theta'=\lambda/(2\tau)$ decreases in the asymptotic past, its integral accumulates logarithmically, so the rotation remains part of the leading solution for any fixed nonzero mixing.

When $c_\varphi\neq c_\sigma$, the same rotation still removes the derivative coupling, but it also rotates the gradient matrix. In particular, the transformed matrix has the off-diagonal entry
\begin{equation}
\left[
R^{\mathsf T}
\begin{pmatrix}
c_\varphi^2&0\\
0&c_\sigma^2
\end{pmatrix}
R
\right]_{12}
=
(c_\sigma^2-c_\varphi^2)\sin\theta\cos\theta.
\label{eq:early-time-rotated-gradients}
\end{equation}
This contribution enters multiplied by $k^2$, so it is part of the leading propagation equations rather than an inverse-time correction. The rotated fields therefore no longer behave as independent free oscillators. The original field directions, on the other hand, are distinguished by their different leading frequencies.

An incoming curvature wave oscillates as $\ee^{-\ii kc_\varphi\tau}$ and induces a smaller entropy component, while an incoming entropy wave oscillates as $\ee^{-\ii kc_\sigma\tau}$ and induces a smaller curvature component. Substituting expansions in inverse powers of $\tau$ into Eq.~\eqref{eq:early-time-coupled-equations} gives a canonically normalized choice with
\begin{align}
\boldsymbol Q_1(k,\tau)
&=
\frac{\ee^{-\ii kc_\varphi\tau}}{\sqrt{2kc_\varphi}}
\begin{pmatrix}
1+\mathcal O((kc_\varphi\tau)^{-1})
\\[2mm]
\displaystyle
\frac{\ii\lambda c_\varphi}
{k(c_\varphi^2-c_\sigma^2)\tau}
+\mathcal O((kc_\varphi\tau)^{-2})
\end{pmatrix},
\label{eq:early-time-curvature-incoming}
\\
\boldsymbol Q_2(k,\tau)
&=
-\frac{\ee^{-\ii kc_\sigma\tau}}{\sqrt{2kc_\sigma}}
\begin{pmatrix}
\displaystyle
\frac{\ii\lambda c_\sigma}
{k(c_\varphi^2-c_\sigma^2)\tau}
+\mathcal O((kc_\sigma\tau)^{-2})
\\[2mm]
1+\mathcal O((kc_\sigma\tau)^{-1})
\end{pmatrix}.
\label{eq:early-time-entropy-incoming}
\end{align}
These expansions are taken at fixed unequal speeds and fixed mass and mixing. The induced component in each mode vanishes as $1/\tau$, and its feedback on the frequency of the leading component begins at order $1/\tau^2$. Integrating these frequency corrections gives inverse-time corrections to the phase, rather than the logarithm appearing in Eq.~\eqref{eq:early-time-degenerate-modes}.

We can also understand this distinction through the relative oscillation of the incoming waves. Once their free plane waves have been factored out, the mixing between them carries factors $\ee^{\pm\ii k(c_\varphi-c_\sigma)\tau}$. Its accumulated early-time contribution therefore involves integrals of the form
\begin{equation}
\int_{-\infty}^{\tau}
\frac{\de\eta}{\eta}\,
\ee^{\pm\ii k(c_\varphi-c_\sigma)\eta}.
\label{eq:early-time-relative-phase-integral}
\end{equation}
For any fixed nonzero speed difference, oscillatory cancellation makes this integral converge at its lower limit, and its magnitude vanishes as $\tau\to-\infty$. At equal speeds the relative oscillation disappears, allowing the $1/\tau$ coupling to accumulate into the logarithmic rotation derived above.

Finally, the factors $c_\varphi^2-c_\sigma^2$ in Eqs.~\eqref{eq:early-time-curvature-incoming} and \eqref{eq:early-time-entropy-incoming} anticipate that separating the incoming curvature and entropy waves requires increasingly early times as the sound speeds approach each other. At a given early time, reducing the speed difference enhances the induced component in each mode, while moving further into the past suppresses it through the factor $1/\tau$. The competition between these two effects determines when the modes can be approximated by waves propagating along the individual field directions. When the speeds coincide, the leading frequencies no longer distinguish these directions, and the derivative coupling instead selects the mixed polarizations and logarithmic phases derived above.

\section{Unequal sound speeds in the Laplace representation}
\label{sec:exact}

We now proceed to solve the coupled equations for unequal sound speeds, $r_c\neq1$, and arbitrary nonzero mixing $\lambda$, taking both sound speeds to be positive and constant. The early-time analysis in Section~\ref{sec:early-time-propagation} identifies the two incoming waves that specify the vacuum modes. We use these asymptotic conditions to extend the Laplace construction introduced in Section~\ref{sec:equal-speed}, fixing the amplitudes of the reconstructed fields through the conserved canonical product.

\subsection{Differential equation in Laplace space}
\label{sec:plane-wave}
\label{sec:tau-reconstruction}
\label{sec:tau-spectral-equation}

To begin with, notice that reconstructing $\varphi_k$ and $\sigma_k$ from the mixed velocity $P_k$ uses only the curvature equation and the definition of $P_k$. The relations in Eqs.~\eqref{eq:F-from-P} and \eqref{eq:S-from-P} therefore remain valid with $c$ replaced by $c_\varphi$, independently of the entropy sound speed. The same replacement applies to the reconstruction operators in Eqs.~\eqref{eq:S-Laplace}, \eqref{eq:P-general} and \eqref{eq:F-Laplace}, with normalization $N_k=H/\sqrt{2c_\varphi^3k^3}$. We consequently retain the curvature sound speed as the scale in the Laplace representation, writing
\begin{equation}
\cI_{\Gamma,k}[K](\tau)
\equiv\frac{1}{2\pi\ii}\int_\Gamma\de u\,
\ee^{-\ii kc_\varphi\tau u}K(u).
\label{eq:I-definition-tau}
\end{equation}
With this choice, the curvature wave operator still multiplies each plane wave by $k^2c_\varphi^2(1-u^2)$, so the denominator $1-u^2$ remains unchanged in the reconstruction of $P_k$ and $\varphi_k$. The entropy wave operator instead gives
\begin{equation}
\left(\partial_\tau^2+k^2c_\sigma^2\right)
\ee^{-\ii kc_\varphi\tau u}
=k^2c_\varphi^2(r_c^2-u^2)
\ee^{-\ii kc_\varphi\tau u}.
\label{eq:unequal-entropy-wave-eigenvalue}
\end{equation}
The ratio $r_c$ thus enters through the entropy equation, while the relation between the spectral kernel and the reconstructed fields retains the form established in the equal-speed case.

Letting $Y(u)$ denote the entropy kernel as before, we substitute the reconstructed entropy field and mixed velocity into Eq.~\eqref{eq:eom-sigma-tau}. The identities in Eq.~\eqref{eq:I-identities}, with $c$ replaced by $c_\varphi$, again allow us to exchange the explicit powers of $\tau$ for derivatives of the kernel. After integrating by parts and collecting terms, the entropy equation is satisfied provided that
\begin{equation}
\frac{\de}{\de u}\left[(r_c^2-u^2)Y'(u)\right]
+\left[\nu^2-\frac14+
\frac{\lambda^2}{1-u^2}\right]Y(u)=0.
\label{eq:Y-equation}
\end{equation}
For unequal sound speeds, this equation has four finite regular singular points, located at $u=\pm1$ and $u=\pm r_c$. The points $u=\pm1$ arise from the mixing term, while $u=\pm r_c$ are the zeros of the coefficient of the highest derivative. Through the exponential in Eq.~\eqref{eq:I-definition-tau}, the two positive endpoints supply precisely the oscillations identified in Section~\ref{sec:early-time-propagation}: $u=1$ gives $\ee^{-\ii kc_\varphi\tau}$, associated with an incoming curvature wave, while $u=r_c$ gives $\ee^{-\ii kc_\sigma\tau}$, associated with an incoming entropy wave. The negative endpoints supply the corresponding negative-frequency oscillations. As $r_c$ approaches unity, the two positive endpoints merge, as do the two negative endpoints, and Eq.~\eqref{eq:Y-equation} reduces to the associated Legendre equation obtained in Section~\ref{sec:equal-speed}.

\subsection{Singular points, contours, and incoming waves}
\label{sec:positive-frequency-contours}

To translate the incoming-wave conditions into conditions on $Y$, we first examine its behavior near the positive singular points. At $u=1$, the coefficient of the highest derivative in Eq.~\eqref{eq:Y-equation} is nonzero, while the mixing term has a simple pole. Substituting a local expansion into the equation gives one solution that is analytic and vanishes linearly, and another with a finite endpoint value accompanied by a logarithmic correction. A general solution around $u=1$ therefore behaves as
\begin{equation}
Y(u)=Y(1)\left[
1+\frac{\lambda^2}{2(r_c^2-1)}
(u-1)\log(u-1)
\right]+\mathcal O(u-1).
\label{eq:local-minus-one}
\end{equation}
Instead, at $u=r_c$, the coefficient of the highest derivative vanishes linearly, with $r_c^2-u^2=-2r_c(u-r_c)+\mathcal O((u-r_c)^2)$. The leading derivative terms are therefore proportional to $(u-r_c)Y''+Y'$, admitting a constant and a logarithm as the two local behaviors. Including the corrections determined by the full equation, a general solution takes the form
\begin{equation}
Y(u)=C_0+C_{\log}\log(u-r_c)
+\mathcal O\!\left((u-r_c)\log(u-r_c)\right),
\label{eq:local-entropy-endpoint}
\end{equation}
where $C_0$ and $C_{\log}$ are the two independent local coefficients. Reflection under $u\to-u$ gives the corresponding behavior at the negative singular points.

Following the convention of Section~\ref{sec:equal-speed}, we place cuts vertically upward from all four finite singular points. Near either positive endpoint, the logarithms are defined by
\begin{equation}
\log z=\ln|z|+\ii\arg z,
\qquad
-\frac{3\pi}{2}<\arg z<\frac{\pi}{2},
\label{eq:unequal-log-branch}
\end{equation}
where $z=u-1$ or $z=u-r_c$. This choice allows us to continue a solution from the origin to either positive endpoint by passing below the real singularities, and fixes its values on both sides of each cut. 

\begin{figure}[htbp]
\centering
\includegraphics[width=\textwidth]{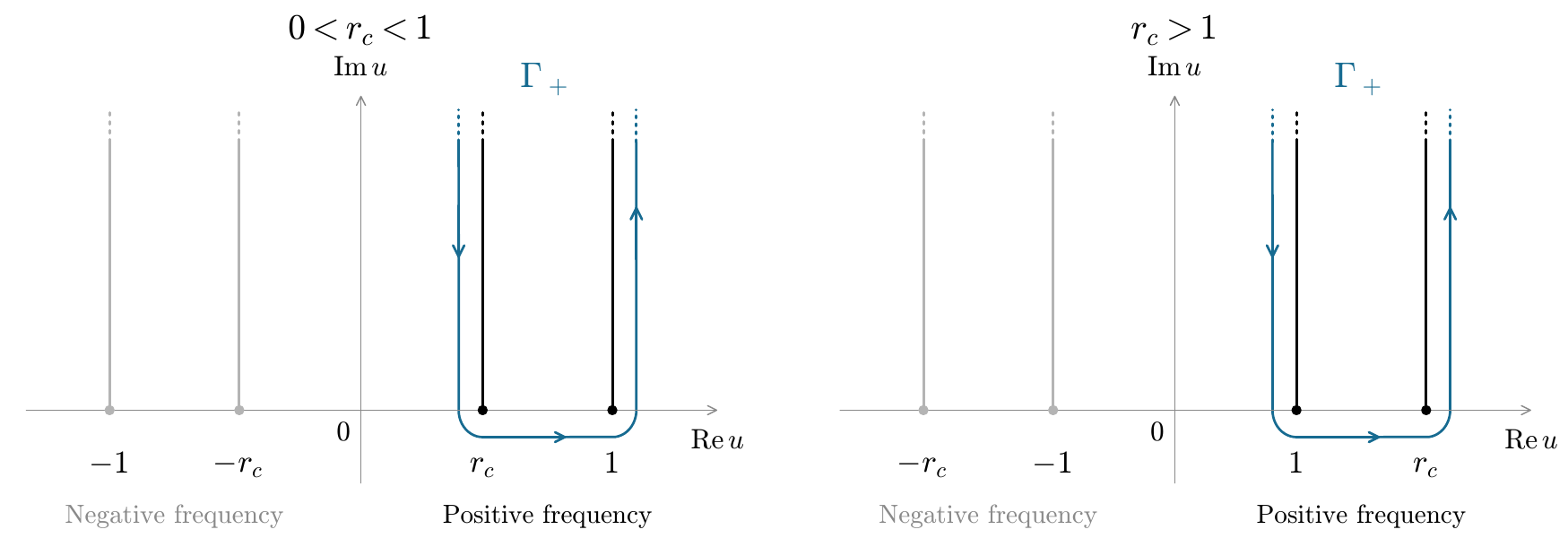}
\caption{The connected positive-frequency contour $\Gamma_+$ for the two possible orderings of $u=1$ and $u=r_c$. All four branch cuts extend upward. The blue contour descends to the left of the two positive endpoints, passes below them, and returns to positive imaginary infinity on their right, with counterclockwise orientation. Both infinite tails extend into the region where $\exp(-\ii kc_\varphi\tau u)$ is damped for real $\tau<0$. The negative-endpoint cuts, shown in gray, remain outside the contour. As the positive endpoints merge at $r_c=1$, this path becomes a contour around the single positive cut of the equal-speed system.}
\label{fig:unequal-speed-operational-contours}
\end{figure}

Repeating the same strategy of Section~\ref{sec:equal-speed}, we select the positive-frequency sector using a single connected contour $\Gamma_+$ that passes below both positive endpoints, as shown in Fig.~\ref{fig:unequal-speed-operational-contours}. For $r_c<1$, the contour descends from positive imaginary infinity to the left of $u=r_c$, bends below this endpoint, continues horizontally toward $u=1$, and then bends upward to return to infinity on the right of its cut. For $r_c>1$, the two endpoints exchange roles. The descending tail lies between the origin and the leftmost positive endpoint, leaving the negative singular points outside the contour. In either ordering, the path has counterclockwise orientation around the two positive endpoints, with the kernel continued along the entire path. Both infinite tails extend into the upper half-plane, where
\begin{equation}
\left|\ee^{-\ii kc_\varphi\tau u}\right|
=\ee^{kc_\varphi\tau\operatorname{Im}u},
\qquad \tau<0.
\label{eq:unequal-contour-damping}
\end{equation}
As in the equal-speed case, the spectral solutions have at most algebraic growth, possibly accompanied by logarithms, so the exponential ensures convergence and suppresses the boundary terms at infinity. 

We now use the local behavior at each positive endpoint to determine the incoming amplitudes, following the same endpoint analysis as in the equal-speed case. The contribution associated with an endpoint is determined by the singular terms in the reconstruction kernels, while terms analytic there give no contribution from that endpoint. Near $u=1$, the finite value of $Y$ produces a simple pole in the reconstruction of the canonical curvature field $\varphi$:
\begin{equation}
\frac{Y(u)}{1-u^2}
=-\frac{Y(1)}{2(u-1)}
+\text{less singular terms}.
\label{eq:curvature-pole}
\end{equation}
Inserting this pole into Eq.~\eqref{eq:F-Laplace}, with $c$ replaced by $c_\varphi$, its residue contributes $Y(1)u_\varphi(k,\tau)$, where $u_\varphi$ is the normalized free curvature mode in Eq.~\eqref{eq:u-phi}. The logarithmic correction in Eq.~\eqref{eq:local-minus-one} contributes only subleading terms in the asymptotic past, both to the curvature mode and to the reconstructed entropy field. Consequently, $Y(1)$ fixes the amplitude of the incoming curvature wave as:
\begin{equation}
\frac{\cF_{\Gamma_+,k}[Y](\tau)}
{u_\varphi(k,\tau)}
\longrightarrow Y(1) , \qquad \tau\to-\infty.
\label{eq:unequal-endpoint-amplitudes-varphi}
\end{equation}

At $u=r_c$, the logarithm instead produces the leading entropy wave. With the branch convention in Eq.~\eqref{eq:unequal-log-branch}, its value on the right side of the upward cut exceeds that on the left by $2\pi\ii$. The discontinuity representation derived in Section~\ref{sec:equal-speed} therefore gives
\begin{equation}
\cI_{\Gamma_+,k}\!\left[\log(u-r_c)\right]
=-\frac{\ii}{kc_\varphi\tau}\,
\ee^{-\ii kc_\sigma\tau}.
\label{eq:entropy-log-transform}
\end{equation}
Multiplying by $C_{\log}$ and by the entropy reconstruction prefactor in Eq.~\eqref{eq:S-Laplace}, with $c$ replaced by $c_\varphi$, gives an incoming entropy amplitude $2\sqrt{r_c}C_{\log}/\lambda$ relative to $u_\sigma(k,\tau)$:
\begin{equation}
\frac{\cS_{\Gamma_+,k}[Y](\tau)}
{u_\sigma(k,\tau)}
\longrightarrow
\frac{2\sqrt{r_c}}{\lambda}C_{\log},
\qquad \tau\to-\infty.
\label{eq:unequal-endpoint-amplitudes-sigma}
\end{equation}
Equations~\eqref{eq:unequal-endpoint-amplitudes-varphi} and \eqref{eq:unequal-endpoint-amplitudes-sigma} provide the spectral counterpart of the incoming waves described in Section~\ref{sec:early-time-propagation}. The separated endpoints distinguish the two incoming excitations, even though each kernel reconstructs both fields. They also account for the inverse-time corrections found there: the discontinuities of the local logarithmic terms are analytic series in the distance from the endpoint, whose Laplace integrals produce inverse powers of $-kc_\varphi\tau$. The mixing-dependent logarithmic phases of the equal-speed case instead arise from its imaginary local powers.

We therefore define $Y_1$ to describe a unit incoming canonical curvature perturbation with no leading entropy wave, and $Y_2$ to describe an incoming entropy perturbation with no leading curvature wave. Choosing the phase of the second mode to match Eq.~\eqref{eq:early-time-entropy-incoming}, its incoming entropy amplitude is $-1$, and the required endpoint conditions must be
\begin{derivationbox}
\begin{equation}
\begin{aligned}
Y_1(1)&=1,
&Y_1(u)&\ \text{analytic at }u=r_c,
\\
Y_2(1)&=0,
&Y_2(u)&=-\frac{\lambda}{2\sqrt{r_c}}
\log(u-r_c)+\mathcal O(1),
\qquad u\to r_c.
\label{eq:Y-normalization}
\end{aligned}
\end{equation}
\end{derivationbox}

\begin{figure}[tbp]
\centering
\includegraphics[width=\textwidth]{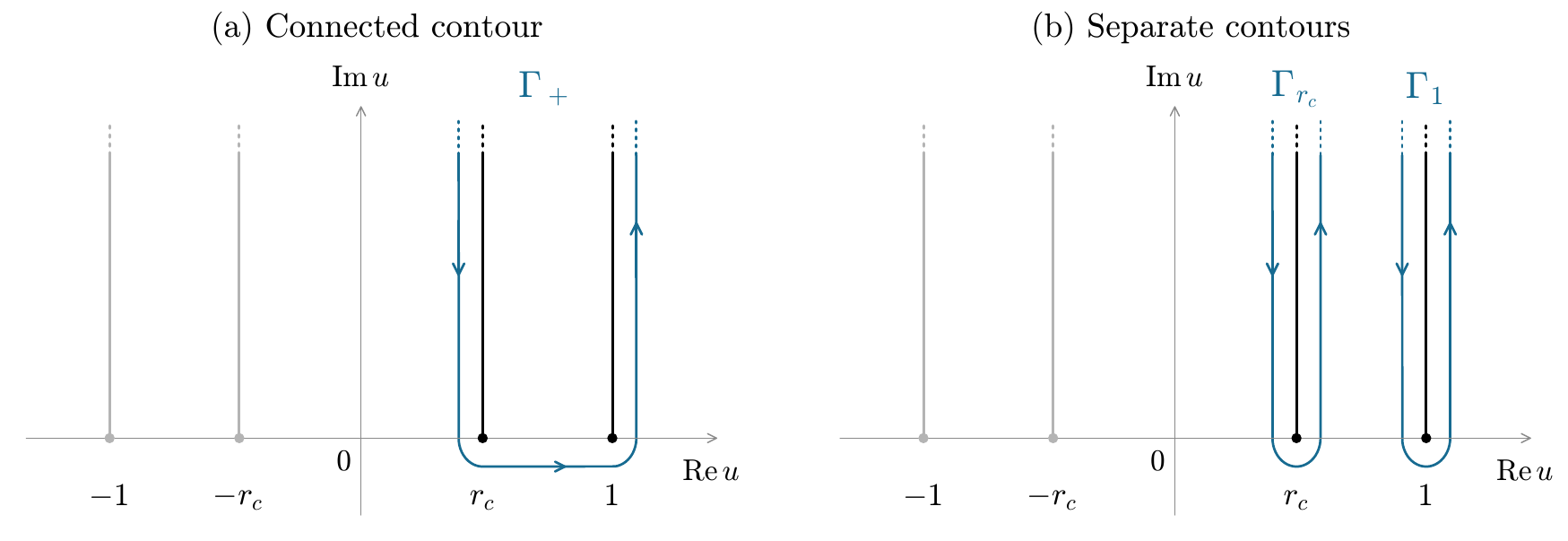}
\caption{Equivalent contour representations for $0<r_c<1$. Left: the connected positive-frequency contour $\Gamma_+$ passes below both positive endpoints. Right: its integral is decomposed into contributions from the counterclockwise contours $\Gamma_{r_c}$ and $\Gamma_1$, surrounding the individual positive cuts. The kernel on each contour is continued on the sheet inherited from $\Gamma_+$. All branch cuts extend upward, and the infinite contour tails lie in the region where $\exp(-\ii kc_\varphi\tau u)$ is damped for real $\tau<0$. The negative-endpoint cuts, shown in gray, remain outside the integration paths. For $r_c>1$, the positive endpoints exchange positions.}
\label{fig:unequal-contour-decomposition}
\end{figure}

To finish this part, notice that we can also describe each incoming perturbation using a contour around its corresponding endpoint. To see this, we deform the connected contour into two counterclockwise Hankel paths, $\Gamma_1$ and $\Gamma_{r_c}$, surrounding the individual positive cuts, as shown in the right panel of Fig.~\ref{fig:unequal-contour-decomposition}. On both paths, the kernel retains the branch inherited from the original contour. The equivalence can be seen by joining the two inner legs at a large positive imaginary height and deforming them through the region between the cuts into the lower connector. The joining contribution vanishes at infinity by Eq.~\eqref{eq:unequal-contour-damping}, giving
\begin{equation}
\cI_{\Gamma_+,k}[K](\tau)
=\cI_{\Gamma_1,k}[K](\tau)
+\cI_{\Gamma_{r_c},k}[K](\tau).
\label{eq:unequal-contour-decomposition}
\end{equation}
The endpoint conditions in Eq.~\eqref{eq:Y-normalization} now determine which contour contributes to each mode. The first mode can be reconstructed using only $\Gamma_1$, since $Y_1$ is analytic at $u=r_c$ and both reconstruction integrands therefore give zero on $\Gamma_{r_c}$. Likewise, the second mode can be reconstructed using only $\Gamma_{r_c}$: its kernel $Y_2$ is analytic at $u=1$, and its zero there cancels the pole in the curvature reconstruction, so neither field receives a contribution from $\Gamma_1$. Each individual contour nevertheless reconstructs both field components, including the response induced by mixing. Using the connected contour $\Gamma_+$ gives a common prescription for the two modes and allows the integration path to remain fixed as the positive endpoints approach each other.

\subsection{The two normalized solutions}
\label{sec:normalized-mode-functions}

We now proceed to find the two spectral solutions satisfying Eq.~\eqref{eq:Y-normalization}. A useful starting point is the change of variable $x=u^2$, which maps the two pairs of finite singular points to $x=1$ and $x=r_c^2$. Expressing $Y$ locally as a function of $x$, Eq.~\eqref{eq:Y-equation} becomes
\begin{equation}
\frac{\de^2Y}{\de x^2}
+\left(\frac{1}{2x}+\frac{1}{x-r_c^2}\right)
\frac{\de Y}{\de x}
+\frac{(\frac14-\nu^2)(x-1)+\lambda^2}
{4x(x-1)(x-r_c^2)}Y=0.
\label{eq:unequal-Heun-equation}
\end{equation}
The transformed equation has four regular singular points, at $x=0,1,r_c^2$ and infinity, and belongs to the general Heun family~\cite{DLMF:Heun}. The singular point at $x=0$ arises from the quadratic change of variable. We use the notation $\HeunG(a,q;\alpha,\beta,\gamma,\delta;z)$ for the local Heun solution analytic at $z=0$ and normalized to one there.

For the first mode, the absence of an incoming entropy wave requires $Y_1$ to be analytic at $u=r_c$. We therefore use the Heun argument $1-u^2/r_c^2$, which places this endpoint at zero, and select the analytic local solution. Dividing this function by its value at $u=1$ then fixes the incoming curvature amplitude to unity. With the accessory parameter
\begin{equation}
Q_1\equiv
\frac{\lambda^2+(\nu^2-\frac14)(1-r_c^2)}
{4r_c^2},
\label{eq:Q1-normalized-Heun}
\end{equation}
the resulting solution is
\begin{derivationbox}
\begin{equation}
Y_1(u)=
\frac{
\HeunG\!\left(
1-r_c^{-2},Q_1;
\frac14-\frac\nu2,\frac14+\frac\nu2,1,\frac12;
1-u^2/r_c^2
\right)}
{
\HeunG\!\left(
1-r_c^{-2},Q_1;
\frac14-\frac\nu2,\frac14+\frac\nu2,1,\frac12;
1-r_c^{-2}
\right)}.
\label{eq:Y1}
\end{equation}
\end{derivationbox}
The denominator is the finite endpoint value reached by continuing the numerator from $u=r_c$ to $u=1$. The interval between these points contains no other singularity, so this value is real for real $\lambda$ and $\nu^2$.

For the second mode, the absence of an incoming curvature wave requires $Y_2(1)=0$, selecting the solution analytic at $u=1$. Factoring out its simple zero through $u^2-1$ leaves a function that can again be expressed as a local Heun solution, now with argument $1-u^2$. Its amplitude must reproduce the logarithmic coefficient at $u=r_c$ specified in Eq.~\eqref{eq:Y-normalization}. We can determine this amplitude from the Wronskian, relating the behavior at the two endpoints without evaluating a Heun connection coefficient. The self-adjoint form of Eq.~\eqref{eq:Y-equation} implies
\begin{equation}
\frac{\de}{\de u}\left[
(r_c^2-u^2)
\left(Y_1Y_2'-Y_1'Y_2\right)
\right]=0,
\label{eq:normalized-Wronskian-conservation}
\end{equation}
where the primes denote derivatives with respect to $u$. At $u=1$, the quantity in brackets approaches $(r_c^2-1)Y_2'(1)$, since $Y_1(1)=1$ and $Y_2$ vanishes linearly. At $u=r_c$, the logarithmic derivative of $Y_2$, together with $r_c^2-u^2\simeq-2r_c(u-r_c)$, instead gives $\lambda\sqrt{r_c}\,Y_1(r_c)$. Equating these values fixes
\begin{equation}
Y_2'(1)=
\frac{\lambda\sqrt{r_c}}{r_c^2-1}Y_1(r_c).
\label{eq:Y2-endpoint-derivative}
\end{equation}
Thus the required entropy amplitude determines the derivative of $Y_2$ at the curvature endpoint, where its analytic Heun representation is defined. Introducing the accessory parameter
\begin{equation}
Q_2\equiv\frac{\lambda^2+6-2r_c^2}{4},
\label{eq:Q2-normalized-Heun}
\end{equation}
we finally obtain
\begin{derivationbox}
\begin{equation}
\begin{aligned}
Y_2(u)={}&
\frac{\lambda\sqrt{r_c}}{2(r_c^2-1)}
Y_1(r_c)(u^2-1)
\\
&\times
\HeunG\!\left(
1-r_c^2,Q_2;
\frac54-\frac\nu2,\frac54+\frac\nu2,2,\frac12;
1-u^2
\right).
\end{aligned}
\label{eq:Y2}
\end{equation}
\end{derivationbox}
The Heun function equals one at $u=1$, while the derivative of $(u^2-1)/2$ equals one there, so the prefactor implements Eq.~\eqref{eq:Y2-endpoint-derivative} directly. The Wronskian relation then ensures that the logarithm at the entropy endpoint has the coefficient required by Eq.~\eqref{eq:Y-normalization}.

Having determined the kernels, we reconstruct the curvature and entropy modes on the connected contour $\Gamma_+$. That is:
\begin{align}
\varphi_b(k,\tau)
&=\cF_{\Gamma_+,k}[Y_b](\tau)
=-2\ii N_k\,
\cI_{\Gamma_+,k}\!\left[
\frac{1+\ii kc_\varphi\tau u}{1-u^2}Y_b(u)
\right],
\label{eq:F-Laplace-tau}
\\
\sigma_b(k,\tau)
&=\cS_{\Gamma_+,k}[Y_b](\tau)
=-\frac{2\ii N_k k^2c_\varphi^2\tau^2}{\lambda}\,
\cI_{\Gamma_+,k}[Y_b],
\label{eq:S-Laplace-tau}
\end{align}
for $b=1,2$, and where $N_k=H/\sqrt{2c_\varphi^3k^3}$. Each kernel determines both field components, and the endpoint conditions give
\begin{equation}
\frac{\bPhi_1(k,\tau)}{u_\varphi(k,\tau)}
\longrightarrow
\begin{pmatrix}1\\0\end{pmatrix},
\qquad
\frac{\bPhi_2(k,\tau)}{u_\sigma(k,\tau)}
\longrightarrow
\begin{pmatrix}0\\-1\end{pmatrix},
\qquad \tau\to-\infty,
\label{eq:normalized-UV}
\end{equation}
with $\bPhi_b=(\varphi_b,\sigma_b)^{\mathsf T}$. To verify canonical normalization, we use the conserved product in Eq.~\eqref{eq:symplectic-product}. The canonical momenta remain $a^2P$ and $a^2\sigma'$, since changing the sound speeds affects only the spatial gradient terms. Evaluating the product in the asymptotic past, the leading terms in Eq.~\eqref{eq:normalized-UV} give the unit norms of the free incoming fields, while the transferred components contribute vanishing corrections. Conservation therefore implies
\begin{equation}
(\bPhi_a,\bPhi_b)=\delta_{ab},
\qquad
(\bPhi_a,\bPhi_b^*)=0.
\label{eq:canonical-normalization}
\end{equation}
For real physical parameters, the field equations have real coefficients, so the complex-conjugate modes supply the negative-frequency partners. Together, these four solutions form a complete basis for the coupled system, with the positive-frequency pair defining the Bunch--Davies vacuum.


\section{Exact power spectrum}
\label{sec:power-exact}

Having obtained the normalized modes for unequal sound speeds, we now proceed to compute the curvature power spectrum. The two incoming excitations identified in Section~\ref{sec:exact} both contribute to the final curvature perturbation, with their amplitudes retaining the effect of the coupled evolution. We continue to assume $\mu^2>0$, so that the entropy perturbation decays and the curvature modes freeze at late times.

\subsection{Late-time curvature amplitudes}
\label{sec:frozen-curvature-amplitudes}

With $\varphi=F_\zeta\zeta$, the dimensionless curvature spectrum is obtained by summing over the two independent vacuum modes:
\begin{equation}
\Delta_\zeta(k,\tau)=\frac{k^3}{2\pi^2F_\zeta^2}
\sum_{b=1}^2|\varphi_b(k,\tau)|^2.
\label{eq:power-definition}
\end{equation}
To extract its late-time value, let us return to the curvature reconstruction in Eq.~\eqref{eq:F-Laplace-tau}. The factor multiplying $Y_b/(1-u^2)$ can be generated by differentiating the plane wave, allowing us to write
\begin{equation}
\varphi_b(k,\tau)
=-2\ii N_k(1-\tau\partial_\tau)
\cI_{\Gamma_+,k}\!\left[\frac{Y_b}{1-u^2}\right](\tau).
\label{eq:curvature-reconstruction-time-derivative}
\end{equation}
To evaluate this limit, we first examine the behavior of the spectral solutions at infinity. For generic $\nu$, Eq.~\eqref{eq:Y-equation} gives
\begin{equation}
Y_b(u)
=C_{b,+}\,u^{-1/2+\nu}\left[1+\mathcal O(u^{-2})\right]
+C_{b,-}\,u^{-1/2-\nu}\left[1+\mathcal O(u^{-2})\right],
\label{eq:large-u-spectral-behavior}
\end{equation}
where the constants $C_{b,\pm}$ depend on the mode and its analytic continuation. At resonant indices, logarithms may accompany these powers, as happens when $\nu=0$ and the two independent behaviors become $u^{-1/2}$ and $u^{-1/2}\ln u$. Including this possibility, the kernel entering the curvature reconstruction satisfies the large-$|u|$ bound
\begin{equation}
\left|\frac{Y_b(u)}{1-u^2}\right|
\lesssim
|u|^{-5/2+\operatorname{Re}\nu}
\left(1+\ln|u|\right).
\label{eq:large-u-curvature-kernel}
\end{equation}
It is therefore integrable at infinity when $\operatorname{Re}\nu<3/2$, which is precisely the condition $\mu^2>0$ on the principal branch.

\begin{figure}[tbp]
\centering
\includegraphics[width=\textwidth]{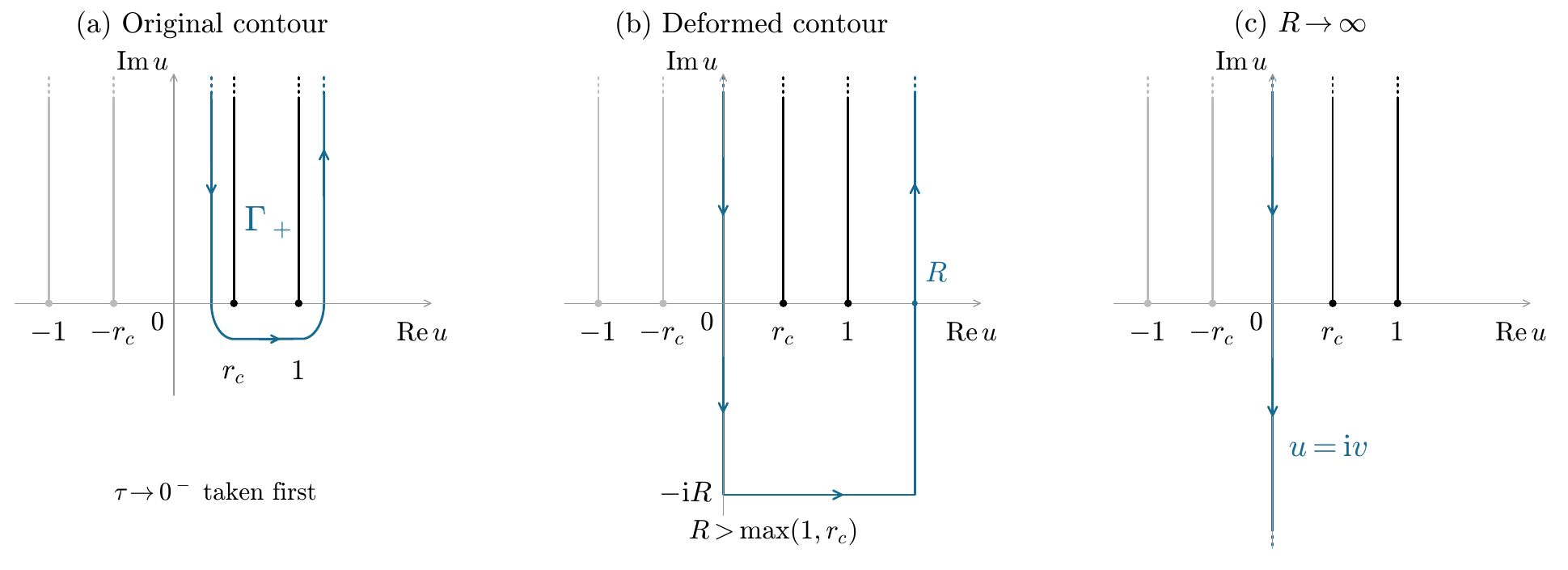}
\caption{Contour deformation for the frozen curvature amplitude, shown for $0<r_c<1$. The limit $\tau\to0^-$ is taken first, leaving the integrand $Y_b(u)/(1-u^2)$. (a) The connected contour $\Gamma_+$ passes below both positive endpoints. (b) Its left leg is moved onto the imaginary axis and extended to $-\ii R$, while the horizontal connector ends at $R-\ii R$ and the right leg rises along $\operatorname{Re}u=R$, with $R>\max(1,r_c)$. The contour remains open throughout the deformation and crosses no branch cut. (c) As $R\to\infty$, the horizontal connector and right leg give vanishing contributions for $\operatorname{Re}\nu<3/2$, leaving the descending imaginary axis. All cuts extend upward; the same deformation applies for $r_c>1$, with the endpoint ordering reversed.}
\label{fig:frozen-contour-deformation}
\end{figure}

On the upward tails of $\Gamma_+$, the plane-wave factor has modulus $\exp(kc_\varphi\tau\,\operatorname{Im}u)\leq1$ for real $\tau<0$, so the bound above allows us to take the late-time limit under the integral. Acting with $\tau\partial_\tau$ introduces the additional factor $-\ii kc_\varphi\tau u$, which being multiplied with the exponential remains uniformly bounded on the tails and tends to zero at every fixed $u$. The finite part of the contour also gives a vanishing contribution to this derivative term, and hence
\begin{equation}
\lim_{\tau\to0^-} (1-\tau\partial_\tau)
\cI_{\Gamma_+,k}\!\left[\frac{Y_b}{1-u^2}\right](\tau)
=
\frac{1}{2\pi\ii}
\int_{\Gamma_+}\frac{Y_b(u)}{1-u^2}\,\de u .
\label{eq:late-time-spectral-integral}
\end{equation}
Having taken this late-time limit, we can deform the remaining integral as shown in Fig.~\ref{fig:frozen-contour-deformation}. 
Starting from the connected contour $\Gamma_+$ in panel (a), we move its left leg onto the imaginary axis and extend it down to $u=-\ii R$, with $R>\max(1,r_c)$. From there, the contour runs horizontally to $u=R-\ii R$ and returns to positive imaginary infinity along $\operatorname{Re}u=R$, as shown in panel (b). The horizontal segment remains below the positive endpoints, while the returning leg lies to the right of both cuts, so this deformation crosses neither a singularity nor a branch cut. Separating the three portions of the deformed contour gives
\begin{equation}
\begin{aligned}
\frac{1}{2\pi\ii}
\int_{\Gamma_+}\frac{Y_b(u)}{1-u^2}\,\de u
={}&-\frac{1}{2\pi}
\int_{-R}^{\infty}
\frac{Y_b(\ii v)}{1+v^2}\,\de v
\\
&+\frac{1}{2\pi\ii}
\int_0^R
\frac{Y_b(x-\ii R)}{1-(x-\ii R)^2}\,\de x
\\
&+\frac{1}{2\pi}
\int_{-R}^{\infty}
\frac{Y_b(R+\ii v)}{1-(R+\ii v)^2}\,\de v.
\end{aligned}
\label{eq:late-time-contour-finite-R}
\end{equation}
The minus sign in the first term follows from the descending orientation of the imaginary-axis segment. Throughout this deformation, $Y_b$ is continued from the original contour. We can now see explicitly what happens as $R$ grows. On the horizontal segment, $|u|\geq R$ and the length is $R$, while on the right vertical leg we have $|u|=(R^2+v^2)^{1/2}$. Applying Eq.~\eqref{eq:large-u-curvature-kernel} and rescaling the integration variables by $R$, we obtain that the second and third contributions in \eqref{eq:late-time-contour-finite-R} therefore vanish for $\operatorname{Re}\nu<3/2$, leaving the descending imaginary axis shown in panel (c) of Fig.~\ref{fig:frozen-contour-deformation}. The order of these steps matters: the late-time limit removes the exponential before the horizontal segment is pushed into the lower half-plane, where that exponential would otherwise grow. Taking $R\to\infty$ in Eq.~\eqref{eq:late-time-contour-finite-R}, we obtain
\begin{equation}
\lim_{\tau\to0^-}
\cI_{\Gamma_+,k}\!\left[\frac{Y_b}{1-u^2}\right](\tau)
=-\frac{1}{2\pi}\int_{-\infty}^{\infty}
\frac{Y_b(\ii v)}{1+v^2}\,\de v.
\label{eq:J-imaginary-axis}
\end{equation}
Here $v$ is real, and $Y_b$ is continued from the origin along the imaginary axis on the sheet specified in Section~\ref{sec:positive-frequency-contours}. Substitution into Eq.~\eqref{eq:curvature-reconstruction-time-derivative} finally yields the frozen curvature amplitude:
\begin{equation}
\varphi_b(k,0)=\frac{\ii N_k}{\pi}
\int_{-\infty}^{\infty}
\frac{Y_b(\ii v)}{1+v^2}\,\de v.
\label{eq:frozen-mode-imaginary-axis}
\end{equation}
To evaluate this expression notice that the symmetric integration range selects the even part of $Y_b(\ii v)$, which can be determined without evaluating the full mode along the contour. Indeed, Eq.~\eqref{eq:Y-equation} is invariant under $u\to-u$, so this even part is fixed by its value at the origin. Let $e(v)$ be the imaginary-axis solution with unit value and zero first derivative there:
\begin{equation}
\bigl[(r_c^2+v^2)e'(v)\bigr]'
-\left[\nu^2-\frac14+\frac{\lambda^2}{1+v^2}\right]e(v)=0,
\qquad e(0)=1,\quad e'(0)=0.
\label{eq:real-period-ODE}
\end{equation}
The even part of each kernel obeys the same equation, with initial value $Y_b(0)$, and therefore
\begin{equation}
\frac{Y_b(\ii v)+Y_b(-\ii v)}{2}=Y_b(0)e(v).
\label{eq:even-part-normalized-kernels}
\end{equation}
The equation and initial data for $e(v)$ are real for physical $\nu^2$, including the heavy-field regime in which $\nu$ is imaginary. The integral entering both frozen modes is consequently real, and we denote it by
\begin{equation}
J\equiv\frac1\pi\int_0^\infty
\frac{e(v)}{1+v^2}\,\de v.
\label{eq:single-real-period}
\end{equation}
Inserting Eq.~\eqref{eq:even-part-normalized-kernels} into the frozen amplitude therefore gives
\begin{equation}
\varphi_b(k,0)=2\ii N_kJY_b(0).
\label{eq:late-phi}
\end{equation}
The late-time effect of each incoming excitation is thus determined by its kernel at the ordinary point $u=0$, multiplied by the same integral $J$. To evaluate this integral, recall that this mode is analytic at the entropy endpoint $u=r_c$ and satisfies $Y_1(1)=1$. The even solution $e(v)$ entering $J$ and the endpoint value $Y_1(r_c)$ obey an integral identity that gives
\begin{equation}
J
=\frac{1}{2P_{\nu-1/2}(0)\,Y_1(r_c)}
=\frac{\Gamma(\frac34-\frac\nu2)
\Gamma(\frac34+\frac\nu2)}
{2\sqrt\pi\,Y_1(r_c)}.
\label{eq:period-normalization-identity}
\end{equation}
Here $P_{\nu-1/2}$ is the Legendre function of the first kind. The identity holds for $\operatorname{Re}\nu<3/2$, the same condition that ensures convergence of the imaginary-axis integral. Its derivation is given in Appendix~\ref{app:period-transform-proof}, where an integral representation connects the even solution on the imaginary axis to the solution analytic at the entropy endpoint.

\subsection{The spectrum in terms of Heun functions}
\label{sec:explicit-Heun-spectrum}

Combining Eqs.~\eqref{eq:late-phi} and \eqref{eq:period-normalization-identity}, we obtain
\begin{equation}
\varphi_b(k,0)=\frac{\ii N_k}{\sqrt\pi}
\Gamma\!\left(\frac34-\frac\nu2\right)
\Gamma\!\left(\frac34+\frac\nu2\right)
\frac{Y_b(0)}{Y_1(r_c)}.
\label{eq:frozen-modes-gamma}
\end{equation}
Inserting this result back into \eqref{eq:power-definition} together with the results \eqref{eq:Y1} and \eqref{eq:Y2}, the mode sum yields the exact spectrum quoted in Eq.~\eqref{eq:intro-power-Heun}:
\begin{equationbox}
\begin{equation}
\begin{aligned}
\frac{\Delta_\zeta}{\Delta_0^{(c_\varphi)}}
={}&\frac{\left|\Gamma(\frac34-\frac\nu2)
\Gamma(\frac34+\frac\nu2)\right|^2}{\pi} \Bigg[
\left|\HeunG\!\left(1-r_c^{-2},Q_1;
\frac14-\frac\nu2,\frac14+\frac\nu2,1,\frac12;1\right)\right|^2\\
&\qquad+\frac{\lambda^2r_c}{4(r_c^2-1)^2}
\left|\HeunG\!\left(1-r_c^2,Q_2;
\frac54-\frac\nu2,\frac54+\frac\nu2,2,\frac12;1\right)\right|^2
\Bigg].
\end{aligned}
\label{eq:power-Heun}
\end{equation}
\end{equationbox}
Here $Q_1$ and $Q_2$ are the accessory parameters defined in Eqs.~\eqref{eq:Q1-normalized-Heun} and \eqref{eq:Q2-normalized-Heun}. The first term is the final curvature power produced by the incoming curvature wave, and the second is the power transferred from the incoming entropy wave. Both include the full coupled evolution, without an expansion in $\lambda$ or $r_c$. Since all momentum dependence of the frozen modes is carried by $N_k\propto k^{-3/2}$, the dimensionless spectrum is independent of $k$ for constant parameters. The Heun functions in Eq.~\eqref{eq:power-Heun} are evaluated by continuing their local solutions from argument zero to argument one along paths just above the real axis, passing above any intervening singularity. This prescription follows from the continuation used to construct the normalized modes.

As a first check, consider the limit $\lambda\to0$ directly in Eq.~\eqref{eq:power-Heun}, keeping $r_c\neq1$ fixed. The accessory parameter of the first Heun function can be written as
\begin{equation}
Q_1=
\left(1-r_c^{-2}\right)
\left(\frac14-\frac\nu2\right)
\left(\frac14+\frac\nu2\right)
+\frac{\lambda^2}{4r_c^2}.
\label{eq:Q1-unmixed-reduction}
\end{equation}
At zero mixing, its parameters therefore satisfy the conditions for the hypergeometric reduction
\begin{equation}
\HeunG\!\left(
a,a\alpha\beta;
\alpha,\beta,\gamma,\alpha+\beta+1-\gamma;z
\right)
={}_2F_1(\alpha,\beta;\gamma;z).
\label{eq:Heun-hypergeometric-reduction}
\end{equation}
In this identity, the singularity at $z=a$ disappears, leaving a hypergeometric function with the same unit normalization at the origin. Applying it to the first term of the power spectrum gives
\begin{equation}
\begin{aligned}
&\lim_{\lambda\to0}
\HeunG\!\left(
1-r_c^{-2},Q_1;
\frac14-\frac\nu2,\frac14+\frac\nu2,1,\frac12;1
\right)
\\
&\qquad =
{}_2F_1\!\left(
\frac14-\frac\nu2,\frac14+\frac\nu2;1;1
\right)
=
\frac{\sqrt\pi}
{\Gamma(\frac34-\frac\nu2)
\Gamma(\frac34+\frac\nu2)}.
\end{aligned}
\label{eq:unmixed-Heun-value}
\end{equation}
The last equality follows from Gauss summation, whose convergence condition is satisfied because the third hypergeometric parameter exceeds the sum of the first two by $1/2$. Notice that the dependence on $r_c$ has disappeared from this contribution. The second Heun value remains finite as $\lambda\to0$ at fixed unequal sound speeds, so its contribution vanishes with the explicit factor $\lambda^2$ in Eq.~\eqref{eq:power-Heun}. Substituting Eq.~\eqref{eq:unmixed-Heun-value} into the surviving term then gives
\begin{equation}
\lim_{\lambda\to0}
\frac{\Delta_\zeta}{\Delta_0^{(c_\varphi)}}
=
\frac{
\left|\Gamma(\frac34-\frac\nu2)
\Gamma(\frac34+\frac\nu2)\right|^2}{\pi}
\left|
\frac{\sqrt\pi}
{\Gamma(\frac34-\frac\nu2)
\Gamma(\frac34+\frac\nu2)}
\right|^2
=1.
\label{eq:unmixed-power-check}
\end{equation}
We thus recover the free curvature spectrum directly from the exact Heun expression, with no remaining dependence on the entropy mass or its propagation speed.

We next consider the equal-speed limit directly in Eq.~\eqref{eq:power-Heun}, keeping $\lambda\neq0$ and $\nu$ fixed. We first approach from $r_c>1$. As the singularities of the Heun functions merge, their expansions contain two oscillatory powers, $(r_c^2-1)^{-\ii w\lambda/2}$ with $w=\pm1$:
\begin{align}
&\HeunG\!\left(
1-r_c^{-2},Q_1;
\frac14-\frac\nu2,\frac14+\frac\nu2,1,\frac12;1
\right)
\nonumber\\
&\qquad =
\sum_{w=\pm1}
\ee^{\pi w\lambda/2}\mathcal C_w
(r_c^2-1)^{-\ii w\lambda/2}
+ o(1),
\label{eq:equal-limit-first-Heun}
\\[2mm]
&\frac{\lambda\sqrt{r_c}}{2(r_c^2-1)}
\HeunG\!\left(
1-r_c^2,Q_2;
\frac54-\frac\nu2,\frac54+\frac\nu2,2,\frac12;1
\right)
\nonumber\\
&\qquad =
-\ii\sum_{w=\pm1}
w\mathcal C_w
(r_c^2-1)^{-\ii w\lambda/2}
+o(1),
\label{eq:equal-limit-second-Heun}
\end{align}
where the common coefficients $\mathcal C_w$ are given by
\begin{equation}
\mathcal C_w=
\frac{
2^{2\ii w\lambda-1}
\Gamma(\frac12+\frac{\ii w\lambda}{2})^2
}{
\sqrt\pi\,
\Gamma(\frac34-\frac\nu2+\frac{\ii w\lambda}{2})
\Gamma(\frac34+\frac\nu2+\frac{\ii w\lambda}{2})
}.
\label{eq:equal-limit-connection-coefficient}
\end{equation}
These coefficients follow from the hypergeometric connection formulas for the merging singularities. The exponential weights in the first expression \eqref{eq:equal-limit-first-Heun} follow from continuing above its positive Heun singularity. Although the powers $(r_c^2-1)^{-\ii w\lambda/2}$ have unit modulus, their phases continue to rotate as $r_c\to1$, so the two contributions must be combined before taking the limit. To do so, we take the absolute square of each expression. The cross term between $w=+1$ and $w=-1$ in the first contribution carries the product $\ee^{\pi\lambda/2}\ee^{-\pi\lambda/2}=1$. In the second contribution, the factors $-\ii w$ instead give $(-\ii)(+\ii)^*=-1$. The leading interference terms therefore cancel, leaving only the diagonal terms:
\begin{equation}
\lim_{r_c\to1^+}
\frac{\Delta_\zeta}{\Delta_0^{(c_\varphi)}}
=
\frac{
\left|\Gamma(\frac34-\frac\nu2)
\Gamma(\frac34+\frac\nu2)\right|^2
}{\pi}
\sum_{w=\pm1}
\left(1+\ee^{\pi w\lambda}\right)
|\mathcal C_w|^2.
\label{eq:equal-limit-diagonal-sum}
\end{equation}
The remaining coefficients can be evaluated using $|\Gamma(\frac12+\ii y)|^2=\pi/\cosh(\pi y)$. For real entropy mass squared, $\nu$ is real or purely imaginary, so the denominator in Eq.~\eqref{eq:equal-limit-connection-coefficient} has the same modulus for both values of $w$. Consequently
\begin{equation}
|\mathcal C_w|^2
=
\frac{\pi}{
4\cosh^2(\frac{\pi\lambda}{2})
\left|
\Gamma(\frac34-\frac\nu2+\frac{\ii\lambda}{2})
\Gamma(\frac34+\frac\nu2+\frac{\ii\lambda}{2})
\right|^2
}.
\label{eq:equal-limit-coefficient-modulus}
\end{equation}
The sum of weights in Eq.~\eqref{eq:equal-limit-diagonal-sum} is $\sum_{w=\pm1}(1+\ee^{\pi w\lambda})=4\cosh^2(\pi\lambda/2)$, cancelling the hyperbolic factor in this expression. Approaching from $r_c<1$ interchanges the placement of the exponential weights between the two Heun contributions, while the oscillatory powers become $|r_c^2-1|^{-\ii w\lambda/2}$. The interference terms again cancel, and the diagonal sum is unchanged. Both approaches therefore give
\begin{equation}
\lim_{r_c\to1}
\frac{\Delta_\zeta}{\Delta_0^{(c_\varphi)}}
=
\left|
\frac{
\Gamma(\frac34-\frac\nu2)
\Gamma(\frac34+\frac\nu2)
}{
\Gamma(\frac34-\frac\nu2+\frac{\ii\lambda}{2})
\Gamma(\frac34+\frac\nu2+\frac{\ii\lambda}{2})
}
\right|^2,
\label{eq:equal-speed-limit-check}
\end{equation}
recovering Eq.~\eqref{eq:equal-speed-power} directly from the Heun expression.

\begin{figure}[t]
\centering
\includegraphics[width=.9\textwidth]{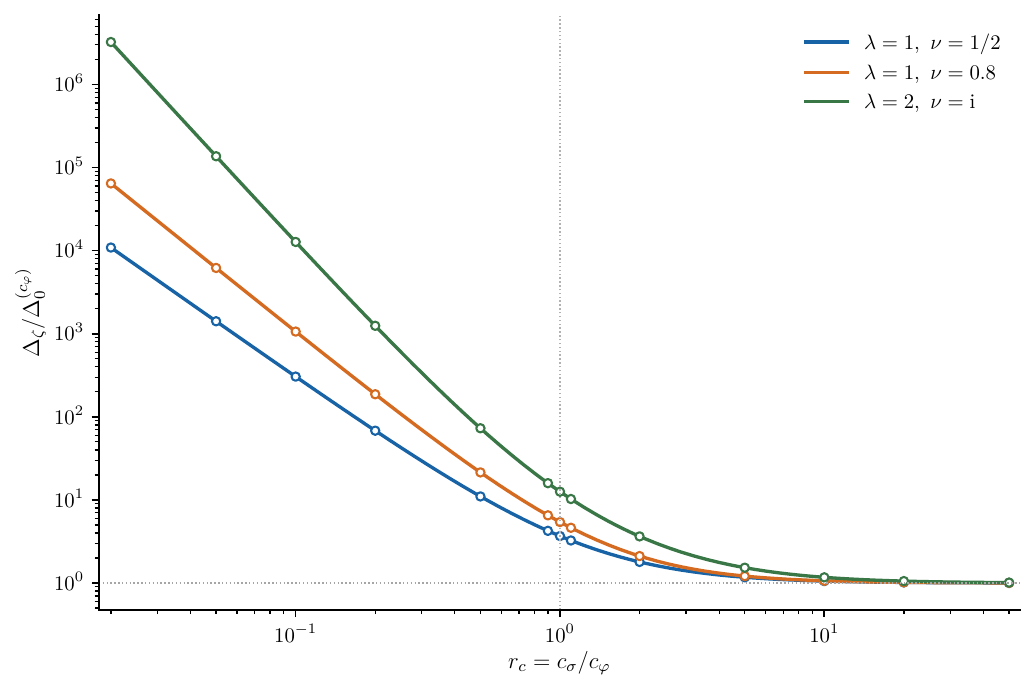}
\caption{Exact curvature power spectrum (solid curves) and independent solutions of the coupled time-domain equations (open circles), for two light entropy fields and one heavy example with $\nu=\ii$. At $r_c=1$, the spectral reference is the equal-speed gamma-function expression. All three examples have $\mu^2>0$ and $\nu^2+\lambda^2>0$. The dotted lines mark equal sound speeds and the unmixed curvature amplitude.}
\label{fig:power-numerical-validation}
\end{figure}

Figure~\ref{fig:power-numerical-validation} compares the exact spectrum with independent numerical solutions of the original coupled field equations, for two light entropy fields and one heavy example. The curves connect continuously across $r_c=1$ and show the different effects of an entropy fluctuation that propagates more slowly or more rapidly than curvature. At six representative points covering both sound-speed orderings and real and imaginary $\nu$, the time-domain spectra agree with the convergent-series evaluation to better than $2\times10^{-11}$ in relative power.\footnote{The numerical modes are initialized deep in the ultraviolet region with independent, canonically normalized Bunch--Davies data, and the leading decaying entropy contributions are removed before extracting the final curvature amplitudes. Reducing the final value of $x=-kc_\varphi\tau$ from $10^{-6}$ to $2\times10^{-7}$ and tightening the relative integration tolerance from $2\times10^{-10}$ to $2\times10^{-12}$ changes the tested spectra by less than $2\times10^{-10}$ in relative magnitude. For the three examples shown, the leading small-$r_c$ expression has relative errors between $0.11\%$ and $0.23\%$ at $r_c=0.02$, while the large-$r_c$ expression through order $r_c^{-3}$ has relative errors between $5.1\times10^{-6}$ and $6.1\times10^{-5}$ at $r_c=50$. The initial-data recurrences, late-time extrapolation, and further comparisons are included in the accompanying numerical material.}

\section{Sound-speed hierarchies of the exact spectrum}
\label{sec:sound-speed-hierarchies}

We now consider the limits $r_c\ll1$ and $r_c\gg1$, in which the entropy fluctuation propagates much more slowly or much more rapidly than the curvature fluctuation, respectively. In the first case, the entropy fluctuation crosses its sound horizon before the curvature fluctuation, whereas the order is reversed in the second case. The interval between these crossings allows the fields to evolve and exchange power under different conditions, whose consequences can be extracted directly from the exact spectrum in Eq.~\eqref{eq:power-Heun}. Throughout this section we keep the mass and mixing fixed, with $\mu^2>0$ so that the curvature perturbation approaches a constant at late times.

To follow how each incoming fluctuation contributes to the final spectrum, we denote the power generated by the incoming curvature and entropy waves by $\Delta_{\zeta,1}$ and $\Delta_{\zeta,2}$, respectively:
\begin{equation}
\Delta_\zeta=\Delta_{\zeta,1}+\Delta_{\zeta,2},
\qquad
\Delta_{\zeta,b}
=\left|\frac{\varphi_b(k,0)}{N_k}\right|^2
\Delta_0^{(c_\varphi)}.
\label{eq:small-r-mode-contributions}
\end{equation}
These are the two contributions already identified in Eq.~\eqref{eq:power-Heun}, each including the full coupled evolution. Their limiting behavior follows from the corresponding Heun values, together with the explicit sound-speed dependence multiplying the second term.

\subsection{Small sound-speed ratio}
\label{sec:small-kappa}
\label{sec:small-r-power}

Consider first $r_c\ll1$. For the first Heun function in Eq.~\eqref{eq:power-Heun}, the singularity parameter $1-r_c^{-2}$ tends to $-\infty$. The ratio $Q_1/(1-r_c^{-2})$ approaches $1/16-(\nu^2+\lambda^2)/4$, so the limiting hypergeometric parameters depend on the combination
\begin{equation}
\chi\equiv\sqrt{\nu^2+\lambda^2},
\label{eq:chi-L}
\end{equation}
where we use the principal branch of the square root. The endpoint value follows by taking the hypergeometric limit and applying Gauss summation~\cite{DLMF:Heun,DLMF:Hypergeometric}:
\begin{equation}
\begin{aligned}
\HeunG\!\left(
1-r_c^{-2},Q_1;
\frac14-\frac\nu2,\frac14+\frac\nu2,1,\frac12;1
\right) &=
{}_2F_1\!\left(
\frac14-\frac\chi2,\frac14+\frac\chi2;1;1
\right)
+\mathcal O\left(r_c^2\right)
\\
& =
\frac{\sqrt\pi}
{\Gamma(\frac34-\frac\chi2)
\Gamma(\frac34+\frac\chi2)}
+\mathcal O\left(r_c^2\right).
\end{aligned}
\label{eq:Heun1-small-r}
\end{equation}
Substitution into the first term of the exact spectrum therefore gives
\begin{equation}
\lim_{r_c\to0}
\frac{\Delta_{\zeta,1}}{\Delta_0^{(c_\varphi)}}
=
\left|
\frac{
\Gamma(\frac34-\frac\nu2)
\Gamma(\frac34+\frac\nu2)
}{
\Gamma(\frac34-\frac\chi2)
\Gamma(\frac34+\frac\chi2)
}
\right|^2.
\label{eq:small-r-curvature-contribution}
\end{equation}
Thus the power generated by the incoming curvature wave remains finite in this hierarchy, although its limiting value depends on the mixing through $\chi$.

The second Heun function behaves differently because its additional singularity, at $1-r_c^2$, approaches the evaluation point. Writing its argument as $z$, the limiting equation outside the narrow region $|1-z|\sim r_c^2$ has solutions with powers $(1-z)^{-1/4-\chi/2}$ and $(1-z)^{-1/4+\chi/2}$. Connecting these solutions through the narrow region determines the endpoint value. For real $\chi>0$, the first power gives the dominant contribution:
\begin{equation}
\begin{aligned}
&\HeunG\!\left(
1-r_c^2,Q_2;
\frac54-\frac\nu2,\frac54+\frac\nu2,2,\frac12;1
\right)
\\
&\qquad \simeq
\frac{
2\sqrt\pi\,
\ee^{\ii\pi(\chi+1/2)/2}
\mathcal C(\lambda,\nu)
}{
\Gamma(\frac34-\frac\nu2)
\Gamma(\frac34+\frac\nu2)
}
r_c^{-\chi-1/2},
\qquad r_c\to0,
\end{aligned}
\label{eq:Heun2-small-r}
\end{equation}
where
\begin{equation}
\mathcal C(\lambda,\nu)=
\frac{
\chi\,\Gamma(\chi)^2
}{
2\,\Gamma(\frac34+\frac\chi2)^2
}
\frac{
\Gamma(\frac34-\frac\nu2)
\Gamma(\frac34+\frac\nu2)
}{
\Gamma(1+\frac{\chi-\nu}{2})
\Gamma(1+\frac{\chi+\nu}{2})
}.
\label{eq:C-small-kappa}
\end{equation}
The connection uses hypergeometric solutions outside the separation of the singularities and Legendre solutions within it, whose matching is given in Appendix~\ref{app:small-kappa-integral}. The phase in Eq.~\eqref{eq:Heun2-small-r} follows from continuing above the intervening singularity in the Heun argument, according to the prescription established in Section~\ref{sec:power-exact}. For real $\chi$, this phase has unit modulus and drops out of the spectrum.

We can now insert Eq.~\eqref{eq:Heun2-small-r} into the second term of Eq.~\eqref{eq:power-Heun}. The mass-dependent gamma functions cancel those in the overall prefactor, while the explicit factor $r_c$ cancels the extra inverse power in the squared Heun value. At fixed nonzero mixing and real $\chi>0$, the resulting entropy contribution grows as $r_c^{-2\chi}$ and eventually exceeds the finite curvature contribution:
\begin{equation}
\frac{\Delta_\zeta}{\Delta_0^{(c_\varphi)}}
\simeq
\frac{\Delta_{\zeta,2}}{\Delta_0^{(c_\varphi)}}
\simeq
\lambda^2
\left|\mathcal C(\lambda,\nu)\right|^2
r_c^{-2\chi},
\qquad r_c\to0,\quad \chi>0.
\label{eq:power-small-kappa}
\end{equation}
This result retains the full dependence on the mixing. At a finite sound-speed ratio, the approximation to the total power requires the transferred contribution to dominate Eq.~\eqref{eq:small-r-curvature-contribution}. The hierarchy needed to reach this regime depends on both the exponent and its coefficient.

The weak-mixing limit illustrates how a modest coupling can nevertheless produce a large enhancement. At fixed real $0<\nu<3/2$, the coefficient approaches
\begin{equation}
\lim_{\lambda\to0}\mathcal C(\lambda,\nu)
=
\frac12\Gamma(\nu)
\frac{
\Gamma(\frac34-\frac\nu2)
}{
\Gamma(\frac34+\frac\nu2)
}.
\label{eq:C-weak}
\end{equation}
Replacing $\chi$ by $\nu$ in the power of $r_c$ additionally requires $(\lambda^2/\nu)|\ln r_c|\ll1$, because even a small change in the exponent can accumulate over a sufficiently large hierarchy. For the conformal mass $\nu=1/2$, this gives
\begin{equation}
\frac{\Delta_\zeta}{\Delta_0^{(c_\varphi)}}
\simeq
\frac{\pi^2\lambda^2}{4r_c},
\qquad
\lambda\ll1,\quad
\lambda^2|\ln r_c|\ll1,\quad
\frac{\lambda^2}{r_c}\gg1.
\label{eq:small-kappa-conformal}
\end{equation}
The last condition ensures that the transferred power dominates the contribution approaching the unmixed spectrum. These conditions also make clear why taking the mixing to zero and taking a large sound-speed hierarchy need not give the same leading approximation.

\subsection{Logarithmic growth and oscillatory dependence}
\label{sec:small-r-nonreal-chi}

The power law in Eq.~\eqref{eq:power-small-kappa} assumes real $\chi>0$, for which one of the two endpoint powers dominates. When $\chi$ vanishes or becomes imaginary, both powers must be retained. Their connection coefficients are obtained from the same gamma-function expression in Eq.~\eqref{eq:C-small-kappa}, evaluated at $\chi$ and $-\chi$, and the combined expression is given in Appendix~\ref{app:small-kappa-integral}.

At $\chi=0$, corresponding to $\nu^2=-\lambda^2$, the two coefficients have opposite simple poles. Their sum remains well defined because the difference between the accompanying powers vanishes at the same rate:
\begin{equation}
\lim_{\chi\to0}
\frac{r_c^{-\chi}-r_c^\chi}{\chi}
=
2\ln\frac1{r_c}.
\label{eq:small-r-logarithmic-limit}
\end{equation}
Consequently, the entropy contribution grows logarithmically in amplitude and eventually dominates the finite curvature contribution. Keeping the residue of the two coefficients gives
\begin{equation}
\frac{\Delta_\zeta}{\Delta_0^{(c_\varphi)}}
\simeq
\lambda^2
\left|
\frac{
\Gamma(\frac34-\frac\nu2)
\Gamma(\frac34+\frac\nu2)
}{
\Gamma(\frac34)^2
\Gamma(1-\frac\nu2)
\Gamma(1+\frac\nu2)
}
\right|^2
\ln^2\!\frac1{r_c},
\qquad
r_c\to0,\quad \chi=0.
\label{eq:small-r-critical-power}
\end{equation}
Here the mixing is held fixed and nonzero. The logarithm follows from combining the two branches before taking their coincident-power limit.

For imaginary $\chi$, we write $\chi=\ii\eta$, with $\eta=\sqrt{-\nu^2-\lambda^2}>0$. The powers $r_c^{\mp\chi}$ now have unit modulus and describe opposite phases. The two connection coefficients are complex conjugates, while their continuation factors become $\ee^{\mp\pi\eta/2}$. Evaluating $\mathcal C(\lambda,\nu)$ on the branch $\chi=\ii\eta$, the entropy contribution therefore takes the form
\begin{equation}
\begin{aligned}
\frac{\Delta_{\zeta,2}}{\Delta_0^{(c_\varphi)}}
={}&
\lambda^2
\left|
\ee^{-\pi\eta/2}\mathcal C(\lambda,\nu)r_c^{-\ii\eta}
+
\ee^{\pi\eta/2}\mathcal C(\lambda,\nu)^*r_c^{\ii\eta}
\right|^2
+\mathcal O(1)
\\
={}&
2\lambda^2
\left[
\cosh(\pi\eta)
\left|\mathcal C(\lambda,\nu)\right|^2
+
\operatorname{Re}\!\left(
\mathcal C(\lambda,\nu)^2r_c^{-2\ii\eta}
\right)
\right]
+ \mathcal O(1),
\qquad r_c\to0.
\end{aligned}
\label{eq:small-r-oscillatory-power}
\end{equation}
The curvature contribution still approaches Eq.~\eqref{eq:small-r-curvature-contribution}, now evaluated at $\chi=\ii\eta$. The total spectrum thus remains bounded and oscillates as a function of $\ln r_c$, with period $\pi/\eta$. These oscillations record the relative phase accumulated between the two sound horizons, following the early-time propagation described in Section~\ref{sec:early-time-propagation}.

\subsection{Large sound-speed ratio}
\label{sec:large-kappa}
\label{sec:large-r-power}

We now turn to $r_c\gg1$. In this hierarchy, it is the first Heun function whose additional singularity approaches the evaluation point, since $1-r_c^{-2}\to1$. Its accessory parameter $Q_1$ approaches $1/16-\nu^2/4$, and the leading endpoint value is the hypergeometric value already obtained in Eq.~\eqref{eq:unmixed-Heun-value}. Substituting this value into the first term of Eq.~\eqref{eq:power-Heun} cancels the gamma-function prefactor and gives the unit unmixed curvature spectrum.

The first correction depends on the small separation $r_c^{-2}$ between the singularities. At fixed mixing, the perturbation is proportional to $\lambda^2/r_c^2$, but its connection to the endpoint produces an additional logarithm. In particular, the solution in the range $r_c^{-2}\ll|1-z|\ll1$ contains a logarithmic term whose continuation into the endpoint region gives $\ln r_c$. The finite part of the same connection is expressed through digamma functions. Together, these contributions give
\begin{equation}
\begin{aligned}
&\left|
\HeunG\!\left(
1-r_c^{-2},Q_1;
\frac14-\frac\nu2,\frac14+\frac\nu2,1,\frac12;1
\right)
\right|^2
\\
&\qquad =
\frac{\pi}{
\left|
\Gamma(\frac34-\frac\nu2)
\Gamma(\frac34+\frac\nu2)
\right|^2
}
\Bigg\{
1+\frac{\lambda^2}{r_c^2}
\left[2\ln r_c+\mathcal B(\nu)\right]
\\
&\hspace{43mm}
+\mathcal O\!\left(
\frac{\lambda^2\ln r_c+\lambda^4\ln^2r_c}{r_c^4}
\right)
\Bigg\},
\end{aligned}
\label{eq:Heun1-large-r-modulus}
\end{equation}
where
\begin{equation}
\begin{aligned}
\mathcal B(\nu)
= 
2-2\ln2+\psi\!\left(\frac12\right)+\psi(1)
-\psi\!\left(\frac34-\frac\nu2\right)
-\psi\!\left(\frac34+\frac\nu2\right).
\end{aligned}
\label{eq:B-large-kappa}
\end{equation}
Here $\psi$ is the digamma function, and the combination above is real for real $\nu^2$, including heavy entropy fields with imaginary $\nu$. The complex Heun value also contains phase corrections. Its term of order $r_c^{-3}$ is purely imaginary relative to the leading value, which explains why no term of that order appears in Eq.~\eqref{eq:Heun1-large-r-modulus}. Appendix~\ref{app:large-kappa-green} gives the connection calculation, including these phases and the stated remainder.

For the second Heun function, the singularity parameter $1-r_c^2$ tends to $-\infty$, while $Q_2/(1-r_c^2)\to1/2$. Its limiting hypergeometric function has parameters $1$, $1/2$ and $2$, so
\begin{equation}
\begin{aligned}
&\HeunG\!\left(
1-r_c^2,Q_2;
\frac54-\frac\nu2,\frac54+\frac\nu2,2,\frac12;1
\right)
\\
&\qquad =
{}_2F_1\!\left(1,\frac12;2;1\right)
+\mathcal O(r_c^{-2})
=
2+\mathcal O(r_c^{-2}).
\end{aligned}
\label{eq:Heun2-large-r}
\end{equation}
The Heun value remains finite, but its contribution to the spectrum is suppressed by the explicit factor $r_c/(r_c^2-1)^2$. Substitution into Eq.~\eqref{eq:power-Heun} gives
\begin{equation}
\frac{\Delta_{\zeta,2}}{\Delta_0^{(c_\varphi)}}
=
\frac{\lambda^2}{\pi r_c^3}
\left|
\Gamma\!\left(\frac34-\frac\nu2\right)
\Gamma\!\left(\frac34+\frac\nu2\right)
\right|^2
+\mathcal O(r_c^{-5}).
\label{eq:entropy-large-kappa}
\end{equation}
Thus the power transferred from the incoming entropy wave is smaller than the leading correction to the curvature contribution by one inverse power of $r_c$, apart from the logarithm. Since the latter has no term of order $r_c^{-3}$, the entropy contribution supplies the complete correction at that order. Adding the two terms gives
\begin{equation}
\begin{aligned}
\frac{\Delta_\zeta}{\Delta_0^{(c_\varphi)}}
&=
1+\frac{\lambda^2}{r_c^2}
\left[2\ln r_c+\mathcal B(\nu)\right]
+
\frac{\lambda^2}{\pi r_c^3}
\left|
\Gamma\!\left(\frac34-\frac\nu2\right)
\Gamma\!\left(\frac34+\frac\nu2\right)
\right|^2
\\
&+
\mathcal O\!\left(
\frac{\lambda^2\ln r_c+\lambda^4\ln^2r_c}{r_c^4}
\right).
\end{aligned}
\label{eq:power-large-kappa}
\end{equation}

For the conformal mass $\nu=1/2$, the digamma terms simplify to $\mathcal B(1/2)=2-2\ln2$, while the gamma-function product is $\sqrt\pi$. The spectrum becomes
\begin{equation}
\begin{aligned}
\frac{\Delta_\zeta}{\Delta_0^{(c_\varphi)}}
= 
1+\frac{2\lambda^2}{r_c^2}
\left(\ln r_c+1-\ln2\right)
+\frac{\lambda^2}{r_c^3}
 +
\mathcal O\!\left(
\frac{\lambda^2\ln r_c+\lambda^4\ln^2r_c}{r_c^4}
\right).
\end{aligned}
\label{eq:large-kappa-conformal-total}
\end{equation}
This expression, like Eq.~\eqref{eq:power-large-kappa}, is an expansion in the sound-speed hierarchy at fixed mixing. All displayed corrections vanish as $r_c\to\infty$, so this hierarchy gives a decoupling limit even when $\lambda$ is large. At finite $r_c$, the approximation requires the corrections to be small, with the necessary hierarchy depending on the mass and mixing. In particular, the limit is not uniform as $\mu^2\to0$, where the final entropy decay becomes arbitrarily slow.

\subsection{Physical interpretation of the two hierarchies}
\label{sec:hierarchy-physical-interpretation}

The different limits can be understood by following the fluctuations through the interval separating their sound-horizon crossings. For $r_c\ll1$, the entropy fluctuation crosses first. Its gradient becomes small while the curvature fluctuation still has a large gradient term, which suppresses the curvature response. During this interval, the entropy fluctuation therefore evolves approximately with the mass coefficient obtained by expanding the square in Eq.~\eqref{eq:action}:
\begin{equation}
m_0^2\equiv\mu^2-\lambda^2H^2,
\qquad
\chi^2=\frac94-\frac{m_0^2}{H^2}.
\label{eq:intermediate-mass-definition}
\end{equation}
In the part of the interval where $c_\varphi k/a$ exceeds $H$ and the relevant mass and mixing frequencies, the two entropy solutions behave approximately as $a^{-3/2\pm\chi}$. At fixed mass and mixing, the duration of this evolution grows as $\ln(1/r_c)$. For real $\chi>0$, the more slowly decaying solution therefore acquires a factor proportional to $r_c^{3/2-\chi}$ between the two crossings. Combining this evolution with the entropy vacuum normalization gives
\begin{equation}
c_\sigma^{-3/2}r_c^{3/2-\chi}
=
c_\varphi^{-3/2}r_c^{-\chi},
\label{eq:small-r-hierarchy-scaling}
\end{equation}
which explains the exponent in Eq.~\eqref{eq:power-small-kappa}. The mass and mixing scales affect the beginning and end of this approximate evolution, and hence its coefficient, while the growing separation between the gradient scales determines its dependence on $r_c$. Once the curvature gradient also becomes negligible, the coupled system enters its final superhorizon regime. The entropy decay is then governed by $\mu$, ensuring that the curvature spectrum freezes for $\mu^2>0$. The distinction between these two stages explains why $\chi$ controls the sound-speed hierarchy even though $\nu$ controls the final entropy behavior.

An enhanced spectrum need not require the entropy fluctuation to grow with time. For $0<m_0^2<9H^2/4$, the intermediate entropy solution decays, but retains enough of its larger vacuum amplitude to enhance the final curvature power. Actual intermediate growth occurs for $m_0^2<0$, corresponding to $\chi>3/2$. At $m_0^2=9H^2/4$, the two intermediate decay powers coincide and their independent solutions include a logarithm. Above this mass, the powers become oscillatory, and the relative phase accumulated over the interval between the two crossings gives the oscillations in $\ln r_c$ found in Eq.~\eqref{eq:small-r-oscillatory-power}.

The possibility of intermediate growth can also be examined through the local dispersion relation. Freezing the background and introducing physical momentum $p=k/a$ and a local frequency $\Omega$, the coupled system gives
\begin{equation}
\bigl(c_\varphi^2p^2-\Omega^2\bigr)
\bigl(c_\sigma^2p^2+m_0^2-\Omega^2\bigr)
-\lambda^2H^2\Omega^2=0.
\label{eq:local-mixed-dispersion}
\end{equation}
For positive squared sound speeds and $\mu^2>0$, the lower frequency-squared branch is negative at nonzero momentum precisely when $c_\sigma^2p^2+m_0^2<0$. This identifies a range of local growth when the mode evolves rapidly compared with the expansion, while the full de Sitter evolution determines how the fluctuation enters and leaves this range. The time-kinetic terms remain positive. Related transient amplification has been studied in sidetracked inflation and in more general multifield systems~\cite{GarciaSaenz:2018sidetracked,Christodoulidis:2023enhanced}. When the entropy mode can be eliminated, this behavior can appear as an imaginary sound speed in the effective curvature theory~\cite{GarciaSaenz:2018imaginary}, even though both propagation speeds in the two-field description are positive.

For $r_c\gg1$, the order of the crossings is reversed. The curvature fluctuation is already nearly frozen while the entropy gradient remains large, suppressing the entropy response to curvature. In this interval, the residual evolution of the free curvature mode is $\dot\varphi_1\simeq-c_\varphi^2p^2\varphi_1/H$. Where $c_\sigma p$ also exceeds the relevant mass and mixing frequencies, balancing the entropy gradient against its curvature source gives
\begin{equation}
\sigma_1
\simeq
-\frac{\lambda H\dot\varphi_1}{c_\sigma^2p^2}
\simeq
\frac{\lambda}{r_c^2}\varphi_1.
\label{eq:large-r-quasistatic-entropy}
\end{equation}
Although small, this approximately constant entropy response continues to source curvature. After the decaying transient in the curvature response has become negligible, the induced correction obeys $\delta\dot\varphi_1\simeq\lambda H\sigma_1$. The interval between the two gradient scales lasts approximately $\Delta N\simeq\ln r_c$ e-folds, so
\begin{equation}
\frac{\delta\varphi_1}{\varphi_1}
\simeq
\frac{\lambda^2}{r_c^2}\Delta N
\simeq
\frac{\lambda^2\ln r_c}{r_c^2}.
\label{eq:large-r-feedback-interpretation}
\end{equation}
Taking the squared amplitude gives the factor of two in the logarithmic power correction. The transitions into and out of this interval, together with the final entropy decay, determine the finite mass-dependent term $\mathcal B(\nu)$.

\begin{figure}[t]
\centering
\includegraphics[width=\textwidth]{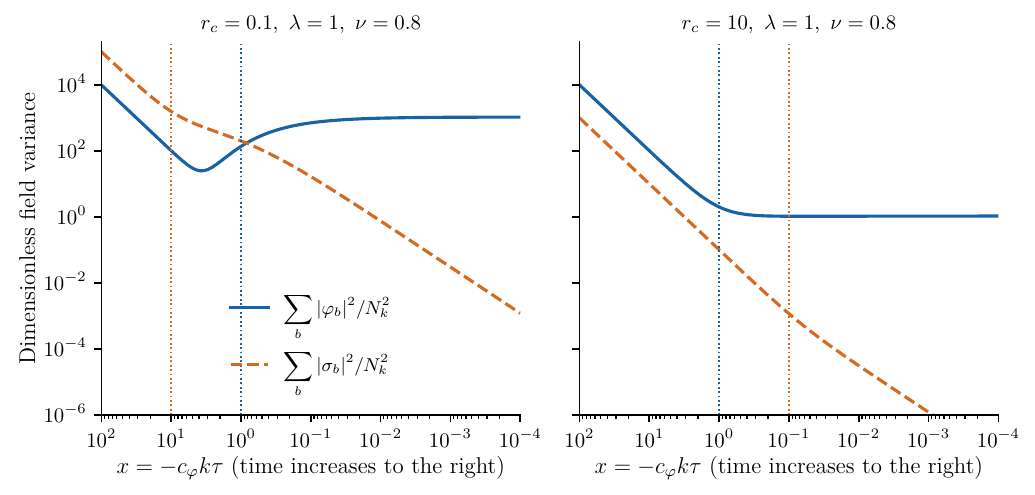}
\caption{Evolution of the curvature and entropy mode sums for $\lambda=1$ and $\nu=0.8$, shown by the solid and dashed curves, respectively, in units of $N_k^2$. Writing $x=-kc_\varphi\tau$, the vertical lines mark the bare sound-horizon scales $x=1$ and $x=1/r_c$. Time increases to the right. For $r_c=0.1$, the entropy fluctuation crosses its sound horizon first and transfers an enhanced amplitude to curvature. For $r_c=10$, the curvature fluctuation crosses first and the correction to its final power is small.}
\label{fig:mode-evolution}
\end{figure}

Figure~\ref{fig:mode-evolution} illustrates the coupled evolution for the two orderings of the sound speeds. When $r_c<1$, the entropy fluctuation crosses first and transfers an enhanced amplitude to curvature. When $r_c>1$, the curvature fluctuation crosses first and its subsequent evolution receives a much smaller correction, in agreement with the decoupling limit of Eq.~\eqref{eq:power-large-kappa}.

\section{Squeezed limit of the bispectrum}
\label{sec:squeezed}

We now use the normalized modes to compute the three-point function of the curvature perturbation, focusing on squeezed configurations in which one momentum is much smaller than the other two. Taking $k_1=k_2=k_S$ and $k_3=k_L$, we characterize this limit by $q\equiv k_L/k_S\ll1$. At fixed mass, mixing and sound-speed ratio, the long fluctuation then reaches its late-time regime before the short fluctuations undergo the evolution relevant to their interaction. This allows us to extract the characteristic dependence on $q$ from the long mode while retaining the full coupled evolution of the short modes. Throughout this section we assume $\mu^2>0$ and work to first order in the cubic interactions, keeping the quadratic mixing exact.

\subsection{Cubic interactions and the exact bispectrum integrals}
\label{sec:squeezed-boundary-kernels}

We consider the three interactions studied in Ref.~\cite{Huenupi:2026aqc}, expressed with the field normalization used in the preceding sections:
\begin{equation}
S_{\rm int}^{(3)}
=
\frac{1}{F_\zeta}
\int\de\tau\,\de^3x
\left[
\frac{a^2\alpha}{2}\sigma(D_\tau\varphi)^2
+\frac{a^3\beta H}{2}\sigma^2D_\tau\varphi
+\frac{a^4\gamma H^2}{3!}\sigma^3
\right].
\label{eq:cubic-action-squeezed}
\end{equation}
Here $\alpha$, $\beta$ and $\gamma$ are real dimensionless couplings, and $D_\tau\varphi=\varphi'+\lambda\sigma/\tau$ is the momentum combination already appearing in the quadratic action. These operators provide a direct comparison with our equal-speed results. A specified inflationary effective theory can also contain spatial-derivative interactions, including operators associated with the mixing and the sound speeds~\cite{Cheung:2007st,GarciaSaenz:2019cubic,Pinol:2024EFT}. We consider the contribution of each interaction in Eq.~\eqref{eq:cubic-action-squeezed} separately. The interaction-picture fields evolve with the complete mixed quadratic Hamiltonian. At cubic order, the Legendre transform gives $H_3=-L_3$, with the velocities in $L_3$ evaluated using this quadratic Hamiltonian. The bispectrum therefore follows from a single insertion of $H_3$ in the in-in expectation value~\cite{Weinberg:2005vy,Adshead:2009cb}, with all mixing effects included in the field modes.

To proceed, we define the bispectrum by
\begin{equation}
\langle
\zeta_{\boldsymbol k_1}
\zeta_{\boldsymbol k_2}
\zeta_{\boldsymbol k_3}
\rangle
=
(2\pi)^3
\delta^{(3)}(\boldsymbol k_1+\boldsymbol k_2+\boldsymbol k_3)
B_\zeta(k_1,k_2,k_3),
\label{eq:bispectrum-definition-unequal}
\end{equation}
and separate the amplitude and shape of each interaction contribution following the convention of Ref.~\cite{Huenupi:2026aqc}:
\begin{equation}
B_I(k_1,k_2,k_3)
=
\frac{(2\pi)^4\Delta_\zeta^2}{(k_1k_2k_3)^2}
\frac9{10}f_{\rm NL}^{(I)}
S_I(k_1,k_2,k_3),
\qquad I=\alpha,\beta,\gamma.
\label{eq:reduced-shape-unequal}
\end{equation}
The shape is normalized at equilateral momenta, so that
\begin{equation}
S_I(k,k,k)=1,
\qquad
f_{\rm NL}^{(I)}
=
\frac{10}{9}
\frac{k^6B_I(k,k,k)}
{(2\pi)^4\Delta_\zeta^2}.
\label{eq:equilateral-bispectrum-normalization}
\end{equation}
Here $\Delta_\zeta$ is the full curvature spectrum obtained in Section~\ref{sec:power-exact}. For constant parameters, $f_{\rm NL}^{(I)}$ is independent of the overall momentum scale, while $S_I$ depends on the shape of the momentum triangle. At the order considered here, the overall cubic coupling is contained in $f_{\rm NL}^{(I)}$, and the normalized shape describes the momentum dependence determined by the mass, quadratic mixing and sound speeds.

To construct the three-point function, each field at the cubic vertex must be contracted with a final curvature perturbation. These contractions are determined by the normalized modes $\varphi_b(k,\tau)$ and $\sigma_b(k,\tau)$ obtained above. In particular, the final curvature amplitudes for unequal sound speeds are
\begin{equation}
\varphi_b(k,0)
=
\frac{\ii N_k}{\sqrt\pi}
\Gamma\!\left(\frac34-\frac\nu2\right)
\Gamma\!\left(\frac34+\frac\nu2\right)
\frac{Y_b(0)}{Y_1(r_c)},
\qquad
N_k=\frac{H}{\sqrt{2c_\varphi^3k^3}},
\label{eq:bispectrum-frozen-amplitudes}
\end{equation}
for both modes $b=1,2$, as derived in Eq.~\eqref{eq:frozen-modes-gamma}. The same Heun values that determine the power spectrum therefore also fix the external curvature amplitudes in the bispectrum. For the entropy field and the differentiated curvature field, the contractions take the form~\cite{Chen:2017ryl}
\begin{align}
\langle
\zeta_{\boldsymbol k}(0)
\sigma_{-\boldsymbol k}(\tau)
\rangle'
&=
\frac{1}{F_\zeta}
\sum_{b=1}^2
\varphi_b(k,0)\sigma_b^*(k,\tau),
\nonumber\\
\langle
\zeta_{\boldsymbol k}(0)
D_\tau\varphi_{-\boldsymbol k}(\tau)
\rangle'
&=
\frac{1}{F_\zeta}
\sum_{b=1}^2
\varphi_b(k,0)
\bigl[D_\tau\varphi_b(k,\tau)\bigr]^*,
\label{eq:boundary-functions-bispectrum}
\end{align}
where the prime removes the momentum-conserving delta function. Each sum includes both independent vacuum fluctuations, weighted by their final curvature amplitudes. These sums are invariant under unitary rotations of the normalized vacuum modes, as is the combination determining the power spectrum,
\begin{equation}
\sum_{b=1}^2|\varphi_b(k,0)|^2
=
N_k^2
\frac{\Delta_\zeta}{\Delta_0^{(c_\varphi)}}.
\label{eq:power-ratio-bispectrum}
\end{equation}

We first retain the full time dependence at all three momenta. For the entropy self-interaction, the three contractions give
\begin{equation}
\begin{aligned}
B_\gamma(k_S,k_S,k_L)
={}&
-\frac{2\gamma H^2}{F_\zeta^4}
\operatorname{Im}
\sum_{b,c,d=1}^2
\varphi_b(k_S,0)
\varphi_c(k_S,0)
\varphi_d(k_L,0)
\\
&\times
\int_{-\infty}^0
\de\tau\,a^4
\sigma_b^*(k_S,\tau)
\sigma_c^*(k_S,\tau)
\sigma_d^*(k_L,\tau).
\end{aligned}
\label{eq:exact-time-gamma}
\end{equation}
The six equivalent contractions cancel the factor $3!$ in the interaction. The early limit of this integral is understood with the Bunch--Davies continuation $\tau=-R(1-\ii\epsilon)$, with $\epsilon>0$ and $R\to+\infty$, taking $\epsilon\to0^+$ after evaluating the integral. This continuation damps the conjugate modes appearing in the integrand. We use the same prescription for the other interactions. For the $\beta$ interaction, the differentiated curvature field can carry either of the two short momenta or the long momentum, giving
\begin{equation}
\begin{aligned}
B_\beta(k_S,k_S,k_L)
={}&
-\frac{2\beta H}{F_\zeta^4}
\operatorname{Im}
\sum_{b,c,d=1}^2
\varphi_b(k_S,0)
\varphi_c(k_S,0)
\varphi_d(k_L,0)
\\
&\times
\int_{-\infty}^0
\de\tau\,a^3
\Bigl[
2\bigl[D_\tau\varphi_b(k_S,\tau)\bigr]^*
\sigma_c^*(k_S,\tau)
\sigma_d^*(k_L,\tau)
\\
&\hspace{31mm}
+
\sigma_b^*(k_S,\tau)
\sigma_c^*(k_S,\tau)
\bigl[D_\tau\varphi_d(k_L,\tau)\bigr]^*
\Bigr].
\end{aligned}
\label{eq:exact-time-beta}
\end{equation}
For the $\alpha$ interaction, the entropy field can carry the long momentum, leaving both differentiated fields on the short legs, or it can carry either short momentum. The result is
\begin{equation}
\begin{aligned}
B_\alpha(k_S,k_S,k_L)
={}&
-\frac{2\alpha}{F_\zeta^4}
\operatorname{Im}
\sum_{b,c,d=1}^2
\varphi_b(k_S,0)
\varphi_c(k_S,0)
\varphi_d(k_L,0)
\\
&\times
\int_{-\infty}^0
\de\tau\,a^2
\Bigl[
\bigl[D_\tau\varphi_b(k_S,\tau)\bigr]^*
\bigl[D_\tau\varphi_c(k_S,\tau)\bigr]^*
\sigma_d^*(k_L,\tau)
\\
&\hspace{20mm}
+
2\sigma_b^*(k_S,\tau)
\bigl[D_\tau\varphi_c(k_S,\tau)\bigr]^*
\bigl[D_\tau\varphi_d(k_L,\tau)\bigr]^*
\Bigr].
\end{aligned}
\label{eq:exact-time-alpha}
\end{equation}
All fields at a vertex are evaluated at the same conformal time, and the momentum dependence of each derivative is contained in $D_\tau\varphi_b(k,\tau)$. At the final endpoint, the imaginary part is taken at a common time cutoff before that cutoff is removed. This preserves the cancellations required by the in-in commutator, which are particularly important for light entropy fields. These expressions apply to every isosceles triangle with $0<q\le2$. Evaluating them at $q=1$ determines $f_{\rm NL}^{(I)}$ through Eq.~\eqref{eq:equilateral-bispectrum-normalization}, while the normalized isosceles shape is
\begin{equation}
S_I(q)
\equiv S_I(k_S,k_S,qk_S)
=
q^2
\frac{B_I(k_S,k_S,qk_S)}
{B_I(k_S,k_S,k_S)}.
\label{eq:isosceles-equilateral-normalization}
\end{equation}
The equilateral amplitude thus follows from the exact time integrals, and the squeezed approximation concerns the behavior of this shape as $q\to0$. The common power-spectrum normalization cancels from this ratio.

\subsection{The long entropy mode and the squeezed coefficients}
\label{sec:bispectrum-long-mode}
\label{sec:unequal-squeezed-shapes}
\label{sec:finite-squeezed-coefficients}

We now extract the squeezed dependence by resolving the long fluctuation into its late-time branches. For a mode of momentum $k$, the linear equations can be written as
\begin{align}
\bigl(D_\tau\varphi_b\bigr)'
-\frac2\tau D_\tau\varphi_b
+k^2c_\varphi^2\varphi_b
&=0,
\label{eq:dimensionless-first-long}
\\
\sigma_b''-\frac2\tau\sigma_b'
+\left[
k^2c_\sigma^2+\frac{\frac94-\nu^2}{\tau^2}
\right]\sigma_b
&=
\frac{\lambda}{\tau}D_\tau\varphi_b.
\label{eq:dimensionless-second-long}
\end{align}
For an entropy branch $\sigma_b\propto(-\tau)^p$, the leading terms in the first equation require $\varphi_b=-\lambda\sigma_b/p+\cdots$. The corresponding contribution to $D_\tau\varphi_b=\varphi_b'+\lambda\sigma_b/\tau$ therefore cancels at this order. Substituting into the entropy equation leaves
\begin{equation}
\left(p-\frac32\right)^2=\nu^2,
\qquad
p_\eta=\frac32-\eta\nu,
\qquad \eta=\pm1.
\label{eq:long-indicial-equation}
\end{equation}
The entropy sound speed multiplies a gradient term that is subleading in this limit. The two powers consequently depend on the final entropy mass, while their coefficients retain the earlier evolution through the two sound horizons.

Away from resonant masses, the contribution of these branches to the entropy contraction can be written as
\begin{equation}
\left.
\sum_{b=1}^2
\varphi_b(k,0)\sigma_b^*(k,\tau)
\right|_{\rm NA}
=
N_k^2
\sum_{\eta=\pm1}
C_\eta
(-kc_\varphi\tau)^{p_\eta}
\left[
1+\mathcal O(k^2c_\varphi^2\tau^2)
\right].
\label{eq:S-late-bispectrum}
\end{equation}
Here $C_+$ and $C_-$ are the coefficients of the two entropy branches in the contraction with the final curvature perturbation. The subscript ${\rm NA}$ selects these branches, which give the characteristic nonanalytic momentum dependence. The complete contraction also contains the adiabatic branches, including a response proportional to $(-kc_\varphi\tau)^2$. The normalized modes fix $C_\pm$, which can be evaluated from conserved Wronskians of specified Heun solutions at an ordinary point using Eq.~\eqref{eq:Xi-explicit-connections}. Appendix~\ref{app:bispectrum-connections} gives the continuation and normalization of the two asymptotic branches entering these coefficients. The differentiated long leg is suppressed relative to the entropy leg. Inserting one entropy branch into Eq.~\eqref{eq:dimensionless-first-long} once more gives
\begin{equation}
D_\tau\varphi_b(k,\tau)
=
\frac{\lambda k^2c_\varphi^2\tau}
{p_\eta(p_\eta-1)}
\sigma_b(k,\tau)+\cdots.
\label{eq:long-D-suppression}
\end{equation}
The factor $k_L^2$ places this contribution two powers of $q$ beyond the leading squeezed term with a long entropy leg in Eqs.~\eqref{eq:exact-time-beta} and \eqref{eq:exact-time-alpha}. At $p_\eta=1$, the corresponding resonant expansion contains a logarithm and gives a $q^2\ln q$ suppression.

To extract the leading coefficient, we also need the imaginary part required by the in-in expectation value. This part of each contraction is fixed by its commutator with the final curvature perturbation. Canonical normalization selects the decaying adiabatic solution in the commutator, giving
\begin{align}
\operatorname{Im}
\sum_{b=1}^2
\varphi_b(k,0)
\bigl[D_\tau\varphi_b(k,\tau)\bigr]^*
&=
\frac{H^2\tau^2}{2}+\cdots,
\nonumber\\
\operatorname{Im}
\sum_{b=1}^2
\varphi_b(k,0)\sigma_b^*(k,\tau)
&=
\frac{\lambda H^4\tau^3}{2\mu^2}+\cdots.
\label{eq:retarded-endpoint-powers}
\end{align}
The entropy-branch contribution in Eq.~\eqref{eq:S-late-bispectrum} is consequently real on the negative real time axis. For real $\nu$, both $C_+$ and $C_-$ are real, while for $\nu=\ii\rho$ they obey $C_-=C_+^*$. We can therefore factor the complete long entropy contribution out of the imaginary part, leaving the imaginary part of the product of the two short contractions. The remaining short-mode evolution enters through three dimensionless integrals. Keeping the normalization factors explicit and taking $k=k_S$, we define
\begin{align}
\mathcal M_{\gamma,\eta}
&=
\frac{kc_\varphi}{N_k^4}
\int_{-\infty}^0
\de\tau\,
(-kc_\varphi\tau)^{p_\eta-4}
\nonumber\\
&\quad\times
\operatorname{Im}\!\left[
\sum_{b,c=1}^2
\varphi_b(k,0)\varphi_c(k,0)
\sigma_b^*(k,\tau)\sigma_c^*(k,\tau)
\right],
\label{eq:M-gamma-unequal}
\\
\mathcal M_{\beta,\eta}
&=
\frac{2}{N_k^4}
\int_{-\infty}^0
\de\tau\,
(-kc_\varphi\tau)^{p_\eta-3}
\nonumber\\
&\quad\times
\operatorname{Im}\!\left[
\sum_{b,c=1}^2
\varphi_b(k,0)\varphi_c(k,0)
\bigl[D_\tau\varphi_b(k,\tau)\bigr]^*
\sigma_c^*(k,\tau)
\right],
\label{eq:M-beta-unequal}
\\
\mathcal M_{\alpha,\eta}
&=
\frac{1}{kc_\varphi N_k^4}
\int_{-\infty}^0
\de\tau\,
(-kc_\varphi\tau)^{p_\eta-2}
\nonumber\\
&\quad\times
\operatorname{Im}\!\left[
\sum_{b,c=1}^2
\varphi_b(k,0)\varphi_c(k,0)
\bigl[D_\tau\varphi_b(k,\tau)\bigr]^*
\bigl[D_\tau\varphi_c(k,\tau)\bigr]^*
\right].
\label{eq:M-alpha-unequal}
\end{align}
The powers containing $p_\eta$ remain outside the imaginary part. This distinction matters for heavy fields, where the two weights are complex conjugates and the corresponding integrals obey $\mathcal M_{I,-}=\mathcal M_{I,+}^*$.

These expressions preserve the late-time cancellation within each integrand. For a light field with $p=3/2-\nu>0$, the slow-branch $\gamma$ and $\beta$ integrands begin as $(-\tau)^{2p-1}$, while the $\alpha$ integrand begins as $(-\tau)^{p+1}$. They are therefore integrable at zero throughout the positive-mass domain. The imaginary parts are formed on the negative real time axis, and the early endpoint is defined by inserting a real Abel factor $\ee^{\epsilon k_Sc_\varphi\tau}$ and taking $\epsilon\to0^+$~\cite{Kaya:2018jdo}. An equivalent contour evaluation continues the two conjugate frequency contributions along their respective damped rays. Appendix~\ref{app:bispectrum-finite-integrals} gives the commutator derivation and a convergent evaluation of these integrals.

Keeping the leading term of each entropy branch in the exact bispectrum integrals now gives
\begin{equationbox}
\begin{equation}
\frac{B_I^{\rm NA}(k_S,k_S,k_L)}
{P_\zeta(k_S)P_\zeta(k_L)}
=
\sum_{\eta=\pm1}
b_{I,\eta}q^{3/2-\eta\nu}+\mathcal O\!\left(q^{7/2-\operatorname{Re}\nu}\right),
\label{eq:squeezed-bispectrum-mode-sum}
\end{equation}
\end{equationbox}
where $P_\zeta(k)=2\pi^2\Delta_\zeta/k^3$. The coefficients for the three interactions are
\begin{align}
b_{\alpha,\eta}
&=
-\alpha
\left(
\frac{\Delta_0^{(c_\varphi)}}{\Delta_\zeta}
\right)^2
C_\eta\mathcal M_{\alpha,\eta},
\nonumber\\
b_{\beta,\eta}
&=
-\beta
\left(
\frac{\Delta_0^{(c_\varphi)}}{\Delta_\zeta}
\right)^2
C_\eta\mathcal M_{\beta,\eta},
\nonumber\\
b_{\gamma,\eta}
&=
-\gamma
\left(
\frac{\Delta_0^{(c_\varphi)}}{\Delta_\zeta}
\right)^2
C_\eta\mathcal M_{\gamma,\eta}.
\label{eq:dimensional-squeezed-coefficients}
\end{align}
Each coefficient contains the transfer of the long entropy fluctuation into curvature through $C_\eta$, the evolution of the interacting short modes through $\mathcal M_{I,\eta}$, and the explicit normalization by the full power spectrum. All three factors retain the dependence on the quadratic mixing and the sound-speed ratio.

To express this result in terms of the equilateral-normalized shape, Eq.~\eqref{eq:reduced-shape-unequal} gives
\begin{equation}
S_I(q)
=
\frac{5}{18f_{\rm NL}^{(I)}q}
\frac{B_I(k_S,k_S,k_L)}
{P_\zeta(k_S)P_\zeta(k_L)}.
\label{eq:shape-isosceles-normalization}
\end{equation}
The squeezed shape therefore takes the form
\begin{equation}
S_I^{\rm NA}(q)
=
\frac{5}{18f_{\rm NL}^{(I)}}
\sum_{\eta=\pm1}
b_{I,\eta}q^{1/2-\eta\nu}+\mathcal O\!\left(q^{5/2-\operatorname{Re}\nu}\right).
\label{eq:finite-squeezed-coefficients}
\end{equation}
The factor $f_{\rm NL}^{(I)}$ is determined by the exact integrals at equilateral momenta. Dividing by it compares the squeezed signal with the equilateral contribution of the same interaction. Both the overall cubic coupling and the explicit power-spectrum normalization cancel from $S_I$, while its dependence on the coupled mode evolution remains.

The displayed expressions retain the leading term of each entropy branch, whose next contributions carry two additional powers of $q$. The error estimates apply at fixed mass, mixing and finite sound-speed ratio, away from resonant masses. The full bispectrum also contains a regular contribution of order $q^2$ in $B_I/[P_\zeta(k_S)P_\zeta(k_L)]$, corresponding to order $q$ in the shape, which is excluded by the superscript ${\rm NA}$. At resonant masses, all terms whose powers coincide must be combined before taking the parameter limit. For example, $\nu=0$ requires combining the two entropy branches, while $\nu=1$ also involves the first descendant of the slower branch and $\nu=1/2$ involves the analytic quadratic response. These combined limits retain the logarithms that replace the separate power laws.

\subsection{Heavy-field oscillations and the sound-speed hierarchy}
\label{sec:evaluated-heavy-signal}

For a heavy entropy field, we write $\nu=\ii\rho$, with $\rho=\sqrt{\mu^2/H^2-9/4}>0$. The two terms in Eq.~\eqref{eq:finite-squeezed-coefficients} are then complex conjugates, and the squeezed shape becomes
\begin{equation}
S_I^{\rm NA}(q)
=
\frac{5\sqrt q}{9f_{\rm NL}^{(I)}}
\operatorname{Re}\!\left[
b_{I,+}\ee^{\ii\rho\ln(1/q)}
\right].
\label{eq:heavy-squeezed-complex}
\end{equation}
The modulus and argument of $b_{I,+}$ determine the amplitude and phase of the oscillatory signal, which together carry information about the particle dynamics and its interactions~\cite{Qin:2022Phase}. Including the equilateral amplitude, we can write this signal as
\begin{equationbox}
\begin{equation}
\frac9{10}f_{\rm NL}^{(I)}S_I^{\rm NA}(q)
=
\frac{|b_{I,+}|}{2}\sqrt q\,
\cos\!\left[
\rho\ln(1/q)+\arg b_{I,+}
\right].
\label{eq:heavy-squeezed-unequal}
\end{equation}
\end{equationbox}
Thus the oscillatory coefficient of $(9/10)f_{\rm NL}^{(I)}S_I$ is $|b_{I,+}|/2$, while the corresponding coefficient in $B_I/[P_\zeta(k_S)P_\zeta(k_L)]$ is $2|b_{I,+}|$. At fixed final entropy mass, varying the sound-speed ratio changes this coefficient and its phase while preserving the logarithmic frequency $\rho$.

We now return to Fig.~\ref{fig:gamma-sound-speed-dependence} in the introduction, which shows $|b_{\gamma,+}|/2$ and $\arg b_{\gamma,+}$ for the $\sigma^3$ interaction with $\gamma=1$ and $\nu=\ii$ at three mixing strengths. At $\lambda=1/2$, $1$ and $3/2$, the combination controlling the linear small-speed hierarchy is $\chi^2=-3/4$, $0$ and $5/4$, respectively. The curves therefore compare the oscillatory, critical and power-law regimes of Section~\ref{sec:sound-speed-hierarchies}, all with the same strict squeezed frequency $\rho=1$. In the last example the intermediate mass coefficient is still positive, $m_0^2/H^2=1$, and the intermediate entropy fluctuation decays.

Across the plotted interval, the $\lambda=1$ and $\lambda=3/2$ signals grow as the entropy sound speed decreases. At $\lambda=1/2$, the plotted amplitude instead turns over in the lower part of the sampled range, taking values of about $0.0153$ at $r_c=0.25$ and $0.0141$ at $r_c=0.15$. The phases also follow different trajectories, although all three curves describe oscillations with the same frequency. For $\lambda=1$, the amplitudes at $r_c=1/2$ and $2$ are $0.09159$ and $0.003616$, a change by a factor of about $25.3$. These values include the exact power-spectrum normalization and keep the cubic coupling fixed.

The squeezed amplitude depends on the interaction region as well as on the frozen power. In particular, the transfer and interaction factor $C_+\mathcal M_{\gamma,+}$ can change at a different rate from $\Delta_\zeta^2$ as $r_c$ varies. The hierarchy laws for the power spectrum alone consequently do not determine the bispectrum amplitude in Eq.~\eqref{eq:heavy-squeezed-unequal}. A calculation of the bispectrum hierarchy must also follow the short-mode integral through the separated sound horizons. Describing the hierarchy dependence of the equilateral-normalized shape additionally requires the equilateral amplitude obtained from the exact three-mode integral.

The separation of the sound horizons also affects the range in which the strict squeezed formula is accurate, a feature that plays an important role in the low-speed collider~\cite{Jazayeri:2022kjy,Jazayeri:2023xcj}. The long mode must have reached its late-time expansion during the times that contribute appreciably to the short-mode integral. Away from resonant indices, its first relative entropy correction is quadratic in $k_Lc_\varphi\tau$. Requiring this correction to remain small gives
\begin{equation}
\left|
\frac{
\lambda^2/[p_\eta(p_\eta-1)]-r_c^2
}{
4p_\eta-2
}
\right|
(k_Lc_\varphi\tau)^2
\ll1,
\label{eq:soft-first-correction}
\end{equation}
over the relevant interaction times. At order-one mass and mixing, requiring the long fluctuation to be outside both sound horizons over the interval between the short-mode crossings gives the rough estimate $q\ll\min(r_c,r_c^{-1})$. Equation~\eqref{eq:soft-first-correction} also shows why the mass and mixing affect this estimate. The exact integrals \eqref{eq:exact-time-gamma}--\eqref{eq:exact-time-alpha} retain the finite-$q$ evolution without assuming that the long mode has already reached its asymptotic form.

Analytic contributions can also affect the approach to the squeezed limit. For the heavy $\gamma$ examples with $\lambda=1$, $\nu=\ii$ and $r_c=1/2,2$, the leading regular term in the shape is proportional to $q$, whereas the oscillatory envelope is proportional to $\sqrt q$. Its relative contribution can therefore begin at order $\sqrt q$, before the corrections from higher powers within the entropy branches become important. Appendix~\ref{app:bispectrum-numerical} subtracts this regular term when comparing the full time integral with the nonanalytic oscillation, providing a test of both the squeezed coefficient and the range of the approximation.

\subsection{The equal-speed limit and independent checks}
\label{sec:bispectrum-equal-speed-checks}

At equal sound speeds, the vacuum modes are naturally expressed in the two polarizations of Section~\ref{sec:equal-speed}. As discussed in Section~\ref{sec:early-time-propagation}, the incoming field directions selected at unequal speeds need not approach these individual polarizations. The complete contractions nevertheless provide a direct comparison, since the sums in Eq.~\eqref{eq:boundary-functions-bispectrum} are invariant under unitary transformations of the normalized vacuum modes. Substituting the equal-speed solutions reproduces the contractions of Ref.~\cite{Huenupi:2026aqc}, and the three time integrals recover its interaction contributions with the same cubic normalization.

The equality of these contractions fixes the bispectrum normalization for every momentum configuration, including the equilateral configuration used to define $f_{\rm NL}^{(I)}$ and $S_I$. The independent numerical comparisons in Appendix~\ref{app:bispectrum-equal-speed} test the squeezed coefficients at $\lambda=1$ and $\nu=\ii$, with the results listed in Table~\ref{tab:bispectrum-equal-check}.

Weak mixing provides another check of the three interaction channels. At fixed positive entropy mass and fixed finite $r_c$, the entropy contraction starts at order $\lambda$, while the differentiated curvature contraction has a nonzero unmixed limit. Since $\Delta_\zeta/\Delta_0^{(c_\varphi)}=1+\mathcal O(\lambda^2)$, the exact equilateral integrals imply
\begin{equation}
f_{\rm NL}^{(\alpha)}
=
\mathcal O(\alpha\lambda),
\qquad
f_{\rm NL}^{(\beta)}
=
\mathcal O(\beta\lambda^2),
\qquad
f_{\rm NL}^{(\gamma)}
=
\mathcal O(\gamma\lambda^3).
\label{eq:bispectrum-weak-mixing-orders}
\end{equation}
Here the cubic coefficients are held fixed as the quadratic mixing is varied. These orders count how many entropy fields at the vertex must be converted into final curvature, recovering the single-, double- and triple-exchange organization of perturbative collider calculations~\cite{Chen:2009we,Noumi:2012vr,Aoki:2024DoubleExchange}. At generic parameter values, dividing out the equilateral amplitude gives a finite weak-mixing limit for each normalized shape. The fixed-mass and fixed-speed conditions keep this check distinct from the singular sound-speed limits of Section~\ref{sec:sound-speed-hierarchies}.

The light-field calculation in Appendix~\ref{app:bispectrum-numerical} tests the late-time cancellations in the coefficient integrals. At $(r_c,\lambda,\nu)=(0.5,0.7,0.8)$, independent real-axis and Euclidean evaluations agree on the three leading short-mode integrals. The full finite-$q$ bispectra also approach the two-branch squeezed expressions, with relative differences below $6\times10^{-4}$ at $q=10^{-3}$ for all three interactions. Using the same exact equilateral amplitude to normalize both expressions leaves these relative comparisons unchanged. Together with the heavy-field and equal-speed calculations, these checks test the mode normalization, the extraction of the long entropy branches and the evaluation of the interaction-time integrals.


\section{Discussion and conclusions}
\label{sec:discussion}

We have obtained the canonically normalized modes, the primordial power spectrum and the squeezed bispectrum of a two-field inflationary system with unequal sound speeds and arbitrary constant quadratic mixing, extending the equal-speed solutions of Ref.~\cite{Huenupi:2026abj} and their application to cosmological-collider bispectra in Refs.~\cite{Huenupi:2026aqc,Pinol:2026xnl,Wang:2026lff,Belrhali:2026uxn,Pinol:2026hey}. The Laplace representation reduces the coupled dynamics to a second-order Heun equation, whose normalized solutions give the exact spectrum in Eq.~\eqref{eq:power-Heun}. This expression involves two specified Heun values and known gamma-function factors, with no remaining time integral or undetermined connection coefficient. It applies to both light and heavy entropy fields and connects continuously to the equal-speed result.

The effects of unequal propagation speeds become particularly clear when the two sound horizons are widely separated. For $r_c\ll1$, the entropy gradient becomes small while the curvature gradient still suppresses the curvature response, allowing an intermediate stage of entropy evolution governed approximately by $\mu^2-\lambda^2H^2$. The final superhorizon decay is instead controlled by $\mu$. This distinction explains the three small-$r_c$ regimes at fixed nonzero mixing: the power has a power-law enhancement when $\nu^2+\lambda^2>0$, grows as $\ln^2(1/r_c)$ when this combination vanishes, and exhibits bounded oscillations in $\ln r_c$ when it is negative. An enhanced curvature spectrum need not involve a growing entropy fluctuation, since the vacuum amplitude and subsequent conversion into curvature also depend on its propagation speed.

For $r_c\gg1$, the entropy gradient suppresses the response to the nearly frozen curvature fluctuation, while the residual feedback accumulates over the interval separating the sound horizons. The leading relative correction scales as $\lambda^2\ln r_c/r_c^2$, so the curvature power approaches its unmixed value at fixed mass and mixing. Both sound-speed limits follow directly from the exact spectrum, which also describes the transition between them.

The bispectrum contains further information about the coupled evolution. Its squeezed coefficients combine the conversion of the long entropy fluctuation into curvature with the interaction of the short modes, together with the exact curvature-power normalization. Their dependence on $r_c$ can substantially change the collider amplitude and need not be monotonic. For the $\sigma^3$ interaction at $\lambda=1$ and $\nu=\ii$, decreasing $r_c$ from $2$ to $1/2$ increases the power-normalized oscillation amplitude by a factor of approximately $25$ at fixed cubic coupling, while the weaker-mixing example in Fig.~\ref{fig:gamma-sound-speed-dependence} exhibits a turnover. These changes in amplitude and phase preserve the nonanalytic squeezed exponents and the heavy-field oscillation frequency fixed by the final entropy mass. The sound-speed hierarchy also affects how closely a configuration must approach the squeezed limit, since the long fluctuation must reach its late-time regime during the interaction of the short modes. The exact isosceles integrals retain this finite-momentum-ratio evolution.

Our results assume constant mass, mixing and sound speeds at leading order in slow variation. Positive entropy mass ensures that the curvature perturbation eventually freezes, although a very light field can require many e-folds to reach this limit. The bispectrum calculation treats three representative cubic interactions at tree level while keeping the quadratic mixing exact. Applications to a specified inflationary model therefore require identifying its additional interaction operators and checking perturbative control in the presence of amplified fluctuations. Slowly varying parameters and a finite duration of inflation provide natural extensions of this analysis.

A broader study of the momentum dependence would connect these results more directly with observations. Extending the squeezed calculation to general triangles would allow a systematic comparison of the equilateral amplitudes and normalized shapes across interaction channels, providing templates for full-shape CMB analyses~\cite{Kumar:2026dih,Philcox:2026Dissecting} and for studies of primordial collider signals in large-scale structure~\cite{Cabass:2024wob,Anbajagane:2026HaloBias}. The dependence of the amplitude and phase on propagation speeds makes this freedom relevant when relating primordial correlations to the underlying particle dynamics.

The normalized mixed modes also provide the propagators needed to investigate quantum corrections. Analytic treatments of massive correlators and dispersive methods~\cite{Xianyu:2025lbk,Liu:2024Dispersive,Li:2026xhw}, together with recent calculations of loop signals~\cite{Qin:2026Triangle,Borinsky:2026OneLoop}, offer useful points of contact for this extension. Following the coupled propagation through loop diagrams would help determine how the sound-speed hierarchy affects quantum corrections and which enhanced regimes remain predictive within a specified effective theory.


\acknowledgments

We are particularly grateful to Claudio Mu\~noz for his help and valuable discussions. We also wish to thank Ana Ach\'ucarro, Sebasti\'an C\'espedes, Perseas Christodoulidis, Gabriel Mar\'in Mac\^edo, Hayden Lee, Qianshu Lu, Anish Pandya, Nicol\'as Parra, Lucas Pinol, Arthur Poisson, S\'ebastien Renaux-Petel, Diederik Roest, Zhong-Zhi Xianyu and Crist\'obal Zenteno for useful discussions and comments on the topic of strongly mixed systems. 
JH acknowledges the hospitality of the Munich Institute for Astro-, Particle and BioPhysics (MIAPbP), which is funded by the Deutsche Forschungsgemeinschaft (DFG, German Research Foundation) under Germany's Excellence Strategy---EXC-2094---390783311. GAP acknowledges support from the Fondecyt Regular project 1251511 (ANID). SS acknowledges support from UBB under grant IN2686604  (Iniciaci\'on). This work was supported in part by a grant of access to OpenAI models through the ChatGPT for Academic Researchers program.


\appendix

\section{Supporting derivations}
\label{app:supporting-derivations}

We collect the calculations needed to connect the exact spectral solutions with the power spectrum and squeezed bispectrum. The first two parts establish the integral identity and the sound-speed limits of the Heun expression. The remaining parts determine the long entropy coefficients and explain the convergence and evaluation of the interaction integrals, including numerical comparisons with the original coupled field equations. Throughout this appendix the mass and mixing are constant, the sound speeds are positive, and $\mu^2>0$.

\subsection{The integral identity for the power spectrum}
\label{app:period-transform-proof}

We now derive the identity relating the imaginary-axis integral $J$ to the normalization of the spectral solution at the entropy endpoint. We work at fixed $r_c\ne1$ and $\operatorname{Re}\nu<3/2$, using the even solution $e(v)$ defined in Eq.~\eqref{eq:real-period-ODE}. It is useful to normalize the first spectral solution at $u=r_c$ by writing
\begin{equation}
\widehat Y_1(u)\equiv\frac{Y_1(u)}{Y_1(r_c)},
\qquad
\widehat Y_1(r_c)=1.
\label{eq:period-regular-endpoint-solution}
\end{equation}
Thus $\widehat Y_1$ is the solution analytic at the entropy endpoint with unit value there, given explicitly by the Heun numerator in Eq.~\eqref{eq:Y1}. For $u$ between the two positive endpoints, introduce
\begin{equation}
\begin{aligned}
V(u)^2&=\frac{r_c^2-u^2}{u^2-1},
\\
s(u,v)&=\frac{V(u)^2-v^2}{1+V(u)^2}
=1-\frac{(u^2-1)(1+v^2)}{r_c^2-1},
\\
K(s)&=s^{-1/2}{}_2F_1\!\left(
-\frac14+\frac\nu2,-\frac14-\frac\nu2;\frac12;s
\right).
\end{aligned}
\label{eq:period-transform-variables}
\end{equation}
We take $V>0$ and the positive square root in $K$ on $0<s<1$. For either ordering of the sound speeds, $V$ tends to zero as $u\to r_c$ and to infinity as $u\to1$. The integral representation connecting these two limits is
\begin{equation}
\widehat Y_1(u)
=\frac2\pi\int_0^{V(u)}
\frac{e(v)K(s(u,v))}{1+v^2}\,\de v.
\label{eq:finite-period-transform}
\end{equation}
We establish this identity by showing that the integral satisfies the spectral equation and is analytic at $u=r_c$ with unit value there.

To carry out the differentiation with respect to $u$, we first replace the real integral by a contour integral whose path can remain fixed as the endpoint $V(u)$ moves. Since $e(v)$ and $s(u,v)$ are even in $v$, we can extend the integration interval to $[-V,V]$. We then place the square-root cut along this segment and take a counterclockwise contour $\mathcal C$ surrounding it, keeping the other singularities at $\pm\ii$ and $\pm\ii r_c$ outside. We choose the branch of $K$ to agree with its value in the real integral on the lower bank, so its value on the upper bank has the opposite sign. The lower bank is traversed from $-V$ to $V$, while the upper bank is traversed in the reverse direction, giving
\begin{equation}
\begin{aligned}
\oint_{\mathcal C}
\frac{e(v)K(s(u,v))}{1+v^2}\,\de v
&=
\int_{-V}^{V}
\frac{e(v)K(s(u,v))}{1+v^2}\,\de v
-
\int_V^{-V}
\frac{e(v)K(s(u,v))}{1+v^2}\,\de v
\\
&=
2\int_{-V}^{V}
\frac{e(v)K(s(u,v))}{1+v^2}\,\de v
=
4\int_0^V
\frac{e(v)K(s(u,v))}{1+v^2}\,\de v.
\end{aligned}
\label{eq:period-contour-collapse}
\end{equation}
The kernels in the real integrals on the right-hand side are evaluated on the original positive-square-root branch. The small arcs around $\pm V$ vanish when the contour is collapsed onto the segment, because their integrands behave as $(v\mp V)^{-1/2}$ and an arc of radius $\epsilon$ therefore contributes only $\mathcal O(\sqrt{\epsilon})$. Comparing with Eq.~\eqref{eq:finite-period-transform}, we obtain
\begin{equation}
\widehat Y_1(u)
=\frac1{2\pi}\oint_{\mathcal C}
\frac{e(v)K(s(u,v))}{1+v^2}\,\de v.
\label{eq:period-transform-closed-contour}
\end{equation}
For a sufficiently small variation of $u$, the contour can be held fixed while both endpoints remain inside it. Moreover, $e$ is analytic in the enclosed region, and continuing $K$ around both square-root endpoints returns it to its initial branch without encircling another singularity. These properties allow us to differentiate under the contour integral and to integrate by parts along the closed path.

We now turn to the differential equation. Multiplying the spectral equation by $1-u^2$ and the equation for $e(v)$ by $-(1+v^2)$ gives the operators
\begin{align}
\mathcal D_u
&=(1-u^2)\partial_u
\bigl[(r_c^2-u^2)\partial_u\bigr]
+\left(\nu^2-\frac14\right)(1-u^2)+\lambda^2,
\nonumber\\
\mathcal D_v
&=-(1+v^2)\partial_v
\bigl[(r_c^2+v^2)\partial_v\bigr]
+\left(\nu^2-\frac14\right)(1+v^2)+\lambda^2.
\label{eq:period-transform-operators}
\end{align}
The function $e$ satisfies $\mathcal D_ve=0$, and we must show that the integral satisfies the corresponding equation in $u$. The connection between the two operators follows from the choice of $K$. Their mixing terms cancel in the difference, and the chain rule gives
\begin{equation}
\begin{aligned}
(\mathcal D_u-\mathcal D_v)K\bigl(s(u,v)\bigr)
&=-(u^2+v^2)\left[
4s(1-s)K''+6(1-s)K'
+\left(\nu^2-\frac14\right)K
\right]
\\
&=0.
\end{aligned}
\label{eq:period-transform-intertwining}
\end{equation}
Here the primes on $K$ denote derivatives with respect to $s$, and the last equality follows from the hypergeometric equation. Thus, when acting on this kernel, $\mathcal D_u$ can be replaced by $\mathcal D_v$. The measure $\de v/(1+v^2)$ cancels the factor multiplying the derivative term in $\mathcal D_v$, allowing us to transfer this operator from $K$ to $e$ by integrating by parts twice. Keeping the resulting total derivative explicit, we find
\begin{equation}
\begin{aligned}
\oint_{\mathcal C}\frac{\de v}{1+v^2}
\left(e\,\mathcal D_vK-K\,\mathcal D_ve\right)
&=
\oint_{\mathcal C}\de v\,\partial_v
\left[
(r_c^2+v^2)
\left(K e'-e\,\partial_vK\right)
\right]
\\
&=0.
\end{aligned}
\label{eq:period-weighted-integration-by-parts}
\end{equation}
In this equation $K=K(s(u,v))$, and $e'$ denotes the derivative with respect to $v$. The total derivative integrates to zero because the functions and their derivatives return to the same values after a complete traversal of $\mathcal C$. Applying these identities to the contour representation gives
\begin{equation}
\begin{aligned}
\mathcal D_u\left[
\frac1{2\pi}\oint_{\mathcal C}
\frac{e(v)K(s(u,v))}{1+v^2}\,\de v
\right]
&=
\frac1{2\pi}\oint_{\mathcal C}
\frac{\de v}{1+v^2}\,e\,\mathcal D_uK
\\
&=
\frac1{2\pi}\oint_{\mathcal C}
\frac{\de v}{1+v^2}\,e\,\mathcal D_vK
\\
&=
\frac1{2\pi}\oint_{\mathcal C}
\frac{\de v}{1+v^2}\,K\,\mathcal D_ve
=0.
\end{aligned}
\label{eq:period-transform-spectral-equation}
\end{equation}

Having established the differential equation, we return to the real integral to determine its normalization at $u=r_c$. Setting $v=V\sin\theta$ gives
\begin{equation}
\frac{\de v}{(1+v^2)\sqrt{s(u,v)}}
=
\frac{\sqrt{1+V^2}}{1+V^2\sin^2\theta}\,\de\theta,
\qquad
0\leq\theta\leq\frac\pi2.
\label{eq:period-transform-shrinking-interval}
\end{equation}
As $V\to0$, the measure approaches $\de\theta$, while $e(v)$ and the hypergeometric factor in $K$ both approach one. The integral consequently tends to $(2/\pi)\int_0^{\pi/2}\de\theta=1$. All its factors have convergent expansions in $V^2$ near zero, so it is also analytic in $u-r_c$. These conditions identify the integral with $\widehat Y_1$, establishing Eq.~\eqref{eq:finite-period-transform} for either ordering of the sound speeds.

We can now take the opposite limit, $u\to1$, in which $V\to\infty$ and $s(u,v)\to1$ at every fixed $v$. Gauss summation gives the limiting kernel
\begin{equation}
K(1)
=
\frac{\sqrt\pi}
{\Gamma(\frac34-\frac\nu2)
\Gamma(\frac34+\frac\nu2)}.
\label{eq:period-kernel-endpoint}
\end{equation}
To justify passing this limit through the integral, split its range at $v=V/2$. On $0\leq v\leq V/2$, the kernel is uniformly bounded for large $V$, and the integrable function $|e(v)|/(1+v^2)$ controls the limit. On $V/2\leq v\leq V$, the square-root singularity remains integrable, while the large-$v$ behavior $e(v)=\mathcal O(v^{-1/2+\operatorname{Re}\nu})$, with logarithmic factors when required, makes this contribution vanish as $V^{-3/2+\operatorname{Re}\nu}$ up to the same logarithms. The condition $\operatorname{Re}\nu<3/2$ therefore allows us to obtain
\begin{equation}
\begin{aligned}
\widehat Y_1(1)
&=
\frac{2K(1)}{\pi}\int_0^\infty
\frac{e(v)}{1+v^2}\,\de v
=2K(1)J,
\\
J
&=
\frac{
\Gamma(\frac34-\frac\nu2)
\Gamma(\frac34+\frac\nu2)
}{2\sqrt\pi}\,
\widehat Y_1(1).
\end{aligned}
\label{eq:period-identity-regular-normalization}
\end{equation}
Finally, using $\widehat Y_1(1)=1/Y_1(r_c)$ recovers Eq.~\eqref{eq:period-normalization-identity}.


\subsection{Connection formulas for the sound-speed limits}
\label{app:small-kappa-integral}
\label{app:large-kappa-green}

The limits in Section~\ref{sec:sound-speed-hierarchies} involve a Heun singularity approaching the point where the spectrum is evaluated. The resulting connection must retain the separation of these points, even though the equations away from them reduce to hypergeometric form. We give the two connection calculations here, keeping the mass and mixing fixed and using the continuation below the positive singularities in the spectral variable $u$.

\paragraph{Small sound-speed ratio.}
For $r_c\ll1$, the nontrivial connection concerns the second Heun value in Eq.~\eqref{eq:power-Heun}. Outside the merging points $u=\pm r_c$, the zero and derivative at $u=1$ specified by Eq.~\eqref{eq:Y2-endpoint-derivative} select
\begin{equation}
\begin{aligned}
\frac{Y_2(u)}{Y_1(r_c)}\simeq{}&
\frac{\lambda\sqrt{r_c}}{2(r_c^2-1)}
(u^2-1)u^{\chi-1/2} \;
{}_2F_1\!\left(
1+\frac{\chi-\nu}{2},1+\frac{\chi+\nu}{2};2;1-u^2\right).
\end{aligned}
\label{eq:Y2-small-r-single-outer}
\end{equation}
For noninteger $\chi$, the hypergeometric connection formula~\cite{DLMF:Hypergeometric} gives its two powers in the overlap region,
\begin{equation}
\frac{Y_2(u)}{Y_1(r_c)}\simeq
\frac{\lambda\sqrt{r_c}}{2}
\sum_{s=\pm\chi}
\frac{\Gamma(s)u^{-1/2-s}}
{\Gamma(1+\frac{s-\nu}{2})
\Gamma(1+\frac{s+\nu}{2})},
\qquad r_c\ll u\ll1.
\label{eq:small-r-two-power-overlap}
\end{equation}
Within the separation, setting $X=u/r_c$ instead gives Legendre's equation at leading order,
\begin{equation}
\frac{\de}{\de X}\left[(1-X^2)\frac{\de Y}{\de X}\right]
+\left(\chi^2-\frac14\right)Y=0.
\label{eq:small-r-Legendre-equation}
\end{equation}
The two overlap powers can be matched to $Q_{s-1/2}(X)$ with $s=\pm\chi$. Taking the branch real for $X>1$ at real $s$ and continuing below $X=1$, the required large-$X$ coefficients and values at the origin are~\cite{DLMF:Legendre}
\begin{align}
Q_{s-1/2}(X)&\sim
\frac{\sqrt\pi\,\Gamma(s+\frac12)}
{2^{s+1/2}\Gamma(1+s)}X^{-1/2-s},
\label{eq:small-r-Q-asymptotic}
\\
Q_{s-1/2}(0-\ii0)&=
\frac{\ii\sqrt\pi}{2}
\ee^{\ii\pi(s-1/2)/2}
\frac{\Gamma(\frac14+\frac s2)}
{\Gamma(\frac34+\frac s2)}.
\label{eq:small-r-Q-origin}
\end{align}
Matching Eq.~\eqref{eq:small-r-Q-asymptotic} to Eq.~\eqref{eq:small-r-two-power-overlap} fixes the coefficient of each Legendre solution. Their origin values can then be inserted directly into Eq.~\eqref{eq:frozen-modes-gamma}. Gamma duplication reduces the result to
\begin{equation}
\begin{aligned}
\frac{\varphi_2(k,0)}{N_k}\simeq
-\lambda\ee^{-\ii\pi/4}
\Bigl\{&
\ee^{\ii\pi\chi/2}\mathcal C(\lambda,\nu)r_c^{-\chi}
+
\left[
\ee^{\ii\pi\chi/2}\mathcal C(\lambda,\nu)r_c^{-\chi}
\right]_{\chi\to-\chi}
\Bigr\}.
\end{aligned}
\label{eq:small-r-two-branch-amplitude}
\end{equation}
Here $\mathcal C$ is the coefficient in Eq.~\eqref{eq:C-small-kappa}, and the substitution in the second term also acts on the gamma functions defining it. Equation~\eqref{eq:Y2} then converts this amplitude into the endpoint expansion of the second Heun function. For real $\chi>0$, the first term gives Eq.~\eqref{eq:Heun2-small-r} and its phase. Equation~\eqref{eq:small-r-two-branch-amplitude} retains the leading coefficient of each branch, rather than all terms up to the order of the second branch. At positive integer $\chi$, the separated coefficients must be combined with the coincident subleading powers before taking the limit. The leading real-$\chi$ coefficient remains regular.

For $\chi=\ii\eta$, the two coefficients are complex conjugates and the continuation factors have moduli $\ee^{\mp\pi\eta/2}$. Both branches therefore survive at leading order, while the omitted terms tend to zero, giving Eq.~\eqref{eq:small-r-oscillatory-power}. At $\chi=0$, the gamma-function coefficient has the residue
\begin{equation}
\lim_{\chi\to0}\chi\mathcal C(\lambda,\nu)
=
\frac{\Gamma(\frac34-\frac\nu2)\Gamma(\frac34+\frac\nu2)}
{2\Gamma(\frac34)^2\Gamma(1-\frac\nu2)\Gamma(1+\frac\nu2)}.
\label{eq:small-r-critical-residue}
\end{equation}
The opposite poles in the two terms of Eq.~\eqref{eq:small-r-two-branch-amplitude} cancel, leaving twice this residue times $\ln(1/r_c)$ in the amplitude. The continuation phases and finite parts contribute bounded terms, giving the leading logarithmic power in Eq.~\eqref{eq:small-r-critical-power}.

\paragraph{Large sound-speed ratio.}
For $r_c\gg1$, the nontrivial connection instead occurs in the first Heun value. Set $v=u/r_c$, $\epsilon=r_c^{-1}$ and $b(v)=Y_1(r_cv)/Y_1(r_c)$, so that $b$ is regular at $v=1$ with $b(1)=1$. The desired Heun value is $b(0)$, and its equation is
\begin{equation}
\bigl[(1-v^2)b'(v)\bigr]'
+\left[\nu^2-\frac14-\frac{\lambda^2\epsilon^2}{v^2-\epsilon^2}\right]b(v)=0.
\label{eq:large-r-scaled-equation}
\end{equation}
Let $P(v)=P_{\nu-1/2}(v)$ and let $O(v)$ be the odd solution of the unmixed Legendre equation, with $O(0)=0$ and $O'(0)=1$. Their weighted Wronskian is $P(0)$, so variation of parameters gives
\begin{equation}
b(v)=P(v)+\lambda^2\epsilon^2\int_v^1
\frac{P(v)O(t)-O(v)P(t)}{P(0)}
\frac{b(t)}{t^2-\epsilon^2}\,\de t,
\label{eq:large-r-Volterra}
\end{equation}
where the path passes below $t=\epsilon$. At $v=0$, the first iteration involves the weight $P(t)O(t)/P(0)=t+[P'(0)/P(0)]t^2+\mathcal O(t^3)$. The linear term gives $\ln(1/\epsilon)+\ii\pi/2+\mathcal O(\epsilon^2)$, and the quadratic term gives the next imaginary correction. Thus
\begin{equation}
\begin{aligned}
\frac{b(0)}{P(0)}={}&1+\lambda^2\epsilon^2\Biggl[
\ln\frac1\epsilon+
\int_0^1\frac{P(t)O(t)/P(0)-t}{t^2}\,\de t
+\frac{\ii\pi}{2}
+\frac{\ii\pi\epsilon}{2}\frac{P'(0)}{P(0)}
\Biggr]
\\
&+\mathcal O\!\left(
\lambda^2\epsilon^4\ln\frac1\epsilon,
\lambda^4\epsilon^4\ln^2\frac1\epsilon
\right).
\end{aligned}
\label{eq:large-r-integral-expansion}
\end{equation}
The subtraction makes the finite integral convergent at zero. To evaluate it, replace $\lambda^2\epsilon^2/(v^2-\epsilon^2)$ in the outer equation by $\delta(1+\delta)/v^2$, where $\delta(1+\delta)=\lambda^2\epsilon^2$ and $\delta\to0$ as $\epsilon\to0$. The solution regular and equal to one at $v=1$ is
\begin{equation}
\begin{aligned}
b_\delta(v)
&=v^{-\delta}{}_2F_1\!\left(
\frac14-\frac\nu2-\frac\delta2,
\frac14+\frac\nu2-\frac\delta2;1;1-v^2\right)
\\
&\simeq
\frac{\Gamma(\frac12+\delta)}
{\Gamma(\frac34-\frac\nu2+\frac\delta2)
\Gamma(\frac34+\frac\nu2+\frac\delta2)}v^{-\delta},
\qquad v\to0.
\end{aligned}
\label{eq:large-r-outer-connection}
\end{equation}
Differentiating this connection at $\delta=0$ gives the finite part accompanying $\ln(1/v)$. The same derivative follows from the first iteration of Eq.~\eqref{eq:large-r-Volterra} with the denominator replaced by $t^2$. Its term proportional to $P(v)$ gives the subtracted integral in Eq.~\eqref{eq:large-r-integral-expansion}, while the term proportional to $O(v)$ approaches $-1$ after division by $P(0)$. Equating the two expressions gives
\begin{equation}
\begin{aligned}
\int_0^1\frac{P(t)O(t)/P(0)-t}{t^2}\,\de t
&=1+\psi\!\left(\frac12\right)
-\frac12\psi\!\left(\frac34-\frac\nu2\right)
-\frac12\psi\!\left(\frac34+\frac\nu2\right)
\\
&=\frac12\mathcal B(\nu),
\end{aligned}
\label{eq:large-r-finite-integral-value}
\end{equation}
where $\mathcal B(\nu)$ is defined in Eq.~\eqref{eq:B-large-kappa}. The last equality uses $\psi(1)-\psi(1/2)=2\ln2$.

The remainder in Eq.~\eqref{eq:large-r-integral-expansion} follows from the same integral equation. Subtracting the linear and quadratic terms of the weight leaves an endpoint correction of order $\epsilon^2\ln(1/\epsilon)$ within the first iteration. To bound further iterations, the path can be taken around $t=\epsilon$ along a lower semicircle of radius $\epsilon/2$. Near zero the Green kernel is proportional to $t-v$, so its integral on the semicircle stays bounded, while the interval $t\gg\epsilon$ contributes at most $1+\ln(1/\epsilon)$. Each additional iteration therefore brings a factor of order $\lambda^2\epsilon^2[1+\ln(1/\epsilon)]$, which vanishes in the stated fixed-parameter limit. Returning to $r_c$, we obtain
\begin{equation}
\begin{aligned}
\frac{b(0)}{P(0)}={}&1+\frac{\lambda^2}{r_c^2}
\left[\ln r_c+\frac12\mathcal B(\nu)+\frac{\ii\pi}{2}\right]
+\frac{\ii\pi\lambda^2}{2r_c^3}\frac{P'(0)}{P(0)}
+\mathcal O\!\left(
\frac{\lambda^2\ln r_c+\lambda^4\ln^2r_c}{r_c^4}
\right).
\end{aligned}
\label{eq:large-r-Heun-complex-expansion}
\end{equation}
For real $\nu^2$, including imaginary $\nu$, the ratio $P'(0)/P(0)$ is real. The term of order $r_c^{-3}$ consequently changes only the phase and is absent from the first-mode power. The second Heun value has the regular limiting equation with solution $\,{}_2F_1(1,1/2;2;z)$, whose endpoint value is $2$. Its corrections are of order $r_c^{-2}$, as in Eq.~\eqref{eq:Heun2-large-r}, so the incoming entropy mode supplies the complete power correction at order $r_c^{-3}$ in Eq.~\eqref{eq:power-large-kappa}.


\subsection{The long entropy coefficients}
\label{app:bispectrum-connections}

The coefficients $C_\pm$ in Eq.~\eqref{eq:S-late-bispectrum} connect the normalized vacuum modes to the two entropy branches at late times. In the Laplace representation, these branches arise from the behavior of the spectral kernels at infinity. We can therefore determine their amplitudes by comparing the normalized kernels with two Heun solutions of unit asymptotic amplitude, using a conserved Wronskian to perform the comparison at an ordinary point.

For this calculation, let $x=-kc_\varphi\tau>0$ and write $\widehat Y_b(v)=[Y_b(v^*)]^*$. Conjugating the entropy reconstruction reflects its contour, and reversing the reflected orientation compensates the conjugation of $1/(2\pi\ii)$. The conjugate modes can consequently be written as
\begin{equation}
\begin{aligned}
\sum_{b=1}^2\varphi_b(k,0)\sigma_b^*(k,\tau)
={}&\frac{2\ii N_kx^2}{\lambda}
\sum_{b=1}^2\varphi_b(k,0)
\frac{1}{2\pi\ii}\int_{\widehat\Gamma}\de v\,
\ee^{-\ii xv}\widehat Y_b(v).
\end{aligned}
\label{eq:connection-conjugate-reconstruction}
\end{equation}
Here $\widehat\Gamma$ surrounds the two positive singularities counterclockwise, with its tails extending into the lower half-plane. To extract the large-$v$ contribution, we deform these tails towards the positive real axis with $x$ continued below it. The contour runs towards increasing $v$ on the lower bank and returns on the upper bank, so its density beyond both endpoints is
\begin{equation}
\rho_\sigma(v)=\frac{1}{\pi\lambda N_k}
\sum_{b=1}^2\varphi_b(k,0)
\left[\widehat Y_b(v-\ii0)-\widehat Y_b(v+\ii0)\right],
\qquad v>\max(1,r_c).
\label{eq:connection-oriented-density}
\end{equation}
The two boundary values are continued from the same germ at the origin, passing below or above the positive singularities. This specifies their relative sheet and fixes the sign of the density. Its normalization follows from Eqs.~\eqref{eq:Y1}, \eqref{eq:Y2} and \eqref{eq:bispectrum-frozen-amplitudes}, and the ratio $\varphi_b(k,0)/N_k$ makes it independent of $k$. Since both lateral values satisfy the spectral equation, the density obeys
\begin{equation}
\bigl[(r_c^2-v^2)\rho_\sigma'(v)\bigr]'
+\left[\nu^2-\frac14+\frac{\lambda^2}{1-v^2}\right]\rho_\sigma(v)=0.
\label{eq:connection-density-ODE}
\end{equation}

The two solutions with unit amplitude at infinity are obtained by setting $Y(v)=v^{1-p_\eta}h(v^{-2})$, where $p_\eta=3/2-\eta\nu$ was defined in the main text. Substituting into Eq.~\eqref{eq:Y-equation} gives
\begin{equation}
\begin{aligned}
T_\eta(v)={}&v^{1-p_\eta}
\HeunG\!\left(
r_c^{-2},\frac{p_\eta(p_\eta-1)-\lambda^2r_c^{-2}}4;
\frac{p_\eta-1}{2},\frac{p_\eta}{2},p_\eta-\frac12,0;v^{-2}
\right),
\\
T_\eta(v)\sim {}& v^{1-p_\eta}.
\end{aligned}
\label{eq:connection-infinity-Heun}
\end{equation}
For $v>\max(1,r_c)$, the Heun argument lies inside its convergence disk, and the prefactor uses real $\ln v$. The normalization at argument zero therefore fixes the two asymptotic amplitudes without a continuation phase. The self-adjoint equation supplies the conserved Wronskian
\begin{equation}
W[f,g]\equiv(r_c^2-v^2)(fg'-f'g),
\qquad W[T_+,T_-]=2\nu,
\label{eq:connection-Wronskian}
\end{equation}
where the second equality follows by evaluating the two powers at infinity. Resolving the density into this basis gives
\begin{equation}
\rho_\sigma(v)=
\frac{W[\rho_\sigma,T_-]}{2\nu}T_+(v)
+\frac{W[T_+,\rho_\sigma]}{2\nu}T_-(v).
\label{eq:connection-density-coefficients}
\end{equation}
Both coefficients can be evaluated at any $v_0>\max(1,r_c)$. The functions $T_\pm$ and their derivatives follow from their local Heun series, while the lateral values of $Y_b$ can be reached by ordinary Taylor continuation along the prescribed paths.

The finite portions of the contour give terms analytic in $x$. To identify the entropy powers, consider one monomial in the expansion at infinity. Its regulated transform is
\begin{equation}
x^2\int_{v_0}^{\infty}\de v\,
\ee^{-(\epsilon+\ii)xv}v^{1-p_\eta}
=
x^{p_\eta}(\epsilon+\ii)^{p_\eta-2}
\Gamma\!\left(2-p_\eta,(\epsilon+\ii)xv_0\right),
\qquad \epsilon>0,
\label{eq:connection-power-transform}
\end{equation}
where $\Gamma(a,z)$ is the upper incomplete gamma function. After taking $\epsilon\to0^+$, its expansion separates the entropy power from the analytic terms associated with the finite lower endpoint:
\begin{equation}
\begin{aligned}
x^{p_\eta}\ee^{-\frac{\ii\pi}{2}(2-p_\eta)}
\Gamma(2-p_\eta,\ii xv_0)
={}&
\ee^{-\frac{\ii\pi}{2}(2-p_\eta)}
\Gamma(2-p_\eta)x^{p_\eta}
-\sum_{n=0}^{\infty}
\frac{(-\ii)^n v_0^{n+2-p_\eta}}
{n!\,(n+2-p_\eta)}x^{n+2}.
\end{aligned}
\label{eq:connection-power-separation}
\end{equation}
The higher inverse powers in $T_\eta$ generate the descendants $x^{p_\eta+2n}$. Comparing the leading terms with Eq.~\eqref{eq:S-late-bispectrum} yields
\begin{equation}
\begin{aligned}
C_+={}&\ee^{-\frac{\ii\pi}{2}(\frac12+\nu)}
\Gamma\!\left(\frac12+\nu\right)
\frac{W[\rho_\sigma,T_-]}{2\nu},
\\
C_-={}&\ee^{-\frac{\ii\pi}{2}(\frac12-\nu)}
\Gamma\!\left(\frac12-\nu\right)
\frac{W[T_+,\rho_\sigma]}{2\nu}.
\end{aligned}
\label{eq:Xi-explicit-connections}
\end{equation}
The transform phases and the oriented discontinuity together determine the relative phase of the two long branches. At nonresonant masses, these formulas give their coefficients directly in terms of specified Heun values and derivatives.

Resonant masses are obtained by combining the coincident powers before taking the parameter limit. At $\nu=0$, the second infinity solution follows from $(T_+-T_-)/(2\nu)$ and contains $\ln v$. At $\nu=1$, the faster entropy power coincides with the first descendant of the slower branch. At $\nu=1/2$, the pole in $\Gamma(1/2-\nu)$ cancels against the analytic $x^2$ term in Eq.~\eqref{eq:connection-power-separation}, leaving $x^2\ln x$. The full expansions must be retained in these limits, including their analytic terms when needed, so that the cancellations remain explicit.


\subsection{Finite interaction integrals and checks}
\label{app:bispectrum-finite-integrals}
\label{app:bispectrum-numerical}
\label{app:bispectrum-equal-speed}

For light entropy fields, the products of mode functions in the interaction integrals can diverge at the final endpoint even though their imaginary parts are integrable. We establish this cancellation by evaluating the commutators with the final curvature field. At a finite final time $\tau_f$, the canonical commutation relations give
\begin{align}
\operatorname{Im}\sum_{b=1}^2
\varphi_b(k,\tau_f)\bigl[D_\tau\varphi_b(k,\tau_f)\bigr]^*
&=\frac{1}{2a(\tau_f)^2},
\nonumber\\
\operatorname{Im}\sum_{b=1}^2\varphi_b(k,\tau_f)\sigma_b^*(k,\tau_f)
&=\operatorname{Im}\sum_{b=1}^2\varphi_b(k,\tau_f)\sigma_b'^*(k,\tau_f)=0.
\label{eq:finite-appendix-canonical-data}
\end{align}
The first mode equation then evolves the momentum commutator as $\tau^2$ at leading order. Taking $\tau_f\to0^-$ fixes
\begin{equation}
\operatorname{Im}\sum_{b=1}^2
\varphi_b(k,0)\bigl[D_\tau\varphi_b(k,\tau)\bigr]^*
=\frac{1}{2a^2}+\mathcal O(\tau^4)
=\frac{H^2\tau^2}{2}+\mathcal O(\tau^4).
\label{eq:finite-appendix-retarded-powers}
\end{equation}
The imaginary part of the entropy equation is then sourced by $\lambda H^2\tau/2$. A particular solution proportional to $\tau^3$ has a vanishing contribution from $\partial_\tau^2-2\partial_\tau/\tau$, so its amplitude is fixed by the mass term. The homogeneous entropy coefficients needed to impose Eq.~\eqref{eq:finite-appendix-canonical-data} scale as $(-\tau_f)^{3-p_\eta}$ and vanish when $\tau_f\to0^-$, since $\operatorname{Re}p_\eta<3$ for $\mu^2>0$. The final-time commutator therefore retains only the particular solution. Integrating the curvature relation $\varphi'=D_\tau\varphi-\lambda\sigma/\tau$ subsequently gives
\begin{align}
\operatorname{Im}\sum_{b=1}^2
\varphi_b(k,0)\sigma_b^*(k,\tau)
&=\frac{\lambda H^4\tau^3}{2\mu^2}
+\mathcal O(\tau^5),
\label{eq:finite-appendix-entropy-commutator}
\\
\operatorname{Im}\sum_{b=1}^2
\varphi_b(k,0)\varphi_b^*(k,\tau)
&=\frac{H^2\tau^3}{6}
\left(1-\frac{\lambda^2H^2}{\mu^2}\right)
+\mathcal O(\tau^5).
\label{eq:finite-appendix-curvature-commutator}
\end{align}
These coefficients are independent of $k$, consistently with $N_k^2(kc_\varphi)^3=H^2/2$. They also provide initial data for a direct integration of the commutator solution, which avoids subtracting nearly real mode sums numerically.

For $0<\nu<3/2$, put $p=3/2-\nu>0$ and use $x=-kc_\varphi\tau$. The leading real entropy contraction scales as $x^p$, while its imaginary part scales as $x^3$. The real contraction containing $D_\tau\varphi_b$ instead begins at order $x$, and its imaginary part at order $x^2$. Combining these powers shows that the integrands defining $\mathcal M_{\gamma,\eta}$ and $\mathcal M_{\beta,\eta}$ scale at worst as $x^{p_\eta+p-1}$, while that of $\mathcal M_{\alpha,\eta}$ scales as $x^{p_\eta+1}$. All are integrable at the origin for both entropy branches. The cancellation also holds in the full bispectrum at fixed nonzero $q$. For example,
\begin{equation}
\!\!\!\frac1{\tau^4}\operatorname{Im}
\sum_{b,c,d=1}^2
\varphi_b(k_S,0)\varphi_c(k_S,0)\varphi_d(k_L,0)
\sigma_b^*(k_S,\tau)\sigma_c^*(k_S,\tau)
\sigma_d^*(k_L,\tau)
\!=\mathcal O\!\left((-\tau)^{2p-1}\right).
\label{eq:finite-full-gamma-endpoint}
\end{equation}
Separating the complex slow-branch integrals before forming the imaginary part would instead produce powers $x^{3p-4}$ and $x^{2p-2}$ in the $\gamma$ and $\beta$ channels. These cease to be integrable at $\nu\ge1/2$ and $\nu\ge1$, respectively. Taking the imaginary part at a common cutoff therefore preserves a cancellation needed for the positive-mass result. For imaginary $\nu$, the entropy powers have real part $3/2$ and the physical endpoint remains integrable. The logarithms at resonant masses follow from the combined limits described above.

The early endpoint can be evaluated with the real Abel regulator specified in the main text. An equivalent procedure keeps the interval $0<x<R$ on the real axis and deforms only the remaining tails, preserving the final-endpoint cancellation. Let $\mathcal C_-$ run from $R$ clockwise along a quarter-circle to $-\ii R$ and then to $-\ii\infty$, and let $\mathcal C_+$ be its complex conjugate. For the $\gamma$ integral, this gives
\begin{equation}
\begin{aligned}
\mathcal M_{\gamma,\eta}
={}&\frac1{N_k^4}\int_0^R\de x\,x^{p_\eta-4}
\operatorname{Im}\sum_{b,c=1}^2
\varphi_b(k,0)\varphi_c(k,0)
\sigma_b^*\!\left(k,-\frac{x}{kc_\varphi}\right)
\sigma_c^*\!\left(k,-\frac{x}{kc_\varphi}\right)
\\
&+\frac1{2\ii N_k^4}\sum_{b,c=1}^2
\bigg[
\varphi_b(k,0)\varphi_c(k,0)
\int_{\mathcal C_-}\de x\,x^{p_\eta-4}
\sigma_b^*\!\left(k,-\frac{x}{kc_\varphi}\right)
\sigma_c^*\!\left(k,-\frac{x}{kc_\varphi}\right)
\\
&\hspace{8mm}
-\varphi_b^*(k,0)\varphi_c^*(k,0)
\int_{\mathcal C_+}\de x\,x^{p_\eta-4}
\sigma_b\!\left(k,-\frac{x}{kc_\varphi}\right)
\sigma_c\!\left(k,-\frac{x}{kc_\varphi}\right)
\bigg].
\end{aligned}
\label{eq:finite-radius-hard-integral}
\end{equation}
The starred modes on $\mathcal C_-$ mean analytic continuation of their negative-real-time boundary values. In particular, their value at $x=-\ii y$ is $[\sigma_b(k,-\ii y/(kc_\varphi))]^*$. The powers of $x$ are continued from the positive real axis along each path, and the two tails are exponentially damped in their respective half-planes. This form applies also when $p_\eta$ is complex, since the weight remains outside the difference of the two frequency contributions. The $\beta$ and $\alpha$ integrals follow by using the products and powers in Eqs.~\eqref{eq:M-beta-unequal} and \eqref{eq:M-alpha-unequal}. Both quarter-circles are retained, making the complete expression independent of $R$ even when the separate complex integrals would diverge at the origin.

At equal speeds, the correspondence with Ref.~\cite{Huenupi:2026aqc} follows directly from the mode functions already identified in Section~\ref{sec:equal-original-solutions}. The unitarity of $A_b^{(w)}$ converts the contractions into
\begin{align}
\sum_{b=1}^2\varphi_b(k,0)\sigma_b^*(k,\tau)
&=N_k\sum_{w=\pm1}\mathcal U_w
\bigl[\cS_U^{(w)}(k,\tau)\bigr]^*,
\nonumber\\
\sum_{b=1}^2\varphi_b(k,0)
\bigl[D_\tau\varphi_b(k,\tau)\bigr]^*
&=N_k\sum_{w=\pm1}\mathcal U_w
\bigl[D_\tau\cF_U^{(w)}(k,\tau)\bigr]^*.
\label{eq:bispectrum-equal-mode-dictionary}
\end{align}
Substitution into the exact time integrals therefore reproduces the original polarization sums and cubic normalization. Their equilateral values fix $f_{\rm NL}^{(I)}$, while division by those values gives $S_I(k,k,k)=1$. The same comparison determines the squeezed coefficients $b_{I,\eta}$ without introducing a separate normalization of the oscillatory signal.

\paragraph{Numerical comparisons.}

We check the squeezed coefficients by integrating the original coupled field equations with canonically normalized early-time data, independently of the Heun construction. The resulting contractions are continued onto an imaginary-time ray, where a linear boundary-value problem matches their values near the origin to the decaying ultraviolet solutions. For light fields we retain the finite contour turn and its arc, as described above. This gives the complete mode sums needed for the interaction integrals without introducing a large exponential hierarchy through a one-sided integration.

Table~\ref{tab:bispectrum-equal-check} collects the heavy-field comparisons. At equal speeds, evaluation of the original hypergeometric mode integrals of Ref.~\cite{Huenupi:2026aqc} agrees with the independent time-domain calculation to better than $3\times10^{-9}$ in relative oscillatory amplitude and $2\times10^{-9}$ radians in phase. For the unequal-speed examples, the Heun expression for $C_+$ agrees with its extraction from the late-time numerical modes to better than $10^{-9}$ in relative magnitude. The entries describe the full combination $(9/10)f_{\rm NL}^{(I)}S_I$, whose oscillatory amplitude is $|b_{I,+}|/2$, rather than the equilateral-normalized shape alone.

\begin{table}[t]
\centering
\begin{tabular}{ccrrr}
\toprule
Interaction $I$ & $r_c$ & $\Delta_\zeta/\Delta_0^{(c_\varphi)}$ & $|b_{I,+}|/2$ & $\arg b_{I,+}$ [rad]\\
\midrule
$\alpha$ & $1$ & $1.8584262471$ & $0.05038704$ & $-0.91615970$\\
$\beta$  & $1$ & $1.8584262471$ & $0.06883331$ & $1.37920877$\\
$\gamma$ & $1$ & $1.8584262471$ & $0.02101035$ & $-2.17144653$\\
$\gamma$ & $0.5$ & $2.7044641621$ & $0.09158695$ & $-2.14763159$\\
$\gamma$ & $2$ & $1.3866907008$ & $0.003615917$ & $-2.3395754$\\
\bottomrule
\end{tabular}
\caption{Heavy-field squeezed coefficients at $\lambda=1$, $\nu=\ii$ and unit cubic couplings. The amplitude and phase use the convention in Eq.~\eqref{eq:heavy-squeezed-unequal}, including the exact curvature-power normalization.}
\label{tab:bispectrum-equal-check}
\label{tab:heavy-gamma}
\end{table}

A comparison at finite $q$ must also include the regular contribution generated by the analytic long mode. For the heavy $\gamma$ examples in the table, its leading term follows directly from the quadratic long-time response:
\begin{equation}
\begin{aligned}
\frac9{10}f_{\rm NL}^{(\gamma)}S_\gamma^{\rm reg}(q)
={}&-\frac{\gamma\lambda q\,k_Sc_\varphi}
{4(\mu^2/H^2-2)N_{k_S}^4}
\frac{\Delta_0^{(c_\varphi)}}{\Delta_\zeta}
\operatorname{Im}\bigg[
\sum_{b,c=1}^2
\varphi_b(k_S,0)\varphi_c(k_S,0)
\\
&\hspace{12mm}\times
\int_{-\infty}^0
\frac{\de\tau}{(-k_Sc_\varphi\tau)^2}
\sigma_b^*(k_S,\tau)\sigma_c^*(k_S,\tau)
\bigg]+\mathcal O(q^2).
\end{aligned}
\label{eq:heavy-regular-q-coefficient}
\end{equation}
The early endpoint uses the same continuation as the exact bispectrum integrals. After subtracting this term, the difference between the full result and the squeezed oscillation is below $7\times10^{-4}$ of its envelope at $q=0.01$, $0.003$ and $0.001$ for both $r_c=1/2$ and $2$. The residual decreases consistently with a contribution of order $q^2$ to $(9/10)f_{\rm NL}^{(\gamma)}S_\gamma$. Comparing with the envelope avoids dividing by the zeros of the oscillation.

Changing the time-domain tolerance, rotation radius, outer boundary and boundary-value tolerance changes the two unequal-speed amplitudes in the table by less than $3\times10^{-10}$ in absolute value and their phases by less than $3\times10^{-8}$ radians. For the scan in Fig.~\ref{fig:gamma-sound-speed-dependence}, repeated evaluations at $r_c=0.15$, $1$ and $6$ change the amplitude by less than $7\times10^{-9}$ fractionally and the phase by less than $3\times10^{-8}$ radians.

Table~\ref{tab:finite-light-check} gives a light-field example for all three interactions. Varying the contour radius from $0.05$ to $0.2$ and the outer Euclidean boundary from $70$ to $90$ leaves the displayed digits unchanged. An independent real-axis evaluation, completed with an integrated ultraviolet expansion, agrees on $\mathcal M_{I,+}$ to better than $10^{-10}$ in relative terms. The last column compares the exact finite-$q$ result with the leading contribution of both entropy branches. Their relative difference is unchanged by dividing both expressions by the same equilateral amplitude, so it also tests the squeezed approximation to $S_I$.

\begin{table}[t]
\centering
\begin{tabular}{lrrr}
\toprule
$I$ & $\mathcal M_{I,+}$ & $b_{I,+}/4$ & Relative difference\\
\midrule
$\alpha$ & $-6.78625621$ & $0.17087517$ & $4.85\times10^{-4}$\\
$\beta$ & $23.45389274$ & $-0.59055947$ & $2.42\times10^{-4}$\\
$\gamma$ & $-12.89498082$ & $0.32469037$ & $5.96\times10^{-4}$\\
\bottomrule
\end{tabular}
\caption{Light-field comparison at $r_c=0.5$, $\lambda=0.7$, $\nu=0.8$ and unit cubic couplings, with $\Delta_\zeta/\Delta_0^{(c_\varphi)}=7.24381348$ and $C_+=5.28497839$. The coefficient $b_{I,+}/4$ multiplies $q^{1/2-\nu}$ in $(9/10)f_{\rm NL}^{(I)}S_I^{\rm NA}$. The last column is $|S_I-S_I^{\rm NA}|/|S_I|$ at $q=10^{-3}$ and includes the regular and higher-power contributions of the full bispectrum.}
\label{tab:finite-light-check}
\end{table}

\FloatBarrier

\end{document}